\documentclass[journal=jacsat,manuscript=article]{achemso}

\usepackage{chemformula} 
\usepackage[T1]{fontenc} 
\usepackage{amssymb}
\usepackage{amsmath}
\usepackage{tikz}
\usepackage{physics}
\usepackage{hyperref}
\usepackage{cleveref}
\usepackage{mathtools}
\usepackage{amssymb}
\usepackage{mathrsfs}
\usepackage{caption}
\usepackage{subcaption}
\usepackage{comment}
\usepackage{bbm}
\usetikzlibrary{calc}

\newtheorem{remark}{Remark}

\usepackage{graphicx}
\usepackage{epstopdf}

\definecolor{apricot}{rgb}{0.98, 0.81, 0.69}
\definecolor{applegreen}{rgb}{0.55, 0.71, 0.0}
\definecolor{amethyst}{rgb}{0.6, 0.4, 0.8}

\newcommand{\bs}{\boldsymbol{s}}
\newcommand{\e}{\mathrm{e}}
\newcommand{\ff}{\mathrm{f}}
\newcommand{\ii}{\mathrm{i}}
\newcommand{\id}{\mathrm{id}}

\newcommand{\sgn}{\mathrm{sgn}}

\definecolor{AlertRed}{RGB}{175, 20, 20} 
\definecolor{AlertOrange}{RGB}{205, 110, 31} 
\definecolor{AlertGreen}{RGB}{60, 180, 75} 
\definecolor{AlertBlue}{RGB}{51, 51, 179} 
\definecolor{AlertYellow}{RGB}{231, 192, 51} 

\definecolor{spingray}{RGB}{206, 206, 206}
\definecolor{spinteal}{RGB}{89, 168, 156}
\definecolor{spinlightteal}{RGB}{181, 209, 17}
\definecolor{spingold}{RGB}{254, 237, 167}
\definecolor{spinred}{RGB}{224, 43, 53}
\definecolor{spinmedblue}{RGB}{142, 193, 218}
\definecolor{spindarkblue}{RGB}{32, 102, 168}
\definecolor{spinlightblue}{RGB}{205, 225, 236}

\author{Yixiao Sun}
\affiliation{Department of Mathematics, National University of Singapore, Block S17, 10 Lower Kent Ridge Road, Singapore 119076}
\email{s.yixiao@u.nus.edu}
\author{Geshuo Wang}
\affiliation{Department of Mathematics, National University of Singapore, Block S17, 10 Lower Kent Ridge Road, Singapore 119076}
\email{geshuowang@u.nus.edu}
\author{Zhenning Cai}
\affiliation{Department of Mathematics, National University of Singapore, Block S17, 10 Lower Kent Ridge Road, Singapore 119076}
\email{matcz@nus.edu.sg}
\title[An \textsf{achemso} demo]
  {Simulation of Spin Chains with off-diagonal Coupling Using Inchworm Method}

\abbreviations{IR,NMR,UV}

\SectionNumbersOn

\begin{document}


\begin{tocentry}

Some journals require a graphical entry for the Table of Contents.
This should be laid out ``print ready'' so that the sizing of the
text is correct.

Inside the \texttt{tocentry} environment, the font used is Helvetica
8\,pt, as required by \emph{Journal of the American Chemical
Society}.

The surrounding frame is 9\,cm by 3.5\,cm, which is the maximum
permitted for  \emph{Journal of the American Chemical Society}
graphical table of content entries. The box will not resize if the
content is too big: instead it will overflow the edge of the box.

This box and the associated title will always be printed on a
separate page at the end of the document.

\end{tocentry}

\begin{abstract}
We study the dynamical simulation of open quantum spin chain with nearest neighboring coupling, where each spin in the chain is associated with a harmonic bath. 
This is an extension of our previous work [G. Wang and Z. Cai, J. Chem. Theory Comput., 19, 8523--8540, 2023] by generalizing the application of the inchworm method and the technique of modular path integrals from diagonally coupled cases to off-diagonally coupled cases.
Additionally, 
to reduce computational and memory cost in long time simulation,
we apply tensor-train representation to efficiently represent the reduced density matrix of the spin chains, and employ the transfer tensor method (TTM) to avoid exponential growth of computational cost with respect to time.
Abundant numerical experiments are performed to validate our method. 
\end{abstract}

\section{Introduction}
In the real world, quantum systems are never completely isolated.
Instead, they are more or less coupled to the environment,
 which leads to quantum dissipation \cite{esposito2010entropy}.
The theory of open quantum systems therefore attracts significant attention. 
Due to the universality of open quantum systems,
 it has wide applications in different areas,
 such as quantum computation \cite{nielsen2002quantum},
 quantum sensing \cite{correa2017enhancement,salado2021spectroscopy}
 and chemical physics \cite{leggett1987dynamics}.
The evolution of the system becomes a non-Markovian process \cite{cerrillo2014nonmarkovian,strathearn2018efficient} because of the existence of the environment.
Since the memory effect decays over time, 
many methods apply a temporal truncation of the memory kernel.
%
Particularly, in the weak coupling limit, 
the Markovian approximation such as the Lindblad equation \cite{lindblad1976generators} can be employed.

For more general cases where the memory effect is not negligible, we have to consider the non-Markovian process,
 which is the most challenging part of the simulation of the open quantum system.
In the Nakajima-Zwanzig equation \cite{nakajima1958quantum,zwanzig1960ensemble},
a memory kernel is used to represent the temporal effect of the process.
With the decay of the memory effect, 
 the transfer tensor method (TTM) \cite{cerrillo2014nonmarkovian} assumes that the memory length is finite,
 so that it is unnecessary to keep a growing log of history as the solution evolves.
%
TTM provides a general framework for kernel-based non-Markovian processes
but does not provide a specific method to construct the transfer tensors for open quantum systems.
The transfer tensors need to be computed based on dynamical maps within the memory length,
which stand for the maps from the initial state to the state at later times.
This means to apply TTM,
 we need to first solve the non-Markovian process for a period of time using a different method,
 and path integrals are one of the commonly used approaches for this job.

Path integral is an important technique for the simulation of the open quantum system.
It offers a classical-like picture of the quantum process.
Many methods are developed based on the idea of path integral.
The quasi-adiabatic propagator path integral (QuAPI) \cite{makri1995numerical,makri1998quantum} uses the influence functional developed by Feynman and Vernon \cite{feynman1963theory} for the non-Markovian process.
A bottleneck of QuAPI is that it requires storage that grows exponentially with respect to the memory length. 
Based on the QuAPI, many researchers propose improved different methods, such as blip-summed path integral (BSPI) \cite{makri2014blip}, differential equation-based path integral (DEBPI) \cite{wang2022differential}, kink sum method \cite{makri2024kink} to reduce memory cost or accelerate the computation. 
Recently, the small matrix path integral (SMatPI) \cite{makri2020smallMatrixPath,makri2021smallMatrixPathIntegralExtended,wang2024tree} reformulates i-QuAPI by small matrices and successfully reduces the exponential memory cost to a linear scale with respect to the memory length.
%
SMatPI bears a resemblance to TTM and can be considered a technique for calculating the memory kernel within the Nakajima-Zwanzig equation.

The quantum Monte Carlo method \cite{erpenbeck2023quantum} is another frequently used path integral-based tool for simulating open quantum systems. 
In these approaches, system-bath coupling is regarded as a perturbation of the uncorrelated system \cite{xu2017convergence}.
A series of high-dimensional integrals, named after Dyson \cite{dyson1949radiation}, is derived based on Wick's theorem \cite{wick1950evaluation}.
Monte Carlo methods are utilized to estimate high-dimensional integrals over simplices.
Direct Monte Carlo sampling for the integrals suffers from numerical sign problem \cite{loh1990sign,cai2023numerical} due to the oscillatory integrands.
The inchworm Monte Carlo method \cite{chen2017inchwormITheory,chen2017inchwormIIBenchmarks,cai2020inchworm}, utilizing the bold line trick \cite{prokof2007bold,prokof2008bold}, is an efficient method to relieve the numerical sign problem \cite{cai2022numerical}. 
Based on the inchworm Monte Carlo method, some techniques are further introduced to reduce the computational cost \cite{boag2018inclusion,yang2021inclusion,cai2022fast,cai2023bold}.

Recently, researchers have focused on more complicated open quantum systems other than single spin.
For example, the open spin chain model has gained attention due to its wide applications \cite{luck1993critical,hopfield1982neural,meier2003quantum,schneidman2006weak}.
In such systems,
except for the non-Markovian intrinsicality,
another difficulty comes from the representation of the state.
For large systems, the tensor network is a practical tool for the representation of the system state and dynamical map \cite{jorgensen2019exploiting,ye2021constructing,bose2022pairwise,erpenbeck2023tensor}.
For example, the multisite tensor network path integral (MS-TNPI)
utilizes matrix product states (MPS) and matrix product operator (MPO) to reduce memory cost \cite{bose2022multisite,bose2022tensor}.
The density matrix renormalization group (DMRG) algorithm \cite{white1992density,white1993density} is an iterative, variational method extensively employed to compute the ground states of quantum many-body systems. Starting from an initial wave function represented as MPS, DMRG iteratively seeks the ground state $\ket{\psi^\star}$ that minimizes the expectation of energy. 
For a spin-chain system, the process involves ``sweeping'' from left to right and then backward,  gauging the orthogonality center of MPS at each site and replacing it with a tensor that minimizes the energy. 
The overall memory usage is controlled, provided that the maximum bond dimension of $\ket{\psi}$ is uniformly bounded.
Some other researchers propose methods avoiding the representation of the whole system states.
For example, the path integral technique is tweaked for complicated systems, which leads to the modular path integral (MPI) \cite{makri2018modular,makri2018communication,kundu2019modular,kundu2020modular,kundu2021efficient}.
MPI can be further modified to be applied in diagonal coupling spin chains, off-diagonal coupling spin chains \cite{kundu2019modular,kundu2020modular,kundu2021efficient} and branched spin chains \cite{makri2018modular,makri2018communication}.
Combining MPI and inchworm algorithm, Wang and Cai develops a method for the simulation of open quantum spin chain \cite{wang2023real}.

In this paper, we further extend the work of Wang and Cai to a general open spin chain where off-diagonal couplings between spins are involved.
Similar to the previous research, 
the dynamic of the spin chain is decoupled to a single spin problem.
Single spin problem is treated by inchworm algorithm and the result on each spin is carefully connected to form the final dynamic of the chain.
Since the spin-spin coupling becomes more complicated than the diagonal coupling case, the computational cost increases significantly.
In this paper, we apply the transfer tensor method (TTM) by constructing dynamical maps and transfer tensors within a specific memory length.
The introduction of TTM provides an efficient way for long-time simulation.

In \cref{sec_model}, we introduce the model for our method, an open spin chain with Heisenberg-type interaction.
\Cref{sec_diagrammatic_representation} gives a framework of our method with path integral tools to decompose the spin chain problem to single spin problems,
which can be simply solved by the inchworm algorithm.
In \cref{sec_spin_connection}, a recurrence method is proposed to connect single spin quantities to the spin chain ones.
To further reduce memory cost, we introduce a memory truncation and employ the transfer tensor method (TTM) in \cref{sec_TTM} 
so that long-time simulation can be completed by short-time transfer tensors. 
Numerical examples are given in \cref{sec_numerical_results} and a simple conclusion is given in \cref{sec_conclusion}.

\section{Open quantum spin chain model}
\label{sec_model}
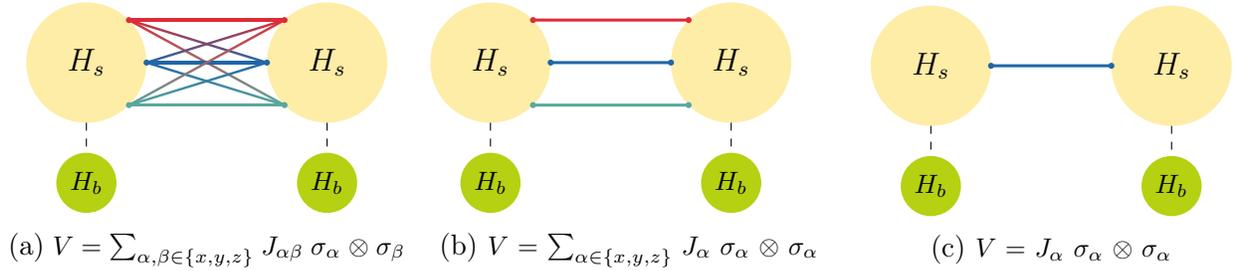
\begin{figure}
\centering
\begin{subfigure}{.32\textwidth}
  \centering
    \begin{tikzpicture}[scale=0.8]
    \pgfmathsetmacro{\offset}{0.02} 
    
        \fill[fill=spingold] (0,0) circle (1cm);
        \node[text=black] at (0,0) {$H_s$};
        \fill[spinred]   ({sqrt(2)*0.5}, {sqrt(2)*0.5}) circle (0.05cm);
        \fill[spindarkblue] (1.0,0.0) circle (0.05cm);
        
        \fill[spinteal]({sqrt(2)*0.5}, {-sqrt(2)*0.5}) circle (0.05cm);

        \fill[fill=spingold] (4,0) circle (1cm);
        \node[text=black] at (4,0) {$H_s$};
        \fill[spinred]({4-sqrt(2)*0.5},{sqrt(2)*0.5}) circle (0.05cm);
        \fill[spindarkblue] (3.0,0.0) circle (0.05cm);
        \fill[spinteal]({4-sqrt(2)*0.5},{-sqrt(2)*0.5}) circle (0.05cm);

    \begin{scope}
        \pgfmathsetmacro{\offset}{0.04} 
        
        \shade[left color = spinred, right color = spinred] ({sqrt(2)*0.5}, {sqrt(2)*0.5-0.5*\offset}) rectangle ({4-sqrt(2)*0.5},{sqrt(2)*0.5+0.5*\offset});
        \shade[rotate around={atan2({-sqrt(2)*0.5}, {3-sqrt(2)*0.5}):({sqrt(2)*0.5},{sqrt(2)*0.5})}, left color=spinred, right color=spindarkblue] ({sqrt(2)*0.5},{sqrt(2)*0.5-0.5*\offset}) rectangle ({sqrt(2)*0.5+sqrt(0.5+(3-sqrt(2)*0.5)*(3-sqrt(2)*0.5))},{sqrt(2)*0.5+0.5*\offset});
        
        \shade[rotate around={atan2({-sqrt(2)}, {4-sqrt(2)}):({sqrt(2)*0.5},{sqrt(2)*0.5})}, left color=spinred, right color=spinteal] ({sqrt(2)*0.5},{sqrt(2)*0.5-0.5*\offset}) rectangle ({sqrt(2)*0.5 + sqrt(2+(4-sqrt(2))*(4-sqrt(2)))},{sqrt(2)*0.5+0.5*\offset});

        \shade[rotate around = {atan2({sqrt(2)*0.5},3-sqrt(2)*0.5):(1,0)}, left color = spindarkblue, right color = spinred] (1, {0-0.5*\offset}) rectangle ({1+sqrt(0.5+(3-sqrt(2)*0.5)*(3-sqrt(2)*0.5))},{0+0.5*\offset});
        \shade[left color = spindarkblue, right color = spindarkblue] (1,{0-0.5*\offset}) rectangle (3,{0+0.5*\offset});
        \shade[rotate around = {atan2({-sqrt(2)*0.5},{3-sqrt(2)*0.5}):(1,0)},left color = spindarkblue, right color = spinteal] (1,-0.5*\offset) rectangle ({1+sqrt((3-sqrt(2)*0.5)*(3-sqrt(2)*0.5)+0.5},0.5*\offset);

        \shade[rotate around = {atan2({sqrt(2)},{4-sqrt(2)}):({sqrt(2)*0.5},{-sqrt(2)*0.5})},left color = spinteal, right color = spinred] ({sqrt(2)*0.5},{-sqrt(2)*0.5-0.5*\offset}) rectangle ({sqrt(2)*0.5 + sqrt(2+(4-sqrt(2))*(4-sqrt(2)))},{-sqrt(2)*0.5+0.5*\offset});
        \shade[rotate around = {atan2({sqrt(2)*0.5},{3-sqrt(2)*0.5}):({sqrt(2)*0.5},{-sqrt(2)*0.5})}, left color = spinteal, right color = spindarkblue] ({sqrt(2)*0.5},{-sqrt(2)*0.5-0.5*\offset}) rectangle ({sqrt(2)*0.5+sqrt(0.5+(3-0.5*sqrt(2))*(3-0.5*sqrt(2))},{-sqrt(2)*0.5+0.5*\offset});
        \shade[left color = spinteal, right color = spinteal]({sqrt(2)*0.5},{-sqrt(2)*0.5-0.5*\offset}) rectangle ({4-sqrt(2)*0.5},{-sqrt(2)*0.5+0.5*\offset});
    \end{scope}
        
        \draw[dashed] (0,-1) -- (0,-2);
        \fill[fill = spinlightteal] (0,-2) circle (0.5cm);
        \node[text=black] at (0,-2) {\footnotesize$H_b$};
        \draw[dashed] (4,-1) -- (4,-2);
        \fill[fill = spinlightteal] (4,-2) circle (0.5cm);
        \node[text=black] at (4,-2) {\footnotesize$H_b$};
        
        \end{tikzpicture}
  \caption{\footnotesize$V$ $=$ $\sum_{\alpha,\beta\in \{x,y,z\}}$ $J_{\alpha\beta}$ $\sigma_{\alpha}$ $\otimes$ $\sigma_{\beta}$}
  \label{diagram_model_9terms}
\end{subfigure}
\hfill
\begin{subfigure}{.32\textwidth}
    \begin{tikzpicture}[scale=0.8]
        \fill[fill=spingold] (0,0) circle (1cm);
        \node[text=black] at (0,0) {$H_s$};
        \fill[spinred]  (0.707106781186548,0.707106781186548) circle (0.05cm);
        \fill[spindarkblue] (1.0,0.0) circle (0.05cm);
        \fill[spinteal](0.707106781186548,-0.707106781186548) circle (0.05cm);
        
        \fill[fill=spingold] (4,0) circle (1cm);
        \node[text=black] at (4,0) {$H_s$};
        \fill[spinred](4-0.707106781186548,0.707106781186548) circle (0.05cm);
        \fill[spindarkblue] (3.0,0.0) circle (0.05cm);
        \fill[spinteal](4-0.707106781186548,-0.707106781186548) circle (0.05cm);
        
        \draw[line width = 0.04cm, color = spinred] (0.707106781186548,0.707106781186548) -- (4-0.707106781186548,0.707106781186548);
        
        \draw[line width = 0.04cm, color = spindarkblue] (1,0) -- (3,0);
    
        \draw[line width = 0.04cm, color = spinteal] (0.707106781186548,-0.707106781186548) -- (4-0.707106781186548,-0.707106781186548);

        \draw[dashed] (0,-1) -- (0,-2);
        \fill[fill = spinlightteal] (0,-2) circle (0.5cm);
        \node[text=black] at (0,-2) {\footnotesize$H_b$};
        \draw[dashed] (4,-1) -- (4,-2);
        \fill[fill = spinlightteal] (4,-2) circle (0.5cm);
        \node[text=black] at (4,-2) {\footnotesize$H_b$};
        
        \end{tikzpicture}
    \caption{\footnotesize$V$ $=$ $\sum_{\alpha\in \{x,y,z\}}$ $J_{\alpha}$ $\sigma_{\alpha}$ $\otimes$ $\sigma_{\alpha}$}
  \label{diagram_model_3terms}
\end{subfigure}
\hfill
\begin{subfigure}{.32\textwidth}
  \centering
  \begin{tikzpicture}[scale=0.8]
        \fill[fill=spingold] (0,0) circle (1cm);
        \node[text=black] at (0,0) {$H_s$};
        \fill[spindarkblue] (1.0,0.0) circle (0.05cm);

        \fill[fill=spingold] (4,0) circle (1cm);
        \node[text=black] at (4,0) {$H_s$};
        \fill[spindarkblue] (3.0,0.0) circle (0.05cm);
        
        \draw[line width = 0.04cm, color = spindarkblue] (1,0) -- (3,0);
       
        \draw[dashed] (0,-1) -- (0,-2);
        \fill[fill = spinlightteal] (0,-2) circle (0.5cm);
        \node[text=black] at (0,-2) {\footnotesize$H_b$};
        \draw[dashed] (4,-1) -- (4,-2);
        \fill[fill = spinlightteal] (4,-2) circle (0.5cm);
        \node[text=black] at (4,-2) {\footnotesize$H_b$};
        
        \end{tikzpicture}
  \caption{\footnotesize$V$ $=$ $J_{\alpha}$ $\sigma_{\alpha}$ $\otimes$ $\sigma_{\alpha}$}
  \label{diagram_model_1term}
\end{subfigure}%

\caption{Illustration of a 2-spin system with different coupling operators.  Each yellow circle stands for a spin and each green circle stands for a harmonic bath. A dashed line between a yellow circle and a green circle stands for the coupling relation between a spin and its bath. The colored dots with solid lines represent the type of $\sigma$'s, red for $\sigma_x$, blue for $\sigma_z$ and green for $\sigma_y$. \cref{diagram_model_9terms} stands for the general form of $V^{(k,k+1)}$ which contains 9 terms, including the ``$\alpha\neq\beta$'' terms (see \cref{rmk_general_model,rmk_connect_general_model}); \cref{diagram_model_3terms} contains 3 terms, which describes the basic interaction form of the Heisenberg model; and \cref{diagram_model_1term} presents a diagonal model where $\sigma_\alpha=\sigma_z$, corresponding to the Ising model discussed in our prior work \cite{wang2023real}.}
\label{diagram_model}
\end{figure}

This section provides a brief introduction to the model studied in this paper, the spin-boson chain, \emph{i.e.}, a list of spin-boson units with nearest-neighbor interactions.
A spin-boson unit is a two-level system coupled with a bosnoic environment, and such a unit is modeled by the Caldeira-Leggett formalism \cite{caldeira1983path}, which offers a simple representation of real dissipative systems, depicting the environment or bath—where the system’s energy dissipates—as an infinite number of harmonic oscillators linearly coupled to the system. When the system alternates between two states, the model is referred to as the spin-boson model, characterized using spin-$\frac{1}{2}$ operators. We consider a spin-boson chain consisting of $K$ spin-boson units with nearest-neighbor couplings between the spins. The Hamiltonian of each spin is given by:
\begin{equation}
    H_s^{(k)}=\epsilon^{(k)} \sigma_z^{(k)}+\Delta^{(k)} \sigma_x^{(k)}
\end{equation}
with $\sigma_x^{(k)}, \sigma_z^{(k)}$ being Pauli matrices for the $k$th spin in the chain. The parameter $\epsilon^{(k)}$ describes the energy difference between two spin states and $\Delta^{(k)}$ is the frequency of the spin flipping of the $k$th spin in the chain. Each spin is associated with its own bosonic bath with the Hamiltonian:
\begin{equation}
\begin{split}
&H_b^{(k)}=\sum_j \frac{1}{2}\left[\left(\hat{p}_j^{(k)}\right)^2+\left(\omega_j^{(k)}\right)^2\left(\hat{q}_j^{(k)}\right)^2\right].
\end{split}
\end{equation}
In this expression, $\hat{p}_j^{(k)}$ and $\hat{q}_j^{(k)}$ are the momentum operator and the position operator of the $j$th harmonic oscillator in the bath of the $k$th spin, respectively, and its oscillating frequency is denoted by $\omega_j^{(k)}$.
The interaction between the spin and its bath is given by the system bath coupling term $W_s^{(k)}\otimes W_b^{(k)}$ with 
\begin{equation}
    W_s^{(k)}=\sigma_z^{(k)}, \quad W_b^{(k)}=\sum_j c_j^{(k)} \hat{q}_j^{(k)}
\end{equation}
Here $c_j^{(k)}$ is the coupling intensity between the $k$th spin and the $j$th oscillator in its bath. Such a coupling causes the dissipation of the energy of each spin to its bath.
These spin-boson systems are combined into a spin chain by nearest-neighbor interactions, described by a coupling operator $V^{(k,k+1)}$ acting on $k$th and $(k+1)$th spins. 
In this paper, we represent it as:
\begin{equation}
\label{eq_neighbor_interaction_general}
    V^{(k,k+1)}=\sum_{\alpha\in A} V_{\alpha}^{(k)}\otimes V_{\alpha}^{(k+1)}
\end{equation}
where $A$ is an index set for the decomposition of the coupling operator.
If $A$ has cardinality 1, the coupling between spins is considered diagonal. 
In particular, when $V^{(k,k+1)}=J_z\sigma_z^{(k)}\otimes \sigma_z^{(k+1)}$, 
the model is generally known as the Ising chain or one-dimensional Ising model in the literature \cite{ising1924beitrag,dani2021quantum,bose2022multisite}.
If the interaction term is not simply a tensor product, \textit{i.e.} $|A|>1$, the coupling is said to be off-diagonal \cite{kundu2019modular}. 
\Cref{diagram_model} shows several types of inter-spin interaction that are commonly considered. 
\Cref{diagram_model_9terms} illustrates the most general case where $|A|=9$;
\Cref{diagram_model_3terms} describes the Heisenberg model \cite{kundu2019modular,kundu2020modular} whose interaction term has the form:
\begin{equation}
\label{eq_heisenberg_coupling}
    V^{(k,k+1)} = J_x\sigma_{x}^{(k)} \otimes \sigma_{x}^{(k+1)}
    + J_y\sigma_{y}^{(k)} \otimes \sigma_{y}^{(k+1)}
    + J_z\sigma_{z}^{(k)} \otimes \sigma_{z}^{(k+1)}
\end{equation}
\Cref{diagram_model_1term} corresponds to the diagonal Ising chain. 
The total Hamiltonian for the whole system-bath is then given by:
\begin{equation}
    \label{eq_H}
    H=
    \sum_{k=1}^K H_0^{(k)}
    +\sum_{k=1}^{K-1} V^{(k,k+1)}
\end{equation}
with 
\begin{equation}
    H_0^{(k)} = H_s^{(k)}
    + H_b^{(k)} + W_s^{(k)} \otimes W_b^{(k)}.
\end{equation}
In this paper, we will mainly focus on the Heisenberg model illustrated by \Cref{diagram_model_3terms}.

We assume that the system does not have any initial entanglement. The initial state of the $k$th spin is $\ket{\varsigma^{(k)}}$, and its associated bath is in a thermal equilibrium state at inverse temperature $\beta^{(k)}$. 
Using $\rho^{(k)}(0) = \rho_s^{(k)}(0) \otimes \rho_b^{(k)}(0)$ to describe the initial density matrix of the $k$th spin-boson unit, the full density matrix of the entire system can be represented by
\begin{equation}
\begin{split}
\rho(0)&=\bigotimes_{k=1}^K \rho^{(k)}(0)=\bigotimes_{k=1}^K\left(\rho_s^{(k)}(0) \otimes \rho_b^{(k)}(0)\right) \\
&=\bigotimes_{k=1}^K\left(\left|\varsigma^{(k)}\right\rangle\left\langle\varsigma^{(k)}\right| \otimes \frac{\exp \left(-\beta^{(k)} H_b^{(k)}\right)}{\operatorname{tr}\left(\exp \left(-\beta^{(k)} H_b^{(k)}\right)\right)}\right).
\end{split}
\end{equation}
The purpose of our work is to obtain the reduced density matrix $\rho_{s}(t) := \tr_b(\rho(t))$, where the operator $\tr_b$ takes the trace over all degrees of freedom of the harmonic oscillators.

\begin{remark}
\label{rmk_general_model}
    In general, the coupling operator $V^{(k,k+1)}$ is required to be Hermitian.
    It can be decomposed into 9 terms as
    \begin{equation}
    \label{eq_spin_interaction_general}
        V^{(k,k+1)} = \sum_{\alpha,\beta \in \{x,y,z\}} 
        J_{\alpha\beta}
        \sigma_\alpha^{(k)} \otimes \sigma_\beta^{(k+1)}.
    \end{equation}
    In this case, a simple idea is to consider $\vert A \vert = 9$ in \cref{eq_neighbor_interaction_general}.
    However, this is not the simplest way to deal with this general model.
    The first step of our method is to transform a spin chain problem into single spin problems.
    On a single spin, there will only be 3 different kinds of interaction operators.
    By allowing the connection of different interaction operators between spins, we will be able to recover this general interaction in \cref{eq_spin_interaction_general}.
    We will also remark on this model in \cref{rmk_connect_general_model} at the end of \cref{sec_spin_connection}.
    However, in this paper, we mainly focus on the Heisenberg model given by \cref{eq_heisenberg_coupling}.
\end{remark}

\begin{remark}
    \label{rmk_multilevel_model}
    For a single spin-$\mathcal{S}$ system, the Hermitian spin operator matrices are of $(2\mathcal{S}+1)\times (2\mathcal{S}+1)$. This system is represented by the Lie algebra $\mathfrak{su}(2S+1)$, which includes $(2S+1)^2 -1$ generators. In these cases, the coupling operator $V^{(k,k+1)}$ can be decomposed into $((2S+1)^2 -1)^2$ terms.
\end{remark}

\section{Inchworm algorithm for spin systems}
\label{sec_diagrammatic_representation}
\subsection{Path-integral formulation of the spin system}
\label{subsec_spinchain_decouple}
In the following analysis,
 we split the total Hamiltonian $H$ into the following two parts:
\begin{equation}
H_0 \coloneqq \sum_{k=1}^K H_0^{(k)}, \quad
V \coloneqq \sum_{k=1}^{K-1} V^{(k,k+1)}.
\end{equation}
Below, we will assume that the interaction between spins $V$ is a perturbation of the unperturbed Hamiltonian $H_0$, and describe the dynamics in the interaction picture.

To this aim, we revisit the integral form of the Liouville-von Neumann equation \cite{breuer2002theory} in the interaction picture:
\begin{equation}\label{eq:liouv_int}
    \rho_I(t)=\rho_I(0)-\mathrm{i} \int_0^t\left[\e^{\ii H_0 s}V \e^{-\ii H_0 s}, \rho_I\left(s\right)\right] \mathrm{d} s
\end{equation}
where $\rho_I(t)=\e^{\ii  H_0 t}\e^{-\ii  H t}\rho(0)\e^{\ii  H t}\e^{-\ii H_0 t}$ denotes the density matrix in the interaction picture, 
and $\e^{\ii H_0 s}V \e^{-\ii H_0 s}$ is the interaction-picture representation of the coupling operator.
We generalize this definition by defining
\begin{equation}\label{op:v_i}
V_I\left(t\right)
    \coloneqq {}  \mathrm{e}^{\mathrm{i} H_0\left|t\right|} V \mathrm{e}^{-\mathrm{i} H_0\left|t\right|} 
    ={}  \sum_{k=1}^{K-1}\sum_{\alpha_k\in A}V_{I,\alpha_k}^{(k)}(t)\otimes V_{I,\alpha_k}^{(k+1)}(t)
\end{equation}
where
\begin{equation}\label{op:v_i_k}
    V_{I,\alpha_k}^{(k)}(s_n):=\mathrm{e}^{\mathrm{i} H_0^{(k)}\left|s_n\right|}V_{\alpha_k}^{(k)}\mathrm{e}^{-\mathrm{i} H_0^{(k)}\left|s_n\right|}.
\end{equation}
Recursively substitute the expression from \cref{eq:liouv_int} back into $\rho_I(t)$ itself, 
we have the following series expansion for $\rho_I(t)$:
\begin{equation}\label{eq:rhoi_dyson}
    \rho_I(t)=\sum_{N=0}^{\infty}
    \int_{-t \leqslant \boldsymbol{s} \leqslant t}
    \left(\prod_{n=1}^N  -\ii \operatorname{sgn}\left(s_n\right)\right) \mathcal{T}\left[V_I\left(s_N\right) \cdots V_I\left(s_1\right) \rho_I(0)\right] \mathrm{d} \boldsymbol{s}
\end{equation}
where $\mathcal{T}$ represents the time-ordering operator that sorts the operators in descending order.
Specifically, given $N$ operators $A_I(t_1),\dots,A_I(t_N)$,
\begin{equation}
    \mathcal{T}[A_I(t_N)\dots A_I(t_1)]
    = A_I(t_{\pi(N)}) \dots A_I(t_{\pi(1)})
\end{equation}
with $\pi$ being the permutation satisfying
\begin{equation}
    t_{\pi(1)} \leqslant t_{\pi(2)} \leqslant \dots \leqslant t_{\pi(N)}.
\end{equation}
The full form of the integrals in (\ref{eq:rhoi_dyson}) is 
\begin{equation}
    \int_{-t\leqslant \boldsymbol{s}\leqslant t} 
    \text{(integrand)}
    \dd\boldsymbol{s}
    = \int_{-t}^{t} 
    \int_{-t}^{s_N}
    \dots
    \int_{-t}^{s_2}
    \text{(integrand)}
    \dd s_1 
    \dots
    \dd s_{N-1}
    \dd s_N
\end{equation}
where the domain is an $N$-dimensional simplex. \Cref{eq:rhoi_dyson} consequently represents a path integral formulation, accumulating contributions from all paths specified by the sequence s$_1, \ldots, s_N$ that connect the initial state to time t, capturing the complete evolution of $\rho_I(t)$.

We can also represent \cref{eq:rhoi_dyson} in the following diagrammatic form
\begin{equation*}
    \rho_I(t)
    = \begin{tikzpicture}[baseline=0]
    \filldraw[fill=black] (-1,-0.1) rectangle (1,0.1);
    \node[text=black,anchor=north] at (-1,0) {$-t$};
    \node[text=black,anchor=north] at (1,0) {$t$};
    \end{tikzpicture}
    +   
    \begin{tikzpicture}[baseline=0]
    \filldraw[fill=black] (-1,-0.1) rectangle (1,0.1);
    \node[text=black,anchor=north] at (-1,0) {$-t$};
    \node[text=black,anchor=north] at (1,0) {$t$};
    \draw plot[only marks, mark=x, mark options={color=applegreen, scale=2, ultra thick}] coordinates {(-0.5,0)};
    \node[text=black,anchor=north] at (-0.5,0) {$s_1$};
    \end{tikzpicture}
    + 
    \begin{tikzpicture}[baseline=0]
    \filldraw[fill=black] (-1,-0.1) rectangle (1,0.1);
    \node[text=black,anchor=north] at (-1,0) {$-t$};
    \node[text=black,anchor=north] at (1,0) {$t$};
    \draw plot[only marks, mark=x, mark options={color=applegreen, scale=2, ultra thick}] coordinates {(-0.5,0)};
    \node[text=black,anchor=north] at (-0.5,0) {$s_1$};
    \draw plot[only marks, mark=x, mark options={color=applegreen, scale=2, ultra thick}] coordinates {(0.3,0)};
    \node[text=black,anchor=north] at (-0.5,0) {$s_1$};
    \node[text=black,anchor=north] at (0.3,0) {$s_2$};
    \end{tikzpicture}
    + \dots
\end{equation*}
where the first diagram corresponding to the term in \cref{eq:rhoi_dyson} with $N=0$.
In the second diagram, the green cross at $s_1$ represents the term with $N=1$ in \cref{eq:rhoi_dyson}.
In the third diagram, two crosses represent the two interaction operators $V_I(s_1)$ and $V_I(s_2)$.
By assuming the initial density matrix is separable, i.e., $\rho(0) = \bigotimes_{k=1}^K \rho^{(k)}(0)$, we can define the density matrix for each specific path:
\begin{equation}
\label{eq_rho_k}
\begin{gathered}
    \rho_I^{(k)}(\boldsymbol{s}^{(k)},\boldsymbol{\alpha}^{(k)})
    = \left(\prod_{n=1}^{N^{(k)}} \sqrt{-\ii\,\sgn(s_{n}^{(k)})} \right)
    \mathcal{T}\left[
    V_{{I,\alpha^{(k)}_{N^{(k)}}}}^{(k)}(s_{N^{(k)}}^{(k)}) \dots V_{I,\alpha_{1}^{(k)}}^{(k)}(s_1^{(k)}) \rho^{(k)}(0)
    \right]\\
    \boldsymbol{\alpha}^{(k)}
    = \left(
        \alpha^{(k)}_{1}, \dots, \alpha^{(k)}_{N^{(k)}}
    \right) \in A^{N^{(k)}}
\end{gathered}
\end{equation}
where $\boldsymbol{s}^{(k)}$ is a subsequence of $\boldsymbol{s} = \left(s_1,s_2,\ldots,s_N\right)$ of length $N^{(k)} \leqslant N$. 
The sequence $\boldsymbol{\alpha}^{(k)}$ has the same length as $\boldsymbol{s}^{(k)}$, with each element $\alpha^{(k)}_{j}\in A$ having $|A|$ possible choices, representing the types of the nearest-spin couplings.
In particular, if $\boldsymbol{s}^{(k)}$ is an empty sequence, we use the notation $\rho_I^{(k)}(\varnothing)\coloneqq \rho^{(k)}(0)$ to denote the above quantity. 
\Cref{eq:rhoi_dyson} sums contributions from all paths, each influenced by $0$ to $\infty$ number of interaction operators.  Additionally, each term in this sum aggregates tensor products of \cref{eq_rho_k}, representing contributions from individual spin systems.
We illustrate the separability of these propagators by showing an example of $N=1$ and $K=4$, where the corresponding summand in (\ref{eq:rhoi_dyson}) has the form:
\begin{equation} \label{eq:rhoi_n1_k4}
\begin{split}
    &\int_{-t}^t
    -\ii \, \sgn(s_1)
    \mathcal{T}\left[V_I(s_1)\rho_{I}(0)\right]
    \dd s_1 \\
    =&\sum_{\alpha_{1}^{(1)}\in A^1}\int_{-t}^t \rho_I^{(1)}(s_1,\alpha_1^{(1)})
     \otimes \rho_I^{(2)}(s_1,\alpha_1^{(1)})
     \otimes \rho^{(3)}(\varnothing)
     \otimes \rho^{(4)}(\varnothing) \dd s_1\\
     +&\sum_{\alpha_1^{(2)}\in A^1}\int_{-t}^t \rho^{(1)}(\varnothing)
     \otimes \rho_I^{(2)}(s_1,\alpha_1^{(2)})
     \otimes \rho_I^{(3)}(s_1,\alpha_1^{(2)})
     \otimes \rho^{(4)}(\varnothing) \dd s_1\\
    +&\sum_{\alpha_1^{(3)}\in A^1}\int_{-t}^t \rho^{(1)}(\varnothing)\otimes\rho^{(2)}(\varnothing)
     \otimes \rho_I^{(3)}(s_1,\alpha_1^{(3)})
     \otimes \rho_I^{(4)}(s_1,\alpha_1^{(3)}) \dd s_1
\end{split}
\end{equation}
We shall use the diagrammatic equations to denote the path integrals as introduced in the previous research \cite{wang2023real} for brevity. 
Consider the scenario where the coupling operators acting on neighboring spins have three distinct types $A=\{a_1,a_2,a_3\}$.
In a four-spin chain, the term with $N=1$ in \cref{eq:rhoi_dyson} can be represented by the following diagrammatic equation. 
\begin{equation} \label{eq_diagrammatic_expansion}
\begin{split}
    \begin{tikzpicture}[baseline=0]
\filldraw[fill=black] (-1,-0.1) rectangle (1,0.1);
\node[text=black,anchor=north] at (-1,0) {$-t$};
\node[text=black,anchor=north] at (1,0) {$t$};
\draw plot[only marks, mark=x, mark options={color=applegreen, scale=2, ultra thick}] coordinates {(-0.5,0)};
\node[text=black,anchor=north] at (-0.5,0) {$s_1$};
\end{tikzpicture}
&= 
\begin{tikzpicture}[baseline=0]
\filldraw[fill=lightgray] (-1,0.55) rectangle (1,0.65);
\filldraw[fill=lightgray] (-1,0.15) rectangle (1,0.25);
\filldraw[fill=lightgray] (-1,-0.25) rectangle (1,-0.15);
\filldraw[fill=lightgray] (-1,-0.65) rectangle (1,-0.55);
\node[text=applegreen] at (-0.5,0.6) {$\times$};
\node[text=applegreen] at (-0.5,0.2) {$\times$};
\draw[applegreen, densely dotted, line width = 1pt] (-0.5,0.6) -- (-0.5,0.2);
\node[text=black,anchor=north] at (-1,-0.6) {$-t$};
\node[text=black,anchor=north] at (1,-0.6) {$t$};
\node[text=black,anchor=north] at (-0.5,-0.6) {$s_1$};
\end{tikzpicture}
+
\begin{tikzpicture}[baseline=0]
\filldraw[fill=lightgray] (-1,0.55) rectangle (1,0.65);
\filldraw[fill=lightgray] (-1,0.15) rectangle (1,0.25);
\filldraw[fill=lightgray] (-1,-0.25) rectangle (1,-0.15);
\filldraw[fill=lightgray] (-1,-0.65) rectangle (1,-0.55);
\node[text=applegreen] at (-0.5,0.2) {$\times$};
\node[text=applegreen] at (-0.5,-0.2) {$\times$};
\draw[applegreen, densely dotted, line width = 1pt] (-0.5,0.2) -- (-0.5,-0.2);
\node[text=black,anchor=north] at (-1,-0.6) {$-t$};
\node[text=black,anchor=north] at (1,-0.6) {$t$};
\node[text=black,anchor=north] at (-0.5,-0.6) {$s_1$};
\end{tikzpicture}
+
\begin{tikzpicture}[baseline=0]
\filldraw[fill=lightgray] (-1,0.55) rectangle (1,0.65);
\filldraw[fill=lightgray] (-1,0.15) rectangle (1,0.25);
\filldraw[fill=lightgray] (-1,-0.25) rectangle (1,-0.15);
\filldraw[fill=lightgray] (-1,-0.65) rectangle (1,-0.55);
\node[text=applegreen] at (-0.5,-0.2) {$\times$};
\node[text=applegreen] at (-0.5,-0.6) {$\times$};
\draw[applegreen, densely dotted, line width = 1pt] (-0.5,-0.6) -- (-0.5,-0.2);
\node[text=black,anchor=north] at (-1,-0.6) {$-t$};
\node[text=black,anchor=north] at (1,-0.6) {$t$};
\node[text=black,anchor=north] at (-0.5,-0.6) {$s_1$};
\end{tikzpicture}
\\
&= 
\begin{tikzpicture}[baseline=0]
\filldraw[fill=lightgray] (-1,0.55) rectangle (1,0.65);
\filldraw[fill=lightgray] (-1,0.15) rectangle (1,0.25);
\filldraw[fill=lightgray] (-1,-0.25) rectangle (1,-0.15);
\filldraw[fill=lightgray] (-1,-0.65) rectangle (1,-0.55);
\node[text=red] at (-0.5,0.6) {$\times$};
\node[text=red] at (-0.5,0.2) {$\times$};
\draw[red, densely dotted, line width = 1pt] (-0.5,0.6) -- (-0.5,0.2);
\node[text=black,anchor=north] at (-1,-0.6) {$-t$};
\node[text=black,anchor=north] at (1,-0.6) {$t$};
\node[text=black,anchor=north] at (-0.5,-0.6) {$s_1$};
\end{tikzpicture}
+
\begin{tikzpicture}[baseline=0]
\filldraw[fill=lightgray] (-1,0.55) rectangle (1,0.65);
\filldraw[fill=lightgray] (-1,0.15) rectangle (1,0.25);
\filldraw[fill=lightgray] (-1,-0.25) rectangle (1,-0.15);
\filldraw[fill=lightgray] (-1,-0.65) rectangle (1,-0.55);
\node[text=blue] at (-0.5,0.6) {$\times$};
\node[text=blue] at (-0.5,0.2) {$\times$};
\draw[blue, densely dotted, line width = 1pt] (-0.5,0.6) -- (-0.5,0.2);
\node[text=black,anchor=north] at (-1,-0.6) {$-t$};
\node[text=black,anchor=north] at (1,-0.6) {$t$};
\node[text=black,anchor=north] at (-0.5,-0.6) {$s_1$};
\end{tikzpicture}
+
\begin{tikzpicture}[baseline=0]
\filldraw[fill=lightgray] (-1,0.55) rectangle (1,0.65);
\filldraw[fill=lightgray] (-1,0.15) rectangle (1,0.25);
\filldraw[fill=lightgray] (-1,-0.25) rectangle (1,-0.15);
\filldraw[fill=lightgray] (-1,-0.65) rectangle (1,-0.55);
\node[text=amethyst] at (-0.5,0.6) {$\times$};
\node[text=amethyst] at (-0.5,0.2) {$\times$};
\draw[amethyst, densely dotted, line width = 1pt] (-0.5,0.6) -- (-0.5,0.2);
\node[text=black,anchor=north] at (-1,-0.6) {$-t$};
\node[text=black,anchor=north] at (1,-0.6) {$t$};
\node[text=black,anchor=north] at (-0.5,-0.6) {$s_1$};
\end{tikzpicture}
+
\\
&\phantom{=} \ \
\begin{tikzpicture}[baseline=0]
\filldraw[fill=lightgray] (-1,0.55) rectangle (1,0.65);
\filldraw[fill=lightgray] (-1,0.15) rectangle (1,0.25);
\filldraw[fill=lightgray] (-1,-0.25) rectangle (1,-0.15);
\filldraw[fill=lightgray] (-1,-0.65) rectangle (1,-0.55);
\node[text=red] at (-0.5,0.2) {$\times$};
\node[text=red] at (-0.5,-0.2) {$\times$};
\draw[black, densely dotted, line width = 1pt] (-0.5,0.2) -- (-0.5,-0.2);
\node[text=black,anchor=north] at (-1,-0.6) {$-t$};
\node[text=black,anchor=north] at (1,-0.6) {$t$};
\node[text=black,anchor=north] at (-0.5,-0.6) {$s_1$};
\end{tikzpicture}
+
\begin{tikzpicture}[baseline=0]
\filldraw[fill=lightgray] (-1,0.55) rectangle (1,0.65);
\filldraw[fill=lightgray] (-1,0.15) rectangle (1,0.25);
\filldraw[fill=lightgray] (-1,-0.25) rectangle (1,-0.15);
\filldraw[fill=lightgray] (-1,-0.65) rectangle (1,-0.55);
\node[text=blue] at (-0.5,0.2) {$\times$};
\node[text=blue] at (-0.5,-0.2) {$\times$};
\draw[blue, densely dotted, line width = 1pt] (-0.5,0.2) -- (-0.5,-0.2);
\node[text=black,anchor=north] at (-1,-0.6) {$-t$};
\node[text=black,anchor=north] at (1,-0.6) {$t$};
\node[text=black,anchor=north] at (-0.5,-0.6) {$s_1$};
\end{tikzpicture}
+
\begin{tikzpicture}[baseline=0]
\filldraw[fill=lightgray] (-1,0.55) rectangle (1,0.65);
\filldraw[fill=lightgray] (-1,0.15) rectangle (1,0.25);
\filldraw[fill=lightgray] (-1,-0.25) rectangle (1,-0.15);
\filldraw[fill=lightgray] (-1,-0.65) rectangle (1,-0.55);
\node[text=amethyst] at (-0.5,0.2) {$\times$};
\node[text=amethyst] at (-0.5,-0.2) {$\times$};
\draw[amethyst, densely dotted, line width = 1pt] (-0.5,0.2) -- (-0.5,-0.2);
\node[text=black,anchor=north] at (-1,-0.6) {$-t$};
\node[text=black,anchor=north] at (1,-0.6) {$t$};
\node[text=black,anchor=north] at (-0.5,-0.6) {$s_1$};
\end{tikzpicture}
+
\\
&\phantom{=} \ \
\begin{tikzpicture}[baseline=0]
\filldraw[fill=lightgray] (-1,0.55) rectangle (1,0.65);
\filldraw[fill=lightgray] (-1,0.15) rectangle (1,0.25);
\filldraw[fill=lightgray] (-1,-0.25) rectangle (1,-0.15);
\filldraw[fill=lightgray] (-1,-0.65) rectangle (1,-0.55);
\node[text=red] at (-0.5,-0.2) {$\times$};
\node[text=red] at (-0.5,-0.6) {$\times$};
\draw[black, densely dotted, line width = 1pt] (-0.5,-0.6) -- (-0.5,-0.2);
\node[text=black,anchor=north] at (-1,-0.6) {$-t$};
\node[text=black,anchor=north] at (1,-0.6) {$t$};
\node[text=black,anchor=north] at (-0.5,-0.6) {$s_1$};
\end{tikzpicture}
+
\begin{tikzpicture}[baseline=0]
\filldraw[fill=lightgray] (-1,0.55) rectangle (1,0.65);
\filldraw[fill=lightgray] (-1,0.15) rectangle (1,0.25);
\filldraw[fill=lightgray] (-1,-0.25) rectangle (1,-0.15);
\filldraw[fill=lightgray] (-1,-0.65) rectangle (1,-0.55);
\node[text=blue] at (-0.5,-0.2) {$\times$};
\node[text=blue] at (-0.5,-0.6) {$\times$};
\draw[blue, densely dotted, line width = 1pt] (-0.5,-0.6) -- (-0.5,-0.2);
\node[text=black,anchor=north] at (-1,-0.6) {$-t$};
\node[text=black,anchor=north] at (1,-0.6) {$t$};
\node[text=black,anchor=north] at (-0.5,-0.6) {$s_1$};
\end{tikzpicture}
+
\begin{tikzpicture}[baseline=0]
\filldraw[fill=lightgray] (-1,0.55) rectangle (1,0.65);
\filldraw[fill=lightgray] (-1,0.15) rectangle (1,0.25);
\filldraw[fill=lightgray] (-1,-0.25) rectangle (1,-0.15);
\filldraw[fill=lightgray] (-1,-0.65) rectangle (1,-0.55);
\node[text=amethyst] at (-0.5,-0.2) {$\times$};
\node[text=amethyst] at (-0.5,-0.6) {$\times$};
\draw[amethyst, densely dotted, line width = 1pt] (-0.5,-0.6) -- (-0.5,-0.2);
\node[text=black,anchor=north] at (-1,-0.6) {$-t$};
\node[text=black,anchor=north] at (1,-0.6) {$t$};
\node[text=black,anchor=north] at (-0.5,-0.6) {$s_1$};
\end{tikzpicture}
\end{split}
\end{equation}%
In the diagrammatic equation above, the solid bold line on the left-hand side represents an operator of all possible types acting on all spins.
The first equality means that the inter-spin interaction can occur in any two neighboring spins, and each gray line stands for a spin.
For the second equality, the crosses with three different colors on the right-hand side denote the coupling operators of different types, $a_1$, $a_2$, and $a_3$. The two gray lines connected by colored crosses represent the two spins involved in the corresponding coupling. Since there are four spins and one-time point $s_1$, each cross can have three possible locations, with each location permitting three choices of coupling types. Consequently, there are $(3\times3)^1$ diagrams on the right-hand side. 
For the $N=2$ term in \cref{eq:rhoi_dyson},
 we may also employ the same idea so that it would be written as the sum of 81 diagrams.

In general, for a chain with $K$ spins, the $N$th term in \cref{eq:rhoi_dyson} can be written as the sum of $\vert A \vert^N (K-1)^N$ diagrams.
In other words, \cref{eq:rhoi_dyson} can be written as the sum of diagrams, and for each diagram, the spin-chain problem is decoupled into the problems on each spin.

In practice, it would be impossible to sum up all the diagrams to infinity.
Since we can now decompose the original problem into problems on a single spin,
we introduce a truncation parameter $\bar{N}$ for the maximum number of interactions neighboring spins, so that each gray line in \cref{eq_diagrammatic_expansion} would have at most $\bar{N}$ crosses of all colors in total.

\subsection{Inchworm algorithm for individual spin systems}
\label{subsec_inchworm_individual_spin}
The total density matrix in the interaction picture $\rho_I(t)$, is decoupled into a sum of spin-specific 
density-matrix-like quantities 
$\rho_I^{(k)}(\boldsymbol{s}^{(k)},\boldsymbol{\alpha}^{(k)})$, 
which exhibit similar structures and are independently characterized for each spin, as indicated \cref{eq:rhoi_dyson}. 
The Hamiltonian in \cref{op:v_i_k} has the form $H_0^{(k)}=H_s^{(k)} + H_b^{(k)} + W_s^{(k)} \otimes W_b^{(k)}$. Given our focus on the dynamics of the reduced density matrix $\rho_{I,s}(t) = \operatorname{tr}_b\left\{\rho_I(t)\right\}$, we apply Dyson series expansion again on $H_0^{(k)}$
to separate the system-bath components within each 
$\rho_I^{(k)}(\boldsymbol{s},\boldsymbol{\alpha})$, assuming $\rho^{(k)}(0)=\rho_s^{(k)}(0)\otimes\rho_b^{(k)}(0)$,
\begin{equation}\label{eq:rhoi_k_dyson}
    \begin{split}
    \rho_{I}^{(k)}(\boldsymbol{s},\boldsymbol{\alpha})
    = &\left(\prod_{n=1}^{N} \sqrt{-\ii\,\sgn(s_{n})} \right)
    \sum_{M=0}^{\infty} (-\mathrm{i})^M\int_{-t\leqslant \boldsymbol{\tau}\leqslant t}\left(\prod_{m=1}^M  \operatorname{sgn}\left(\tau_m\right)\right)\\
    &\times \mathcal{U}_0^{(k)}(\boldsymbol{s},\boldsymbol{\alpha},\boldsymbol{\tau}) \otimes \mathcal{T}\left[W_{b,I}^{(k)}(\tau_1)\ldots W_{b,I}^{(k)}(\tau_M)\rho_b^{(k)}(0)\right] \mathrm{d} \boldsymbol{\tau}
    \end{split}
\end{equation}
 where $\boldsymbol{\tau} = (\tau_1,\dots,\tau_M)$ is an ascending time sequence and
\begin{equation}
    \mathcal{U}_0^{(k)}(\boldsymbol{s},\boldsymbol{\alpha},\boldsymbol{\tau})
    =
    \mathcal{T}\left[
        V_{s,I, \alpha_1}^{(k)}\left(s_1\right) 
        \ldots 
        V_{s, I, \alpha_N}        ^{(k)}\left(s_{N}\right) 
        W_{s,I}^{(k)}\left(\tau_1\right) 
        \ldots 
        W_{s,I}^{(k)}\left(\tau_M\right) \rho_{s}^{(k)}(0)
    \right]
\end{equation}
with
\begin{align}
    V_{s, I,\alpha_{n}}^{(k)}(s)&=\mathrm{e}^{\mathrm{i} H_{s}^{(k)}|s|} V_{\alpha_n}^{(k)} \mathrm{e}^{-\mathrm{i} H_s^{(k)}|s|}\\
    W_{s, I}^{(k)}(\tau)&=\mathrm{e}^{\mathrm{i} H_s^{(k)}|\tau|} W_s^{(k)} \mathrm{e}^{-\mathrm{i} H_s^{(k)}|\tau|}\\
    W_{b,I}^{(k)}(\tau)&=\e^{\ii H_b^{(k)} |\tau|}W_b^{(k)} \mathrm{e}^{-\mathrm{i} H_b^{(k)}|\tau|}
\end{align}
After tracing out the bath components, we obtain the reduced density matrix in the interaction picture represented by the Dyson series:
\begin{equation}
\label{eq:rhos_dyson}
    \begin{split}
    \rho_{s,I}^{(k)}
    \left(\boldsymbol{s},\boldsymbol{\alpha}\right)
    =&\operatorname{tr}_b \left\{\rho_I^{(k)}(\boldsymbol{s},\boldsymbol{\alpha})\right\}\\
    = &
    \left(\prod_{n=1}^{N} \sqrt{-\ii\,\sgn(s_{n})} \right)
    \sum_{M=0}^{\infty} (-\mathrm{i})^M \int_{-t \leqslant \boldsymbol{\tau} \leqslant t}\left(\prod_{m=1}^M \operatorname{sgn}\left(\tau_m\right)\right) \\
    & \times \mathcal{U}_0^{(k)}(\boldsymbol{s},\boldsymbol{\alpha},\boldsymbol{\tau}) \mathcal{L}_b^{(k)}(\boldsymbol{\tau}) \mathrm{d} \boldsymbol{\tau}
    \end{split}
\end{equation}
where $\mathcal{L}_b^{(k)}(\boldsymbol{\tau})$ is the bath influence functional, defined as
\begin{equation}
    \mathcal{L}_b^{(k)}(\boldsymbol{\tau})=\operatorname{tr}_b \left\{\mathcal{T}\left[W_{b,I}^{(k)}(\tau_1)\ldots W_{b,I}^{(k)}(\tau_M)\rho_b^{(k)}(0)\right]\right\}
\end{equation}
It can be rewritten in terms of two-point correlation functions according to Wick's theorem \cite{Negele1988}
\begin{equation}\label{eq:BIF}
\mathcal{L}_b^{(k)}\left(\tau_1, \ldots, \tau_M\right)= 
\begin{dcases}0, & \text { if } M \text { is odd } \\ 
\sum_{\mathfrak{q} \in \mathcal{Q}_M} \prod_{\left(i, j\right) \in \mathfrak{q}} B^{(k)}(\tau_i,\tau_j), & \text { if } M \text { is even }\end{dcases}
\end{equation}
Here $B^{(k)}$ is the two-point correlation function to be specified later in our test cases, and the set $\mathcal{Q}_M$ contains all possible pairings of integers $\{1,2,\cdots, M\}$.
For example, $\mathcal{Q}_4=\left\{\left\{(1,2),(3,4)\right\},\left\{(1,3),(2,4)\right\}\right\}$.
Readers can refer to the literature \cite{dyson1949radiation,cai2020inchworm,cai2023bold,wang2023real} for more information of the the set $\mathcal{Q}_M$.
\footnote{Note that the two-point correlation function in this work is different but related to the two-point correlation in the previous research \cite{cai2020inchworm,cai2022fast,cai2023bold,wang2023real}.
The source of the difference comes from the location of the initial density in the Dyson expansion.
We will further elaborate on their relation later in the appendix.}

Our objective is to evaluate \cref{eq:rhos_dyson} for all spins, all possible time sequence $\boldsymbol{s}$, and corresponding types of coupling operators $\boldsymbol{\alpha}$.
We assume for a fixed non-descending time sequence $\boldsymbol{s}=(s_1,\ldots,s_{N})$ and a specific combination  $\boldsymbol{\alpha}=(\alpha_1,\ldots,\alpha_{N})$, the single propagator $\rho^{(k)}_{s, I}(\boldsymbol{s},\boldsymbol{\alpha})$ can be solved efficiently by the employing the integro-differential equation developed by Wang and Cai \cite{wang2023real}. 

We first generalize the definition of system-associated operator $\mathcal{U}_0^{(k)}(\boldsymbol{s},\boldsymbol{\alpha}, \boldsymbol{\tau})$ to a general form $\Xi_0^{(k)}(s_\ii, \boldsymbol{s},\boldsymbol{\alpha}, \boldsymbol{\tau},s_\ff)$ for $s_\ii <s_\ff$:
\begin{equation}\label{eq_U_generalized}
    \Xi_0^{(k)}(s_\ii,\boldsymbol{s},\boldsymbol{\alpha},\boldsymbol{\tau},s_\ff)
    =\begin{cases}
        \mathcal{T}[V_{s,I,\alpha_1}^{(k)}(s_1) \dots V_{s,I,\alpha_{N}}^{(k)}(s_{N}) W_{s,I}^{(k)}(\tau_1) \dots W_{s,I}^{(k)}(\tau_M) \rho_s^{(k)}(0)],
        &\text{~if~} 0\in[s_\ii,s_\ff], \\
        \mathcal{T}[V_{s,I,\alpha_1}^{(k)}(s_1) \dots V_{s,I,\alpha_{N}}^{(k)}(s_{N}) W_{s,I}^{(k)}(\tau_1) \dots W_{s,I}^{(k)}(\tau_M)],
        &\text{~if~} 0\notin[s_\ii,s_\ff].
    \end{cases}
\end{equation}
Similarly, the density-matrix-like quantity $\rho_I^{(k)}(\boldsymbol{s},\boldsymbol{\alpha})$ is redefined as $\Phi^{(k)}(s_\ii,\boldsymbol{s},\boldsymbol{\alpha},s_\ff)$ by
\begin{equation}
\label{eq_rho_single_spin_generalized}
        \begin{split}
    \Phi^{(k)}\left(s_\ii,\boldsymbol{s},\boldsymbol{\alpha},s_\ff\right)
    = &\left(\prod_{n=1}^{N} \sqrt{-\ii\,\sgn(s_{n})} \right)\sum_{M=0}^{\infty} (-\mathrm{i})^M \int_{s_\ii \leqslant \boldsymbol{\tau} \leqslant s_\ff}\left(\prod_{m=1}^M \operatorname{sgn}\left(\tau_m\right)\right) \\
    & \times \Xi_0^{(k)}(s_\ii,\boldsymbol{s},\boldsymbol{\alpha},\boldsymbol{\tau},s_\ff) \mathcal{L}_b^{(k)}(\boldsymbol{\tau}) \mathrm{d} \boldsymbol{\tau}
    \end{split}
\end{equation}

For any $s_\ii < s_{\text{mid}} < s_\ff$ and the sequences $\boldsymbol{s}_L,\boldsymbol{s}_R,\boldsymbol{\tau}_L,\boldsymbol{\tau}_R$ specified by $s_\ii < \boldsymbol{s}_L, \boldsymbol{\tau}_L < s_{\text{mid}}< \boldsymbol{s}_R, \boldsymbol{\tau}_R < s_{\text{R}}$, we use the notation $\boldsymbol{s} = [\boldsymbol{s}_L, \boldsymbol{s}_R]$ and $\boldsymbol{\tau} = [\boldsymbol{\tau}_L, \boldsymbol{\tau}_R]$ to represent the combined ordered sequences.
By definition, one can observe the following concatenation property of $\Xi^{(k)}_0$:
\begin{equation}
    \Xi_0^{(k)}(s_\ii,[\boldsymbol{s}_L,\boldsymbol{s}_R],[\boldsymbol{\alpha}_L,\boldsymbol{\alpha}_R],[\boldsymbol{\tau}_L,\boldsymbol{\tau}_R],s_\ff) = \Xi_0^{(k)}(s_{\text{mid}},\boldsymbol{s}_{\text{R}},\boldsymbol{\alpha}_{\text{R}},\boldsymbol{\tau}_{\text{R}},s_\ff)\Xi_0^{(k)}(s_\ii,\boldsymbol{s}_{\text{L}},\boldsymbol{\alpha}_{\text{L}},\boldsymbol{\tau}_{\text{L}},s_{\text{mid}})
\end{equation}
Consequently, $\Phi^{(k)}(s_{\text{mid}},\boldsymbol{s}_R,\boldsymbol{\alpha}_R,s_\ff)\Phi^{(k)}(s_\ii,\boldsymbol{s}_L,\boldsymbol{\alpha}_L,s_{\text{mid}})$ is a partial sum of $\Phi^{(k)}(s_\ii,[\boldsymbol{s}_L,\boldsymbol{s}_R],[\boldsymbol{\alpha}_L,\boldsymbol{\alpha}_R],s_\ff)$.


The inchworm algorithm \cite{chen2017inchwormITheory,chen2017inchwormIIBenchmarks,cai2020inchworm,wang2023real} is inspired by this property. It computes longer path integrals by combining shorter path integrals and adding the missing parts. When the extended length is infinitesimal, the inchworm method can be written as an integro-differential equation for $\Phi^{(k)}$, 
\begin{equation}
    \label{inchworm_differential_equation}
    \pdv{\Phi^{(k)}(s_\ii,\boldsymbol{s},\boldsymbol{\alpha},s_\ff)}{s_\ff}
    = \mathcal{K}^{(k)}(s_\ii,\bs,\boldsymbol{\alpha},s_\ff),
\end{equation}
where
\begin{equation}
\label{eq_K}
\begin{split}
    \mathcal{K}^{(k)}(s_\ii,\boldsymbol{s},\boldsymbol{\alpha},s_\ff) = & \sum_{\substack{M=1\\M\text{~is~odd}}}^{\infty} (-\ii)^{M+1}
    \int_{s_\ii \leqslant \tau_1 \leqslant \dots \leqslant \tau_{M} \leqslant s_\ff} 
    \dd \tau_1 \dots \dd \tau_M
    \left( \prod_{m=1}^{M+1}  \, \sgn (\tau_m) \right)\\
    &\times W_s^{(k)}(s_\ff)
    \Xi^{(k)}(s_\ii,\boldsymbol{s},\boldsymbol{\alpha},\boldsymbol{\tau},s_\ff) \mathcal{L}_b^{c(k)}(\boldsymbol{\tau}),
\end{split}
\end{equation}
the sequence $\boldsymbol{\tau}=\left(\tau_1,\ldots,\tau_M,\tau_{M+1}=s_\ff \right)$ is increasing and $\boldsymbol{s}=\left(\boldsymbol{s}_0,\ldots,\boldsymbol{s}_M\right)$ is divided such that their extended sequence $\left(s_\ii,\boldsymbol{s}_0,\tau_1,\boldsymbol{s}_1,\tau_2,\ldots,\tau_{M-1},\boldsymbol{s}_M,\tau_M\right)$ is increasing.
In \cref{eq_K}, the system associated operator $\Xi^{(k)}(s_\ii,\boldsymbol{\alpha},\boldsymbol{s},\boldsymbol{\tau},s_\ff)$ is defined by
\begin{equation}
\label{eq_xi}
    \begin{split}
        \qquad\Xi^{(k)}(s_\ii,\boldsymbol{s},\boldsymbol{\alpha},\boldsymbol{\tau},s_\ff)={}&\Phi^{(k)}(\tau_M,\boldsymbol{s}_M,\boldsymbol{\alpha}_{M},s_\ff)W_s^{(k)}(\tau_M)\Phi^{(k)}(\tau_{M-1},\boldsymbol{s}_{M-1},\boldsymbol{\alpha}_{M-1},\tau_M)\\
        &\times W_s^{(k)}(\tau_{M-1})\cdots W_s^{(k)}(\tau_{1})\Phi^{(k)}(s_\ii,\boldsymbol{s}_0,\boldsymbol{\alpha}_{0},\tau_1),
    \end{split}
\end{equation}
and the bath influence functional $\mathcal{L}_b^{c(k)}$ is given by
\begin{equation}
    \label{eq_L}
    \mathcal{L}_b^{c(k)}(\tau_1,\dots,\tau_{M+1})
    = \sum_{\mathfrak{q}\in\mathcal{Q}_{M+1}^c}
    \prod_{(j,{j'})\in\mathfrak{q}}
    B(\tau_j,\tau_{j'}),
\end{equation}
where $\mathcal{Q}_{M+1}^c$ is the subset of $\mathcal{Q}_{M+1}$ consisting of pairs that are ``connected'', or inchworm proper in some literature \cite{cai2020inchworm,cai2023bold,wang2023real}.
For example, $\mathcal{Q}_{4}^c = \left\{\left\{(1,3),(2,4)\right\}\right\}$. For a fixed $s_\ii$ and $\boldsymbol{s}$, solving \cref{inchworm_differential_equation} requires the following conditions:
\begin{enumerate}
    \item Initial condition: for $s_\ii \not=0$,
    \begin{equation}\label{init1}
    \Phi^{(k)}(s_\ii, s_\ff = s_\ii) = \id^{(k)}.
\end{equation}
    \item Discontinuous condition: for $s_\ff=s_N\not=0$,
    \begin{equation}\label{init2}
\begin{split}
        &\qquad \Phi^{(k)}\left(s_\ii, \left(s_1, \cdots, s_{N}\right),\left(\alpha_{1},\ldots,\alpha_{N}\right), s_\ff = s_{N}\right)\\
        &= \sqrt{-\ii\, \sgn(s_{N})} V_{s,I,\alpha_{N}}^{(k)}(s_{N}) \Phi^{(k)}\left(s_\ii, \left(s_1,\cdots,s_{N-1}\right),\left(\alpha_{1},\ldots,\alpha_{N-1}\right), s_\ff = s_{N}\right).
\end{split}
\end{equation}
    \item Jump condition: at the origin,
    \begin{equation}\label{jump_cond}
    \lim_{s_\ff\rightarrow0^+}
    {\Phi}^{(k)}(s_\ii,\boldsymbol{s},\boldsymbol{\alpha},s_\ff)
    = \rho_s^{(k)}(0)\lim_{s_\ff\rightarrow 0^-}{\Phi}^{(k)}(s_\ii,\boldsymbol{s},\boldsymbol{\alpha},s_\ff).
\end{equation}
\end{enumerate}


 We apply Heun's method for the derivative of $s_\ff$
 and trapezoidal quadrature rule 
 for the high dimensional integrals to solve \cref{inchworm_differential_equation} numerically.

\subsection{Analysis of the computational cost}
\label{subsec_inchworm_comp_cost}
This method offers a computationally efficient framework for evaluating the density matrices. 
While the intricate computational details are beyond the scope of this text, the reader may refer to the referenced work \cite{wang2023real} for an in-depth explanation.
The major difference between our work and previous work \cite{wang2023real} is that we may have more than one type of interaction operator, \emph{i.e.}, different colors of crosses.
The integro-differential equation \cref{inchworm_differential_equation} reveals that the computation of $\Phi$'s with longer time sequences depends on the knowledge of the ones with shorter sequences. An evolution method needs to be applied to compute $\rho_I(t)$. Taking a uniform time step $\Delta t$, the time points $s_n$ and $\tau_m$ are multiple of $\Delta t$. Suppose the simulation of $\rho_I(t)$ is up to $t = \Delta t,\ldots, L\Delta t$ for a given positive integer $L$, and assume that all the diagrams of length less than $l\Delta t$ are already computed. Then for the computational cost of evaluating $\Phi^{(k)}(s_\ii,\boldsymbol{s},\boldsymbol{\alpha},s_\ii + l\Delta t)$ from $\mathcal{K}^{(k)}(s_\ii,\boldsymbol{s},\boldsymbol{\alpha},s_\ii + (l-1)\Delta t)$ 
is $\displaystyle\sum_{M=1,M\text{ is odd}}^{\bar{M}}C_M \binom{M+l}{M}$, 
where $\bar{M}$ is a truncation parameter chosen in evaluating the series \cref{eq_rho_single_spin_generalized} 
and $C_M$ is the computational cost for computing each integrand which is assumed to be a constant since we have $\bar{M}\leqslant 5$ in all our tests. 

The total number of different diagrams of length $l\Delta t$ $\leqslant$ $L\Delta t$ is 
 \begin{equation}
     \sum_{N=0}^{\bar{N}}(2L+1-l)\binom{N+l}{N}|A|^N\leqslant |A|^{\bar{N}}(2L+1-l)\binom{\bar{N}+l+1}{\bar{N}}
 \end{equation}
where $\bar{N}$ is the maximum number of couplings for each spin. The computational cost for the single-spin full propagators of all lengths is 
\begin{equation}
    \sum_{l=1}|A|^{\bar{N}}(2L+1-l)\binom{\bar{N}+l+1}{\bar{N}}\sum_{M=1,\text{ M odd}}^{\bar{M}}C_M\binom{\bar{M}+l}{\bar{M}}\lesssim |A|^{\bar{N}}L^{\bar{M}+\bar{N}+2}.
\end{equation}
In practice, $|A|$, $\bar{M}$ and $\bar{N}$ are relatively small. 
The computational cost scales as $O(L^{\bar{M}+\bar{N}+2})$.
 The computational cost becomes $O(KL^{\bar{M}+\bar{N}+2})$ if the $K$ spins have different parameters.
 
\section{Summation of propagators}
\label{sec_spin_connection}
\subsection{Summation Algorithm}
\label{subsec_summation}
In the previous section,
 an algorithm to decouple a spin chain problem to single-spin problems is proposed.
In this section,
 we connect the single spin information to the full spin chain information by an iterative algorithm as follows:
\begin{align}
    &\mathbbm{p}^{[1]}(-t,\boldsymbol{s}^{(1)},\boldsymbol{\alpha}^{(1)}, t) = \rho_{s,I}^{(1)}(-t,\boldsymbol{s}^{(1)},\boldsymbol{\alpha}^{(1)},t)\quad \text{for } \boldsymbol{\alpha}^{(1)} \in A^{N} \label{eq_summation_propagators_1st_spin} \\
    &\mathbbm{p}^{[k+1]}(-t,\boldsymbol{s}^{(k+1)}, \boldsymbol{\alpha}^{(k+1)},t) = \sum_{N^{(k)}=0}^\infty \sum_{\boldsymbol{\alpha}^{(k)} \in A^{N^{(k)}}} \int_{-t \leqslant \boldsymbol{s}^{(k)}\leqslant t} \mathbbm{p}^{[k]}(-t, \boldsymbol{s}^{(k)},\boldsymbol{\alpha}^{(k)},t) \nonumber \\
    &\qquad\qquad\qquad\qquad\qquad \otimes \rho_{s,I}^{(k+1)}(-t,[\boldsymbol{s}^{(k)},\boldsymbol{s}^{(k+1)}],[\boldsymbol{\alpha}^{(k)},\boldsymbol{\alpha}^{(k+1)}],t)\dd \boldsymbol{s}^{(k)}, \nonumber \\
    &\qquad\qquad\qquad\qquad\qquad \text{for } k=1,\ldots,K-2 \text{ and for } \boldsymbol{\alpha}^{(k+1)} \in A^{N^{(k+1)}} \label{eq_summation_propagators_middle_spins}  \\
    &\mathbbm{p}^{[K]}(-t,t) = \sum_{N^{(K-1)}=0}^\infty \sum_{\boldsymbol{\alpha}^{(K-1)}\in A^{N^{(K-1)}}} \quad \int_{-t\leqslant \boldsymbol{s}^{(K-1)}\leqslant t}\mathbbm{p}^{[K-1]}(-t,\boldsymbol{s}^{(K-1)},\boldsymbol{\alpha}^{(K-1)}, t) \nonumber\\
    &\qquad\qquad\qquad\otimes \rho_{s,I}^{(K)}(-t,\boldsymbol{s}^{(K-1)},\boldsymbol{\alpha}^{(K-1)}, t)\dd \boldsymbol{s}^{(K-1)} \label{eq_summation_propagators_last_spin} 
\end{align}
In this context, $\boldsymbol{s}^{(k)}=\left(s_1^{(k)},\ldots,s_{N^{(k)}}^{(k)}\right)$ for $k=1,\ldots,K-1$ are non-descending sequences. $\boldsymbol{\alpha}^{(k)}$ are sequences of corresponding coupling indices, where the $j$th component of $\boldsymbol{\alpha}^{(k)}$ is the index of the coupling operator acting on $k$th spin at time $s_j$. 


The computational process can be interpreted as iteratively concatenating with adjacent propagators, thereby extending the spin chain. Consider the concatenation of the 2nd spin, for $\boldsymbol{\alpha}^{(2)} \in A^{N^{(2)}}$ and $|A|=3$,
\begin{equation}\label{eq:concate_2nd}
\begin{split}
    \mathbbm{p}^{[2]}(-t,\boldsymbol{s}^{(2)},\boldsymbol{\alpha}^{(2)}, t)
    &=\sum_{N^{(1)}=0}^\infty \sum_{\boldsymbol{\alpha}^{(1)} \in A^{N^{(1)}}}
    \int_{-t \leqslant \boldsymbol{s}^{(1)}\leqslant t} 
    \mathbbm{p}^{[1]}(-t,\boldsymbol{s}^{(1)},\boldsymbol{\alpha}^{(1)}, t) \\
    &\otimes\rho_{s,I}^{(2)}(-t,[\boldsymbol{s}^{(1)},\boldsymbol{s}^{(2)}],[\boldsymbol{\alpha}^{(1)},\boldsymbol{\alpha}^{(2)}],t)\dd \boldsymbol{s}^{(1)},
\end{split}
\end{equation}
For the case of $\boldsymbol{s}^{(2)}=(s_1)$ and $\boldsymbol{\alpha}^{(2)}=(a_1)$, then we can express \cref{eq:concate_2nd} diagrammatically as:
\input{Path_Integral/concate_2nd}%
Here, the open dashed line on the second spin indicates that it will be connected to the third spin in the subsequent iteration.
Two short black lines binding the bold lines indicate that all connections between the first two spins are taken into account.
On the right side, the first diagram is the term for $N^{(1)}=0$ in \cref{eq:concate_2nd};
the second to the fourth diagrams represent the term with $N^{(1)}=1$
and the other nine diagrams stand for the case when $N^{(1)}=2$.
Similarly, for larger $N^{(1)}$, there are $|A|^{N^{(1)}}$ diagrams to be summed.  
We continue this concatenation process iteratively until we connect the last spin as \cref{eq_summation_propagators_last_spin}, 
\begin{equation}
\begin{aligned}
     \begin{tikzpicture}[baseline=0]
        \filldraw[fill=lightgray] (-1,0.55) rectangle (1,0.65);
        \filldraw[fill=lightgray] (-1,0.15) rectangle (1,0.25);
        \filldraw[fill=lightgray] (-1,-0.55) rectangle (1,-0.65);
        \filldraw[fill=lightgray] (-1,-0.15) rectangle (1,-0.25);
        \draw[black, line width=1.5pt] (-1,0.65) -- (-1,-0.65);
        \draw[black, line width=1.5pt] (+1,0.65) -- (+1,-0.65);
    \end{tikzpicture}&= \quad
    \begin{tikzpicture}[baseline=0]
        \filldraw[fill=lightgray] (-1,0.55) rectangle (1,0.65);
        \filldraw[fill=lightgray] (-1,0.15) rectangle (1,0.25);
        \filldraw[fill=lightgray] (-1,-0.15) rectangle (1,-0.25);
        \filldraw[fill=lightgray] (-1,-0.55) rectangle (1,-0.65);
        \draw[black, line width=1.5pt] (-1,0.65) -- (-1,-0.25);
        \draw[black, line width=1.5pt] (+1,0.65) -- (+1,-0.25);
    \end{tikzpicture}
    +
    \begin{tikzpicture}[baseline=0]
        \filldraw[fill=lightgray] (-1,0.55) rectangle (1,0.65);
        \filldraw[fill=lightgray] (-1,0.15) rectangle (1,0.25);
        \filldraw[fill=lightgray] (-1,-0.15) rectangle (1,-0.25);
        \filldraw[fill=lightgray] (-1,-0.55) rectangle (1,-0.65);
        \draw[black, line width=1.5pt] (-1,0.65) -- (-1,-0.25);
        \draw[black, line width=1.5pt] (+1,0.65) -- (+1,-0.25);
        \node[text=red] at (-0.5,0.2-0.4) {$\times$};
        \node[text=red] at (-0.5,-0.2-0.4) {$\times$};
        \draw[black, densely dotted, line width=1pt] (-0.5,0.2-0.4) -- (-0.5,-0.2-0.4);
    \end{tikzpicture}
    +
    \begin{tikzpicture}[baseline=0]
        \filldraw[fill=lightgray] (-1,0.55) rectangle (1,0.65);
        \filldraw[fill=lightgray] (-1,0.15) rectangle (1,0.25);
        \filldraw[fill=lightgray] (-1,-0.15) rectangle (1,-0.25);
        \filldraw[fill=lightgray] (-1,-0.55) rectangle (1,-0.65);
        \draw[black, line width=1.5pt] (-1,0.65) -- (-1,-0.25);
        \draw[black, line width=1.5pt] (+1,0.65) -- (+1,-0.25);
        \node[text=blue] at (-0.5,0.2-0.4) {$\times$};
        \node[text=blue] at (-0.5,-0.2-0.4) {$\times$};
        \draw[black, densely dotted, line width=1pt] (-0.5,0.2-0.4) -- (-0.5,-0.2-0.4);
    \end{tikzpicture}
    +
    \begin{tikzpicture}[baseline=0]
        \filldraw[fill=lightgray] (-1,0.55) rectangle (1,0.65);
        \filldraw[fill=lightgray] (-1,0.15) rectangle (1,0.25);
        \filldraw[fill=lightgray] (-1,-0.15) rectangle (1,-0.25);
        \filldraw[fill=lightgray] (-1,-0.55) rectangle (1,-0.65);
        \draw[black, line width=1.5pt] (-1,0.65) -- (-1,-0.25);
        \draw[black, line width=1.5pt] (+1,0.65) -- (+1,-0.25);
        \node[text=amethyst] at (-0.5,0.2-0.4) {$\times$};
        \node[text=amethyst] at (-0.5,-0.2-0.4) {$\times$};
        \draw[black, densely dotted, line width=1pt] (-0.5,0.2-0.4) -- (-0.5,-0.2-0.4);
    \end{tikzpicture}\\
    &\quad +
    \begin{tikzpicture}[baseline=0]
        \filldraw[fill=lightgray] (-1,0.55) rectangle (1,0.65);
        \filldraw[fill=lightgray] (-1,0.15) rectangle (1,0.25);
        \filldraw[fill=lightgray] (-1,-0.15) rectangle (1,-0.25);
        \filldraw[fill=lightgray] (-1,-0.55) rectangle (1,-0.65);
        \draw[black, line width=1.5pt] (-1,0.65) -- (-1,-0.25);
        \draw[black, line width=1.5pt] (+1,0.65) -- (+1,-0.25);
        \node[text=red] at (-0.5,0.2-0.4) {$\times$};
        \node[text=red] at (-0.5,-0.2-0.4) {$\times$};
        \draw[black, densely dotted, line width=1pt] (-0.5,0.2-0.4) -- (-0.5,-0.2-0.4);
        \node[text=red] at (0.2,0.2-0.4) {$\times$};
        \node[text=red] at (0.2,-0.2-0.4) {$\times$};
        \draw[black, densely dotted, line width=1pt] (0.2,0.2-0.4) -- (0.2,-0.2-0.4);
    \end{tikzpicture}
    +
    \begin{tikzpicture}[baseline=0]
        \filldraw[fill=lightgray] (-1,0.55) rectangle (1,0.65);
        \filldraw[fill=lightgray] (-1,0.15) rectangle (1,0.25);
        \filldraw[fill=lightgray] (-1,-0.15) rectangle (1,-0.25);
        \filldraw[fill=lightgray] (-1,-0.55) rectangle (1,-0.65);
        \draw[black, line width=1.5pt] (-1,0.65) -- (-1,-0.25);
        \draw[black, line width=1.5pt] (+1,0.65) -- (+1,-0.25);
        \node[text=red] at (-0.5,0.2-0.4) {$\times$};
        \node[text=red] at (-0.5,-0.2-0.4) {$\times$};
        \draw[black, densely dotted, line width=1pt] (-0.5,0.2-0.4) -- (-0.5,-0.2-0.4);
        \node[text=blue] at (0.2,0.2-0.4) {$\times$};
        \node[text=blue] at (0.2,-0.2-0.4) {$\times$};
        \draw[black, densely dotted, line width=1pt] (0.2,0.2-0.4) -- (0.2,-0.2-0.4);
    \end{tikzpicture}
    +
    \begin{tikzpicture}[baseline=0]
        \filldraw[fill=lightgray] (-1,0.55) rectangle (1,0.65);
        \filldraw[fill=lightgray] (-1,0.15) rectangle (1,0.25);
        \filldraw[fill=lightgray] (-1,-0.15) rectangle (1,-0.25);
        \filldraw[fill=lightgray] (-1,-0.55) rectangle (1,-0.65);
        \draw[black, line width=1.5pt] (-1,0.65) -- (-1,-0.25);
        \draw[black, line width=1.5pt] (+1,0.65) -- (+1,-0.25);
        \node[text=red] at (-0.5,0.2-0.4) {$\times$};
        \node[text=red] at (-0.5,-0.2-0.4) {$\times$};
        \draw[black, densely dotted, line width=1pt] (-0.5,0.2-0.4) -- (-0.5,-0.2-0.4);
        \node[text=amethyst] at (0.2,0.2-0.4) {$\times$};
        \node[text=amethyst] at (0.2,-0.2-0.4) {$\times$};
        \draw[black, densely dotted, line width=1pt] (0.2,0.2-0.4) -- (0.2,-0.2-0.4);
    \end{tikzpicture}
    +
    \begin{tikzpicture}[baseline=0]
        \filldraw[fill=lightgray] (-1,0.55) rectangle (1,0.65);
        \filldraw[fill=lightgray] (-1,0.15) rectangle (1,0.25);
        \filldraw[fill=lightgray] (-1,-0.15) rectangle (1,-0.25);
        \filldraw[fill=lightgray] (-1,-0.55) rectangle (1,-0.65);
        \draw[black, line width=1.5pt] (-1,0.65) -- (-1,-0.25);
        \draw[black, line width=1.5pt] (+1,0.65) -- (+1,-0.25);
        \node[text=blue] at (-0.5,0.2-0.4) {$\times$};
        \node[text=blue] at (-0.5,-0.2-0.4) {$\times$};
        \draw[black, densely dotted, line width=1pt] (-0.5,0.2-0.4) -- (-0.5,-0.2-0.4);
        \node[text=red] at (0.2,0.2-0.4) {$\times$};
        \node[text=red] at (0.2,-0.2-0.4) {$\times$};
        \draw[black, densely dotted, line width=1pt] (0.2,0.2-0.4) -- (0.2,-0.2-0.4);
    \end{tikzpicture}+\cdots\\
    &\quad +    \begin{tikzpicture}[baseline=0]
        \filldraw[fill=lightgray] (-1,0.55) rectangle (1,0.65);
        \filldraw[fill=lightgray] (-1,0.15) rectangle (1,0.25);
        \filldraw[fill=lightgray] (-1,-0.15) rectangle (1,-0.25);
        \filldraw[fill=lightgray] (-1,-0.55) rectangle (1,-0.65);
        \draw[black, line width=1.5pt] (-1,0.65) -- (-1,-0.25);
        \draw[black, line width=1.5pt] (+1,0.65) -- (+1,-0.25);
        \node[text=red] at (-0.5,0.2-0.4) {$\times$};
        \node[text=red] at (-0.5,-0.2-0.4) {$\times$};
        \draw[black, densely dotted, line width=1pt] (-0.5,0.2-0.4) -- (-0.5,-0.2-0.4);
        \node[text=red] at (0.2,0.2-0.4) {$\times$};
        \node[text=red] at (0.2,-0.2-0.4) {$\times$};
        \draw[black, densely dotted, line width=1pt] (0.2,0.2-0.4) -- (0.2,-0.2-0.4);
        \node[text=red] at (0.7,0.2-0.4) {$\times$};
        \node[text=red] at (0.7,-0.2-0.4) {$\times$};
        \draw[black, densely dotted, line width=1pt] (0.7,0.2-0.4) -- (0.7,-0.2-0.4);
    \end{tikzpicture}
    +    \begin{tikzpicture}[baseline=0]
        \filldraw[fill=lightgray] (-1,0.55) rectangle (1,0.65);
        \filldraw[fill=lightgray] (-1,0.15) rectangle (1,0.25);
        \filldraw[fill=lightgray] (-1,-0.15) rectangle (1,-0.25);
        \filldraw[fill=lightgray] (-1,-0.55) rectangle (1,-0.65);
        \draw[black, line width=1.5pt] (-1,0.65) -- (-1,-0.25);
        \draw[black, line width=1.5pt] (+1,0.65) -- (+1,-0.25);
        \node[text=red] at (-0.5,0.2-0.4) {$\times$};
        \node[text=red] at (-0.5,-0.2-0.4) {$\times$};
        \draw[black, densely dotted, line width=1pt] (-0.5,0.2-0.4) -- (-0.5,-0.2-0.4);
        \node[text=red] at (0.2,0.2-0.4) {$\times$};
        \node[text=red] at (0.2,-0.2-0.4) {$\times$};
        \draw[black, densely dotted, line width=1pt] (0.2,0.2-0.4) -- (0.2,-0.2-0.4);
        \node[text=blue] at (0.7,0.2-0.4) {$\times$};
        \node[text=blue] at (0.7,-0.2-0.4) {$\times$};
        \draw[black, densely dotted, line width=1pt] (0.7,0.2-0.4) -- (0.7,-0.2-0.4);
    \end{tikzpicture}
    +    \begin{tikzpicture}[baseline=0]
        \filldraw[fill=lightgray] (-1,0.55) rectangle (1,0.65);
        \filldraw[fill=lightgray] (-1,0.15) rectangle (1,0.25);
        \filldraw[fill=lightgray] (-1,-0.15) rectangle (1,-0.25);
        \filldraw[fill=lightgray] (-1,-0.55) rectangle (1,-0.65);
        \draw[black, line width=1.5pt] (-1,0.65) -- (-1,-0.25);
        \draw[black, line width=1.5pt] (+1,0.65) -- (+1,-0.25);
        \node[text=red] at (-0.5,0.2-0.4) {$\times$};
        \node[text=red] at (-0.5,-0.2-0.4) {$\times$};
        \draw[black, densely dotted, line width=1pt] (-0.5,0.2-0.4) -- (-0.5,-0.2-0.4);
        \node[text=red] at (0.2,0.2-0.4) {$\times$};
        \node[text=red] at (0.2,-0.2-0.4) {$\times$};
        \draw[black, densely dotted, line width=1pt] (0.2,0.2-0.4) -- (0.2,-0.2-0.4);
        \node[text=amethyst] at (0.7,0.2-0.4) {$\times$};
        \node[text=amethyst] at (0.7,-0.2-0.4) {$\times$};
        \draw[black, densely dotted, line width=1pt] (0.7,0.2-0.4) -- (0.7,-0.2-0.4);
    \end{tikzpicture}
    + \begin{tikzpicture}[baseline=0]
        \filldraw[fill=lightgray] (-1,0.55) rectangle (1,0.65);
        \filldraw[fill=lightgray] (-1,0.15) rectangle (1,0.25);
        \filldraw[fill=lightgray] (-1,-0.15) rectangle (1,-0.25);
        \filldraw[fill=lightgray] (-1,-0.55) rectangle (1,-0.65);
        \draw[black, line width=1.5pt] (-1,0.65) -- (-1,-0.25);
        \draw[black, line width=1.5pt] (+1,0.65) -- (+1,-0.25);
        \node[text=red] at (-0.5,0.2-0.4) {$\times$};
        \node[text=red] at (-0.5,-0.2-0.4) {$\times$};
        \draw[black, densely dotted, line width=1pt] (-0.5,0.2-0.4) -- (-0.5,-0.2-0.4);
        \node[text=blue] at (0.2,0.2-0.4) {$\times$};
        \node[text=blue] at (0.2,-0.2-0.4) {$\times$};
        \draw[black, densely dotted, line width=1pt] (0.2,0.2-0.4) -- (0.2,-0.2-0.4);
        \node[text=red] at (0.7,0.2-0.4) {$\times$};
        \node[text=red] at (0.7,-0.2-0.4) {$\times$};
        \draw[black, densely dotted, line width=1pt] (0.7,0.2-0.4) -- (0.7,-0.2-0.4);
    \end{tikzpicture}
    + \begin{tikzpicture}[baseline=0]
        \filldraw[fill=lightgray] (-1,0.55) rectangle (1,0.65);
        \filldraw[fill=lightgray] (-1,0.15) rectangle (1,0.25);
        \filldraw[fill=lightgray] (-1,-0.15) rectangle (1,-0.25);
        \filldraw[fill=lightgray] (-1,-0.55) rectangle (1,-0.65);
        \draw[black, line width=1.5pt] (-1,0.65) -- (-1,-0.25);
        \draw[black, line width=1.5pt] (+1,0.65) -- (+1,-0.25);
        \node[text=red] at (-0.5,0.2-0.4) {$\times$};
        \node[text=red] at (-0.5,-0.2-0.4) {$\times$};
        \draw[black, densely dotted, line width=1pt] (-0.5,0.2-0.4) -- (-0.5,-0.2-0.4);
        \node[text=blue] at (0.2,0.2-0.4) {$\times$};
        \node[text=blue] at (0.2,-0.2-0.4) {$\times$};
        \draw[black, densely dotted, line width=1pt] (0.2,0.2-0.4) -- (0.2,-0.2-0.4);
        \node[text=blue] at (0.7,0.2-0.4) {$\times$};
        \node[text=blue] at (0.7,-0.2-0.4) {$\times$};
        \draw[black, densely dotted, line width=1pt] (0.7,0.2-0.4) -- (0.7,-0.2-0.4);
    \end{tikzpicture}+\cdots
\end{aligned}
\end{equation}


The process is similar to previous research \cite{wang2023real}; however, there are two main differences:
\begin{itemize}
    \item In the previous research \cite{wang2023real}, a specific observable is computed, and taking a trace is therefore required.
    In particular, taking the trace over the full spin chain can be separated into the traces of each spin, involving scalar operations with a memory cost of $\mathcal{O}(1)$.
    In this research, however, we use a propagator for each spin so that multiplications become Kronecker products in our work, leading to an exponential increase in memory cost, $\mathcal{O}(4^K)$. 
    The change, although it brings extra computational and memory cost, 
    allows further application of the transfer tensor method (TTM), 
    which will be introduced in \cref{sec_TTM}.
    \item In addition to the match of time sequences $\boldsymbol{s}$, the types of couplings $\boldsymbol{\alpha}$ should also be matched and summed due to the off-diagonal coupling.
    In particular, when $\vert A \vert = 1$, \emph{i.e.}, the diagonal coupling case, the indices $\boldsymbol{\alpha}$ will automatically match each other.
\end{itemize}

\begin{remark}
\label{rmk_connect_general_model}
    In \cref{rmk_general_model}, we discussed the most general inter-spin couplings. 
    In prevalent spin-$\frac{1}{2}$ chain models, such as the Heisenberg and the Frenkel-Exciton models, the interactions typically take the form of $\sigma_{\alpha}^{(k)} \otimes\sigma_{\alpha}^{(k+1)}$, as depicted in \cref{diagram_model_3terms}. Consequently, it is unnecessary to account for nine different colors of “crosses” for each spin, as only three types of interactions occur.
    However, our framework allows for the connection of differently colored crosses during summation steps. Specifically, the $V_I(s_n)$ in \cref{op:v_i} should be included with another subscript $\beta_{k+1}$ to reflect the couplings with the preceding spin, 
    \begin{equation*}
        V_I(s_n)=\sum_{k=1}^{K-1}\sum_{\alpha_k\in A}\sum_{\beta_{k+1}\in A}V_{I,\alpha_k}^{(k)}(s_n)\otimes V_{I,\beta_{k+1}}^{(k+1)}(s_n)
    \end{equation*}
    Similarly, the single propagators $\rho^{(k)}(\boldsymbol{s}^{(k)},\boldsymbol{\alpha}^{(k)})$ in \cref{eq_rho_k} are extended to $\rho^{(k)}(\boldsymbol{s}^{(k)},\boldsymbol{\alpha}^{(k)},\boldsymbol{\beta}^{(k)})$. 
    The concatenation process in \cref{eq_summation_propagators_middle_spins} should be modified accordingly, another summation over $\boldsymbol{\beta}^{(k+1)}$ is required, and $ \mathbbm{p}^{[k]}(-t, \boldsymbol{s}^{(k)},\boldsymbol{\alpha}^{(k)},t)$ should be tensor-product with $\rho_{s,I}^{(k+1)}(-t,$ $[\boldsymbol{s}^{(k+1)},\boldsymbol{s}^{(k)}],$ $[\boldsymbol{\alpha}^{(k)},\boldsymbol{\alpha}^{(k+1)}],$ $[\boldsymbol{\beta}^{(k+1)}],t)$. This ensures that the coupling $\sigma_{\alpha^{(k)}}^{(k)} \otimes \sigma_{\beta^{(k+1)}}^{(k+1)}$ along sequence $\boldsymbol{s}^{(k)}$ is matched between the $k$th and the $(k+1)$th spin.
    This adaptation allows our approach to be applied to general coupling cases.
\end{remark}
\subsection{Analysis of the computational cost}
\label{subsec_summation_cost}
The computational cost of propagators $\rho_{s,I}^{(k)}(-t,\boldsymbol{s},\boldsymbol{\alpha},t)$ has been estimated as $O(L^{\bar{N}+\bar{M}+2})$, as discussed at the end of \Cref{sec_diagrammatic_representation}. We now evaluate the computational cost for calculating all possible $\mathbbm{p}^{[k+1]}(-t,\boldsymbol{s}^{(k+1)},\boldsymbol{\alpha}^{(k+1)},t)$ at time $t=l\Delta t$ for sequences $\boldsymbol{s}^{(k+1)}$ of length at most $\bar{N}$. This is based on the already computed $\rho_{s,I}^{(k+1)}(-t,[\boldsymbol{s}^{(k)},\boldsymbol{s}^{(k+1)}],[\boldsymbol{\alpha}^{(k)},\boldsymbol{\alpha}^{(k+1)}],t)$ for $N'\leqslant \bar{N}-N$, as outlined in \cref{eq_summation_propagators_middle_spins},
\begin{equation}
\quad\sum_{N=0}^{\bar{N}}\binom{2l+N}{N}|A|^{N}\sum_{N'=0}^{\bar{N}-N}\binom{2l+N'}{N'}|A|^{N'}\lesssim  \bar{N}|A|^{\bar{N}}l^{\bar{N}}
\end{equation}
Given that $|A|$ is typically small, the cost to compute $\mathbbm{p}^{[K]}(-t,t)$, which approximates $\rho_{s,I}(t)$ up to $t=L\Delta t$, is $O(L^{\bar{N}+1})$. Consequently, the overall computational cost of our inchworm framework is $O(L^{\bar{N}+\bar{M}+2})$.

\begin{figure}
    \centering
    \begin{subfigure}[b]{0.3\textwidth}
         \centering
         \includegraphics[width=\textwidth]{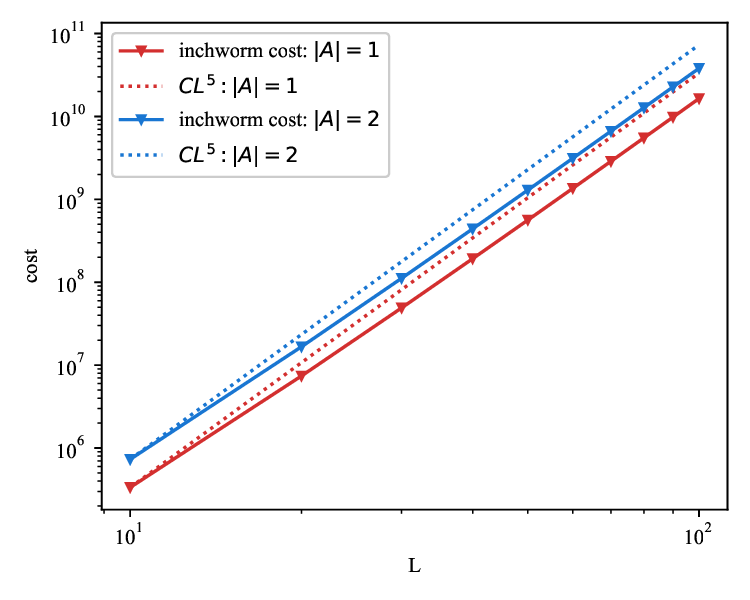}
         \caption{Costs of inchworm.}
         \label{fig_comp_cost_inchworm}
     \end{subfigure}
     \hfill
    \begin{subfigure}[b]{0.3\textwidth}
         \centering
         \includegraphics[width=\textwidth]{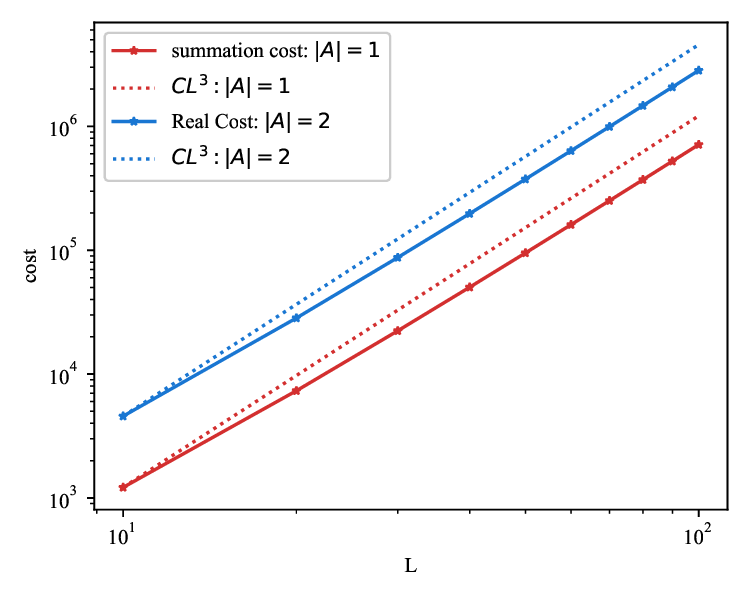}
         \caption{Costs of summation.}
         \label{fig_comp_cost_summation}
     \end{subfigure}
     \hfill
   \begin{subfigure}[b]{0.3\textwidth}
         \centering
         \includegraphics[width=\textwidth]{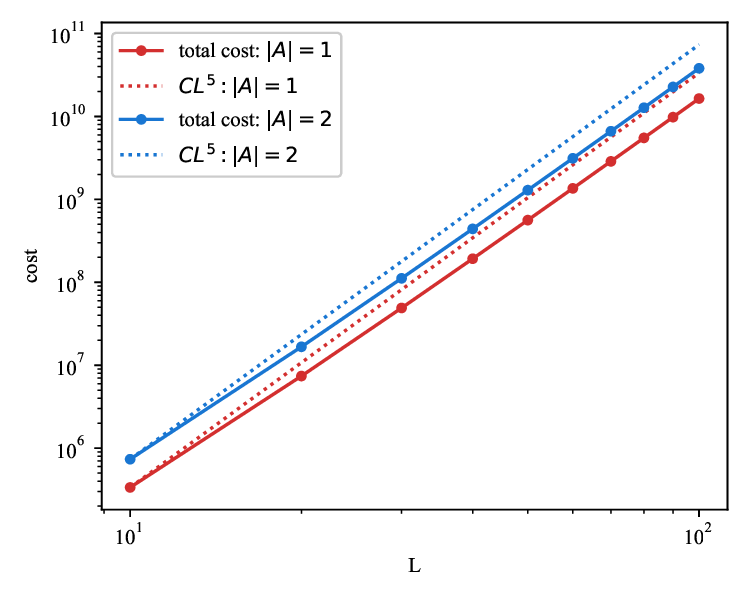}
         \caption{Total costs.}
         \label{fig_comp_cost_total}
     \end{subfigure}
    \caption{Computational costs for different $L$ ($\bar{M}=1$, $\bar{N}=2$, $|A|=1$ and $|A|=2$).}
    \label{fig_comp_cost}
\end{figure}

We demonstrate this asymptotic behavior in the following experiment \cref{fig_comp_cost_summation}. Fixing $\bar{M}=1$ and $\bar{N}=2$, we evaluate the computational cost for a $2$-spin chain with identical spin parameters. This requires running the inchworm algorithm once to solve the \cref{inchworm_differential_equation} for a single spin, followed by concatenating these two spins using \cref{eq_summation_propagators_last_spin}.
We compare the costs for different numbers of non-diagonal spin coupling operators $|A|=1$ and $|A|=2$. The costs associated with the inchworm algorithm, the summation process, and the overall computational costs are displayed in \cref{fig_comp_cost_inchworm}, \cref{fig_comp_cost_summation}, and \cref{fig_comp_cost_total}, respectively. The logarithmic costs plotted for both $|A|=1$ and $|A|=2$ are parallel, indicating a consistent increase with $L$ for both scenarios, showing that the trend of growth aligns with our analysis.

\begin{remark}
    \label{rmk_computation_cost}
    We would like to comment on the difference in the computational cost of our method with the previous work on diagonal coupling cases. The only increase is the cost of matching the ``crosses'' with different colors. For diagonal coupling cases $|A|=1$, the total computational cost scales as $O(L^{\bar{M}+\bar{N}+2})$. In our experiments, $|A|$ is at most 3.
    While the presence of $|A|$ increases the computational cost, it does so without altering the polynomial scaling with $L$, as the primary complexity still scales with $L^{\bar{M}+\bar{N}+2}$, maintaining a similar order of magnitude.
    We also compare the computational cost of our method and that of the factorized MPI methodology \cite{kundu2021efficient}. 
    The procedure involves sequential matrix-vector multiplications and evaluates the path sum separately. For diagonal coupling, the number of operations per unit is $(4L+5)2^{2L+2}$; for the non-diagonal case, where the linking of non-terminal units is predominant, which needs $4(L+4)2^{2L+4}$ per unit\cite{kundu2021efficient}. Such costs scale exponentially as the number of time steps increases. Our method mitigates this exponential increase by employing truncation in path integrals and limiting computations to paths where the number of inter-spin interactions does not exceed $\bar{N}$. However, our approach relies on the tensor-product decomposition of non-diagonal couplings. As the number of decompositions, denoted by $|A|$, increases, so does the computational cost. Unlike our method, the computational cost of the MPI does not change with different types of non-diagonal couplings. 
\end{remark}

\section{Long Time Simulation Using Inchworm Method Coupled with Transfer Tensor Method} 
\label{sec_TTM}
In previous sections,
 a method for the computation of reduced density matrix
 based on the inchworm algorithm is introduced.
This method can be regarded as a natural extension of Wang and Cai's work \cite{wang2023real} to the off-diagonal coupling case.
Its computational cost, however, increases fast with respect to simulation time.
For long-time simulation, the estimation of integrals will be expensive.
The Transfer Tensor Method (TTM) introduced by Cerrillo and Cao \cite{cerrillo2014nonmarkovian} offers a general approach for managing the simulation of large quantum systems over extended periods. TTM enables long-time evolution simulations by utilizing previously extracted exact method data. The computational cost of TTM is comparable to solving the discrete Nakajima-Zwanzig equation, which describes the relationship between the reduced density matrix and the ``memory kernel'' due to the system-bath interactions. Given that system-bath interactions are generally not excessively strong in most cases, TTM's nature of saving computational costs makes it highly compatible with various numerically exact methods, such as the exact hierarchical equations of motion (HEOM) method \cite{gelzinis2017applicability} and trotter-based surface-hopping (TBSH)-based methods \cite{kananenka2016accurate}.


In general, TTM can be applied in any open quantum system
 so we first consider a general open quantum system without any specific setting. The non-Markovian open quantum trajectories generate dynamical maps that propagate the reduced density matrix
\begin{equation}
    \rho_s(t_k) = \mathcal{E}_k \rho_s(0).
\end{equation}
Here $t_k = k\delta t$ with $\delta t$ being the sums of simulation time steps.
In particular, we have $\mathcal{E}_0 = \id$.
The dynamical maps are then further decomposed into tensor products
\begin{equation}
\label{eq_transfer_tensors_En}
    \mathcal{E}_k = 
    \sum_{m=0}^{k-1} \mathcal{T}_{k-m}\mathcal{E}_m
\end{equation}
where $\mathcal{T}_{k}$ are called transfer tensors.
Since the bath correlation effect decays with respect to time,
 a memory length can be introduced such that $\mathcal{T}_k$ can be neglected when $k>K_{\mathrm{max}}$ for some preselected $K_{\max}$. 
Therefore, the reduced density matrix can be computed by
\begin{equation}
\label{eq_propagate}
    \rho_s(t_m) = \sum_{k=1}^{K_{\mathrm{max}}} \mathcal{T}_k \rho_s(t_{m-k})
\end{equation}
for $m>K_{\mathrm{max}}$.

In the previous section,
 we propose an algorithm for the computation of $\rho_s(t_k)$ for based on inchworm algorithm
 for a specific initial density.
By choosing a basis of the Liouville space as initial density,
 we can obtain the full dynamical map $\mathcal{E}_k$ by computing $\rho_s(t_k)$ for different initial densities matrix,
 in particular, a basis of the Liouville space.
With the dynamical maps within the memory length,
 the transfer tensor can be constructed by \cref{eq_transfer_tensors_En}.
For long-time simulation, \cref{eq_propagate} is sufficient to get an accurate result with proper choice of time step $\delta t$ and memory truncation parameter $K_{\max}$. The step size $\delta t$ corresponding to the discretized step size of the Nakajima-Zwanzig equation may differ from the inchworm step size $\Delta t$ used for the input $\rho_s(t_k)$. Using a larger $\delta t$ reduces the number of transfer tensors needed but can lead to increased errors.

In the case of a spin chain,
 it would be costly to compute the dynamical maps and transfer tensors for all possible initial reduced density matrix $\rho_s(0)$.
For a spin chain with $K$ spins, Liouville space has dimension $2^{2K}$, increasing exponentially with respect to the chain length $K$.
A better choice is to compute the ``dynamical maps'' for each spin with ``crosses'' and the process of connecting ``density matrices'' in \cref{sec_spin_connection} becomes the connection of ``dynamical maps''. 
Taking the initial reduced matrices in Liouville space as $\bigotimes_{k=1}^{K}e_{j_k}$, it follows that $\hat{\rho}_{t_n}^{j_1,\ldots,j_K}=\mathcal{E}_n \left(\bigotimes_{k=1}^{K}e_{j_k}\right)$, which corresponds to one of the columns of $\mathcal{E}_n$, denoted as $E^{j_1,\ldots,j_K}_n$. 
For a fixed combination $(j_1,\ldots,j_K)$, $E^{j_1,\ldots,j_K}_n$ can be calculated using \cref{eq_summation_propagators_1st_spin,eq_summation_propagators_middle_spins,eq_summation_propagators_last_spin} with vectorized propagators $\mathbbm{p}^{[1],j_1}$, $\mathbbm{p}^{[k+1],j_1,\ldots,j_{k+1}}$ and $\mathbbm{p}^{[K],j_1,\ldots,j_K}$. 
Noticing that the computation of $\mathbbm{p}^{[k+1],j_1,\ldots,j_k}$ are independent among superscripts $j_1,\ldots,j_k$, we can calculate $\rho_{s,I}^{(k),j_k}(-t,\boldsymbol{s},\boldsymbol{\alpha}^{(k)},t)$ for $j_k=1,\ldots,4$ at the same time in \cref{eq_summation_propagators_1st_spin,eq_summation_propagators_middle_spins,eq_summation_propagators_last_spin}. 
We use $E_n^{j_k}(\boldsymbol{s}^{(k)},\boldsymbol{\alpha}^{(k)})$ to represent the vectorized propagators in \cref{eq:rhoi_k_dyson} at time $t=t_n$ with the initial reduced density matrix taken as $\hat{\rho}_0^{j_k}=e_{j_k}$, and $E_n^{(k)}(\boldsymbol{s}^{(k)},\boldsymbol{\alpha}^{(k)})$ to represent a $4\times 4$ matrix of which the columns are $E_n^{j_k}(\boldsymbol{s}^{(k)},\boldsymbol{\alpha}^{(k)})$, i.e., $E_n^{(k)}(\boldsymbol{s}^{(k)},\boldsymbol{\alpha}^{(k)})=\left[E_n^{j_k}(\boldsymbol{s}^{(k)},\boldsymbol{\alpha}^{(k)})\right]_{j_k=1,\ldots,4}$. In analogy to \cref{eq_summation_propagators_1st_spin,eq_summation_propagators_middle_spins,eq_summation_propagators_last_spin}, we implement a similar summation process,
\begin{align}
\label{eq_summation_dynmaps}
    &\mathbbm{E}_n^{[1]}(\boldsymbol{s}^{(1)},\boldsymbol{\alpha}^{(1)})=E_n^{(1)}(\boldsymbol{s}^{(1)},\boldsymbol{\alpha}^{(1)}),\quad\text{for }\boldsymbol{\alpha}^{(1)}\in A^N\\
    &\mathbbm{E}_n^{[k+1]}(\boldsymbol{s}^{(k+1)},\boldsymbol{\alpha}^{(k+1)})=\sum_{N^{(k)}=0}^\infty \sum_{\boldsymbol{\alpha}^{(k)} \in A^{N^{(k)}}} \int_{-t \leqslant \boldsymbol{s}^{(k)}\leqslant t} \mathbbm{E}_n^{[k]}(\boldsymbol{s}^{(k)},\boldsymbol{\alpha}^{(k)} \nonumber\\
    &\qquad\qquad\qquad\qquad\qquad \otimes E_n^{(k+1)}(\left[\boldsymbol{s}^{(k)},\boldsymbol{s}^{(k+1)}\right],\left[\boldsymbol{\alpha}^{(k)},\boldsymbol{\alpha}^{(k+1)}\right])\dd \boldsymbol{s}^{(k)}, \nonumber\\
    &\qquad\qquad\qquad\qquad\qquad \text{for } k=1,\ldots,K-2 \text{ and for } \boldsymbol{\alpha}^{(k+1)} \in A^{N^{(k+1)}}  \\
    &\mathbbm{E}_n^{[K]} = \sum_{N^{(K-1)}=0}^\infty \sum_{\boldsymbol{\alpha}^{(K-1)}\in A^{N^{(K-1)}}} 
    \int_{-t\leqslant \boldsymbol{s}^{(K-1)}\leqslant t} \mathbbm{E}_n^{[K-1]}(\boldsymbol{s}^{(K-1)},\boldsymbol{\alpha}^{(K-1)})\otimes E_n^{(K)}(\boldsymbol{s}^{(K-1)},\boldsymbol{\alpha}^{(K-1)})\dd \boldsymbol{s}^{(K-1)}.
\end{align}
In a $K$-spin chain, the reduced density matrix $\rho_s$ is a $2^K\times 2^K$ matrix.
Correspondingly, a dynamical map $\mathcal{E}_n$ or a transfer tensor $\mathcal{T}_n$ contains ${(2^K\times 2^K)^2 = 2^{4K}}$ elements, which is not affordable for real computation.
For example, in a chain with 10 spins, the full storage required by a single transfer tensor is $16 \times 2^{4\times 10} = 2^{44}$ bytes $\approx 17.6$ TB.
In this estimation, we use \texttt{complex double} for a single element in the transfer tensor, each of which requires 16 bytes.
To reduce the memory cost,
 we employ the matrix product state (MPS) and matrix product operator (MPO) representation for dynamical maps and transfer tensors.

MPS and MPO naturally reflect the structure of many-body quantum systems and are widely used in numerical simulations of spin-chain models.
To compute the ground state of a $K$-spin chain, DMRG represents the wave function $\ket{\psi}$ in MPS and maintains an overall computational cost of $O(KL^3)$ where $L$ is the maximum bond dimension of $\ket{\psi}$.

In our model, we exploit these efficiencies by representing the vectorized $\rho_s(t_n)$, $\mathcal{E}$ and $\mathcal{T}$ in Liouville space as $K$-dimensional tensors
\begin{align}
    &\rho^t = \sum_{i_1,\dots,i_K} \rho_{i_1,\dots,i_K}^t \ket{i_1,\dots,i_K}; \\
    &\mathcal{E}^t = \sum_{i_1,\dots,i_K} 
    \sum_{i_1',\dots,i_K'} 
    \mathcal{E}_{i_1,\dots,i_K}^{t,i_1',\dots,i_K'}
    \dyad{i_1,\dots,i_K}{i_1',\dots,i_K'}; \\
    &\mathcal{T}^t = \sum_{i_1,\dots,i_K} 
    \sum_{i_1',\dots,i_K'} 
    \mathcal{T}_{i_1,\dots,i_K}^{t,i_1',\dots,i_K'}
    \dyad{i_1,\dots,i_K}{i_1',\dots,i_K'}.
\end{align}
In this expression, we use $\ket{i_k}$ to represent the basis of the Liouville space for the $k$th spin.
The coefficients $\rho_{i_1,\dots,i_K}^t, \mathcal{E}_{i_1,\dots,i_K}^{t,i_1',\dots,i_K'}, \mathcal{T}_{i_1,\dots,i_K}^{t,i_1',\dots,i_K'}$
are decomposed into MPS or MPO,
\begin{align}
    &\rho_{i_1,\dots,i_K}^t 
    = \sum_{l_1,\dots,l_{K-1}}
    P_{i_1;l_1}^{t,1}
    P_{i_2;l_1l_2}^{t,2}
    \dots
    P_{i_{K-1};l_{K-2}l_{K-1}}^{t,K-1}
    P_{i_K;l_{K-1}}^{t,K} \\
    &\mathcal{E}_{i_1,\dots,i_K}^{t,i_1',\dots,i_K'}
    = \sum_{l_1,\dots,l_{K-1}}
    E_{i_1 i_1';l_1}^{t,1}
    E_{i_2 i_2';l_1l_2}^{t,2}
    \dots
    E_{i_{K-1} i_{K-1}';l_{K-2}l_{K-1}}^{t,K-1}
    E_{i_K i_K';l_{K-1}}^{t,K} \\
    &\mathcal{T}_{i_1,\dots,i_K}^{t,i_1',\dots,i_K'}
    = \sum_{l_1,\dots,l_{K-1}}
    T_{i_1 i_1';l_1}^{t,1}
    T_{i_2 i_2';l_1l_2}^{t,2}
    \dots
    T_{i_{K-1} i_{K-1}';l_{K-2}l_{K-1}}^{t,K-1}
    T_{i_K i_K';l_{K-1}}^{t,K}
\end{align}
Here, the indices $l_j$, $j=1,\ldots, K-1$, denote the auxiliary bonds connecting matrix products. The number of components in MPS and MPO can be controlled by adjusting the bond dimension, given by $\displaystyle L = \max_{j=1,\ldots, K-1}\operatorname{dim}(l_j)$. In our computation, the summation and multiplication of tensors $\rho_s$, $\mathcal{E}$, and $\mathcal{T}$ are executed using summation and contraction techniques on the MPS and MPO. To control the bond dimension, we employ a tensor compression technique after performing summation operations. Specifically, by setting a predefined accuracy threshold $\eta$, we approximate the original MPS $A$ with a new MPS $B$ with a reduced bond dimension, satisfying $\|A-B\|_F \leqslant \eta\|A\|_F$. 
In our numerical experiments, we call a \texttt{C++} package \textsc{ITensor} \cite{fishman2022itensor} that provides an MPS/MPO data structure and their basic operations in our implementation of tensor calculations.

In \cref{eq_summation_dynmaps}, the concatenation and summation of ``dynamical maps'' are interpreted within the tensor network paradigm as extensions and summations of MPOs, respectively.
The second line of \cref{eq_summation_dynmaps} can be visualized as,
\begin{equation}
\begin{tikzpicture}[baseline =0]
        \filldraw[spinlightblue, thick] (0,0) circle (0.3);
        \draw[black,line width = 1.5pt](0,0.3) -- (0,0.6);
        \node[text = black, font=\tiny] at (-0.2,0.4) {$n_1$};
        \draw[black,line width = 1.5pt](0,-0.3) -- (0,-0.6);
        \node[text = black, font=\tiny] at (-0.2,-0.4) {$n_1'$};
        \draw[gray,line width = 4pt](0.3,0) -- (0.6,0.0);
        \filldraw[spinlightblue, thick] (0.9,0) circle (0.3);
        \draw[black,line width = 1.5pt](0.9,0.3) -- (0.9,0.6);
        \node[text = black, font=\tiny] at (0.7,0.4) {$n_2$};
        \draw[black,line width = 1.5pt](0.9,-0.3) -- (0.9,-0.6);
        \node[text = black, font=\tiny] at (0.7,-0.4) {$n_2'$};
        \draw[gray,line width = 4pt](1.2,0) -- (1.5,0.0);
        \node[text=black] at (1.75,0) {$\ldots$};
        \draw[gray,line width = 4pt](2,0) -- (2.3,0.0);
        \filldraw[spinlightblue, thick] (2.6,0) circle (0.3);
        \draw[black,line width = 1.5pt](2.6,0.3) -- (2.6,0.6);
        \node[text = black, font=\tiny] at (2.4,0.4) {$n_k$};
        \draw[black,line width = 1.5pt](2.6,-0.3) -- (2.6,-0.6);
        \node[text = black, font=\tiny] at (2.4,-0.4) {$n_k'$};
        \draw[gray,line width = 4pt](2.9,0) -- (3.2,0.0);
        \filldraw[spinlightblue, thick] (3.5,0) circle (0.3);
        \node[text = black, font=\tiny] at (3.2,0.4) {$n_{k+1}$};
        \draw[black,line width = 1.5pt](3.5,0.3) -- (3.5,0.6);
        \node[text = black, font=\tiny] at (3.2,-0.4) {$n_{k+1}'$};
        \draw[black,line width = 1.5pt](3.5,-0.3) -- (3.5,-0.6);
        \end{tikzpicture}
    =\sum_{N'=0}^\infty \sum_{\boldsymbol{\alpha'}_k \in A^{N'}} \int_{-t \leqslant \boldsymbol{s'}\leqslant t}
    \left[
    \begin{aligned}
        &\begin{tikzpicture}[baseline =0]
        \filldraw[spinlightblue, thick] (0,0) circle (0.3);
        \draw[black,line width = 1.5pt](0,0.3) -- (0,0.6);
        \node[text = black, font=\tiny] at (-0.2,0.4) {$n_1$};
        \draw[black,line width = 1.5pt](0,-0.3) -- (0,-0.6);
        \node[text = black, font=\tiny] at (-0.2,-0.4) {$n_1'$};
        \draw[gray,line width = 2.5pt](0.3,0) -- (0.6,0.0);
        \filldraw[spinlightblue, thick] (0.9,0) circle (0.3);
        \draw[black,line width = 1.5pt](0.9,0.3) -- (0.9,0.6);
        \node[text = black, font=\tiny] at (0.7,0.4) {$n_2$};
        \draw[black,line width = 1.5pt](0.9,-0.3) -- (0.9,-0.6);
        \node[text = black, font=\tiny] at (0.7,-0.4) {$n_2'$};
        \draw[gray,line width = 2.5pt](1.2,0) -- (1.5,0.0);
        \node[text=black] at (1.75,0) {$\ldots$};
        \draw[gray,line width = 2.5pt](2,0) -- (2.3,0.0);
        \filldraw[spinlightblue, thick] (2.6,0) circle (0.3);
        \draw[black,line width = 1.5pt](2.6,0.3) -- (2.6,0.6);
        \node[text = black, font=\tiny] at (2.4,0.4) {$n_k$};
        \draw[black,line width = 1.5pt](2.6,-0.3) -- (2.6,-0.6);
        \node[text = black, font=\tiny] at (2.4,-0.4) {$n_k'$};
        \draw[gray,line width = 1pt](2.9,0) -- (3.5,0.0);
        \node[text = gray,font = \tiny] at (3.25,0.2){$l_k=1$};
        \filldraw[spingray, thick] (3.8,0) circle (0.3);
        \node[text = black, font=\tiny] at (3.5,0.4) {$n_{k+1}$};
        \draw[black,line width = 1.5pt](3.8,0.3) -- (3.8,0.6);
        \draw[black,line width = 1.5pt](3.8,-0.3) -- (3.8,-0.6);
        \node[text = black, font=\tiny] at (3.5,-0.4) {$n_{k+1}'$};
        \end{tikzpicture}
    \end{aligned}
    \right]\dd \boldsymbol{s}'
\end{equation}
The ``$\otimes$'' operation is equivalent to connecting the $(k+1)$th site with an auxiliary link of bond dimension $1$ to the MPO consisting of $k$ sites. 
The tensor contraction and summation involved in calculating the transfer tensors \cref{eq_transfer_tensors_En} are now represented as multiplication and summation over MPOs. 
Compression is introduced to control the growth of bond dimensions, thereby reducing memory cost.
Suppose the maximum bond dimension for $\mathcal{E}_n$ is $L$, and the number of elements contained in $\mathcal{E}_n$ is upper bounded by $16KL^2$.
Compared to the original storage $2^{4K}$,
 the memory cost is reduced significantly when $L$ is relatively small.
We will present more details on the bond dimension in \cref{sec_numerical_results}.
\section{Numerical Results}
\label{sec_numerical_results}
In this section, we evaluate the performance of the inchworm method and TTM by investigating the evolution of observables $\langle O_k(t)\rangle$ where $O_k=\sigma_z^{(k)}$ for $k=1,\ldots, K$, respectively. In \cref{eq:BIF},the two-point correlation functions $B^{(k)}(\tau_i,\tau_j)$ are identical for each index $k$,
\begin{equation}\label{bath_function_B}
    B^{(k)}(\tau_i,\tau_j)=\begin{cases}
        \overline{B^*(\tau_i,\tau_j)},&\text{ if } \tau_i\tau_j> 0\\
        B^* (\tau_i,\tau_j),& \text{ if } \tau_i\tau_j \leq 0
    \end{cases}
\end{equation}
where $B^*(\tau_i,\tau_j)$ is defined as
\begin{equation}
\label{B_star}
    B^*(\tau_i,\tau_j) 
    =\frac{1}{\pi} \int_0^{\infty}
    J(\omega)
    \left[
    \coth\left(\frac{\beta\omega}{2}\right) \cos(\omega\Delta \tau)
    - \ii \sin(\omega \Delta \tau)
    \right] \dd \omega
\end{equation}
where $\Delta \tau = |\tau_i|-|\tau_j|$ and $J(\omega)$ is the spectral density of the harmonic oscillators in the bath, defined as
\begin{align}
    &J(\omega) = \frac{\pi}{2} \sum_{l=1}^L \frac{c_l^2}{\omega_l} \delta(\omega - \omega_l) \\
    &\omega_l = -\omega_c \ln \left(1- \frac{l}{L}[1-\exp(-\omega_{\max}/\omega_c)]\right), \\
    & c_l = \omega_l \sqrt{ \frac{\xi \omega_c}{L}[1-\exp(-\omega_{\max}/\omega_c)]},
\end{align}
where $l=1,\dots,L$.

The initial state of the spins in the spin chain is prepared to be $\ket{\zeta^{(k)}}=\ket{+1}$ for $k=2,\ldots, K$ throughout this section, and we first let $\ket{\zeta^{(1)}}=\ket{+1}$.  
The parameters of $B^{(k)}(\tau_1,\tau_2)$ are set as
$$\beta=5.0,\quad \xi =0.2, \quad \omega_c=2.5,\quad \omega_{\max}=4\omega_c,\quad L=400$$

\subsection{Convergence tests of inchworm method}
We first validate the convergence of the inchworm method in \cref{sec_diagrammatic_representation,sec_spin_connection} on truncation parameters $\bar{M}$, $\bar{N}$, time step $\Delta t$ and the accuracy threshold $\eta$. 
 The parameters for this experiment are set as 
$$\epsilon^{(k)}=1.0,\quad \Delta^{(k)} = 1.0,\quad V^{(k,k+1)}=J^{(k)}_z\sigma_z^{(k)}\otimes \sigma_z^{(k+1)}$$
with $J_z^{(k)}=0.04$.
The interaction term $V^{(k,k+1)}$ characterizes the interactions in the Ising model.

\begin{figure}
    \centering
    \begin{subfigure}[b]{0.3\textwidth}
         \centering
         \includegraphics[width=\textwidth]{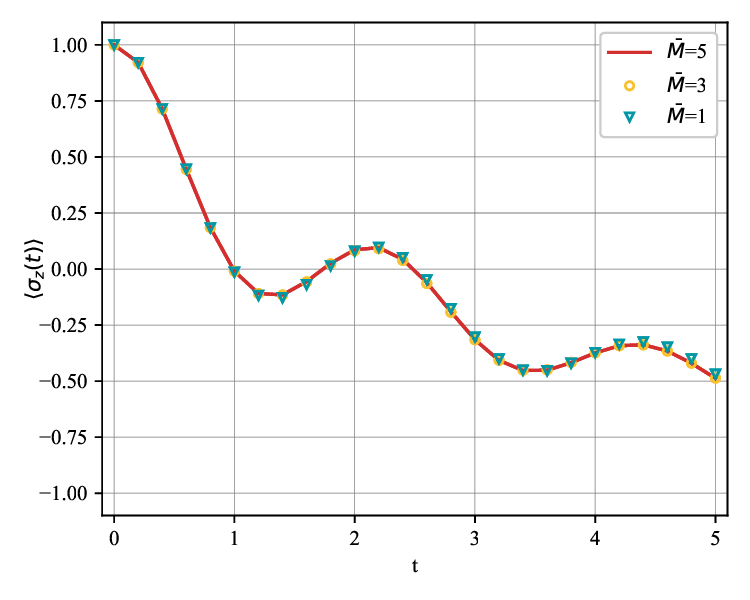}
         \caption{Spins 1 and 5}
         \label{fig_convergence_M_spin1}
     \end{subfigure}
     \hfill
    \begin{subfigure}[b]{0.3\textwidth}
         \centering
         \includegraphics[width=\textwidth]{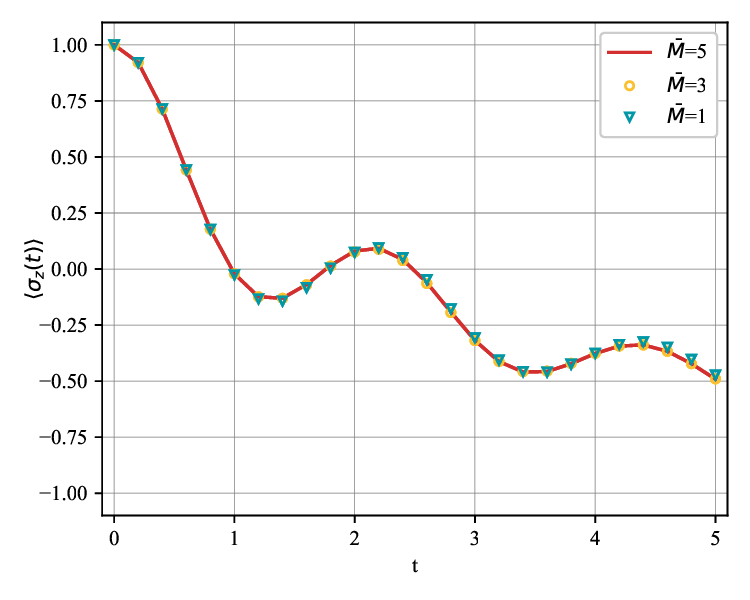}
         \caption{Spins 2 and 4}
         \label{fig_convergence_M_spin2}
     \end{subfigure}
     \hfill
     \begin{subfigure}[b]{0.3\textwidth}
         \centering
         \includegraphics[width=\textwidth]{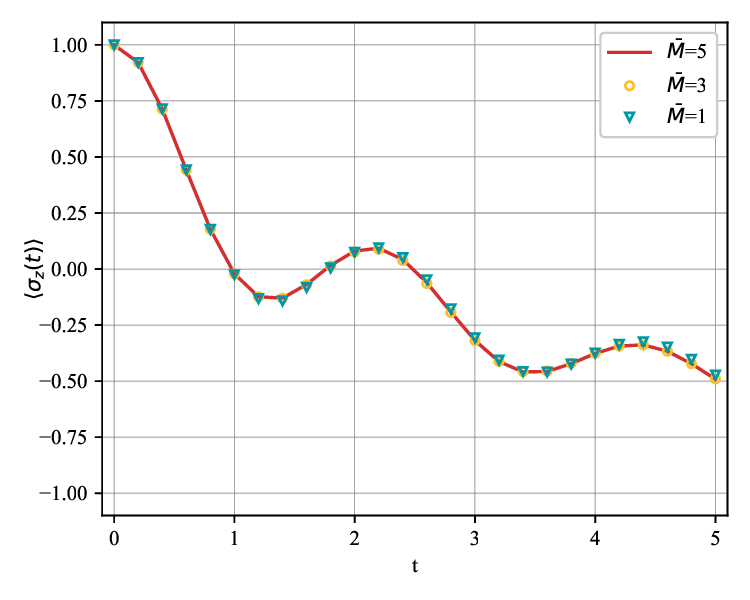}
         \caption{Spin 3}
         \label{fig_convergence_M_spin3}
     \end{subfigure}
    \caption{Convergence of the method with respect to $\bar{M}$.}
    \label{fig_convergence_M}
\end{figure}

Here we first conduct a numerical test for the convergence of $\bar{M}$ on a $5$-spin system. We choose $\bar{N}=2$ and the time step to be $\Delta t=0.2$, the numerical results for $\bar{M}=1,3,5$ are given in \cref{fig_convergence_M}, which shows the evolution of $\langle \sigma_z^{(k)}\rangle$ for $k=1,\ldots,5$. The curves for $\bar{M}=3$ and $\bar{M}=5$ almost coincide while the curve for $\bar{M}=1$ has a slight deviation, suggesting less accuracy.

\begin{figure}
    \centering
    \begin{subfigure}[b]{0.3\textwidth}
         \centering
         \includegraphics[width=\textwidth]{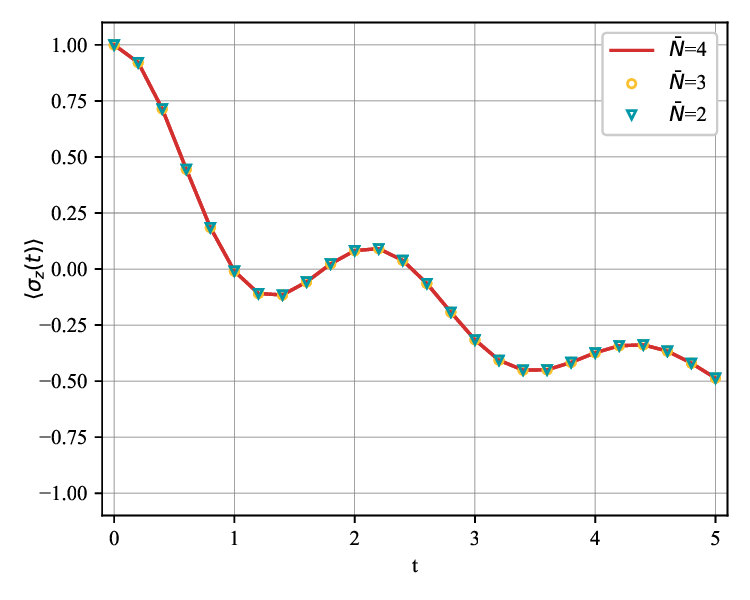}
         \caption{Spins 1 and 5}
         \label{fig_convergence_N_spin1}
     \end{subfigure}
     \hfill
    \begin{subfigure}[b]{0.3\textwidth}
         \centering
         \includegraphics[width=\textwidth]{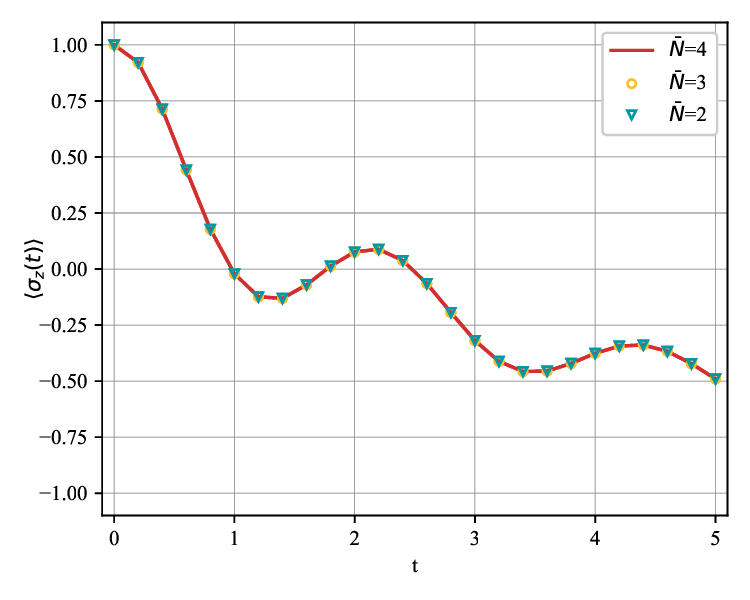}
         \caption{Spins 2 and 4}
         \label{fig_convergence_N_spin2}
     \end{subfigure}
     \hfill
     \begin{subfigure}[b]{0.3\textwidth}
         \centering
         \includegraphics[width=\textwidth]{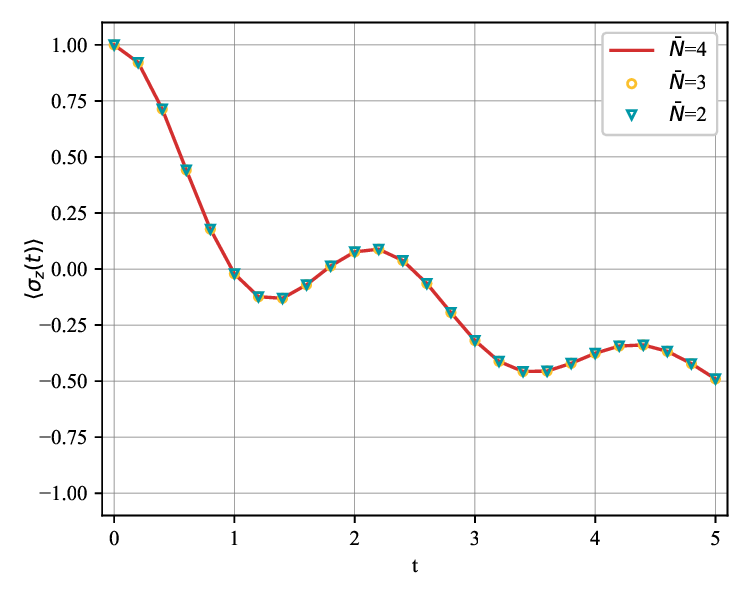}
         \caption{Spin 3}
         \label{fig_convergence_N_spin3}
     \end{subfigure}
    \caption{Convergence of the method with respect to $\bar{N}$.}
    \label{fig_convergence_N}
\end{figure}

We now fix $\bar{M}=3$, $\Delta t= 0.2$ and consider the convergence with respect to $\bar{N}$, the results for $\bar{N}=2,3,4$ are given in \cref{fig_convergence_N}.
In the test, the curves for different $\bar{N}$ are almost identical, suggesting that $\bar{N}=2$ is sufficient in the example. 

\begin{figure}
    \centering
    \begin{subfigure}[b]{0.3\textwidth}
         \centering
         \includegraphics[width=\textwidth]{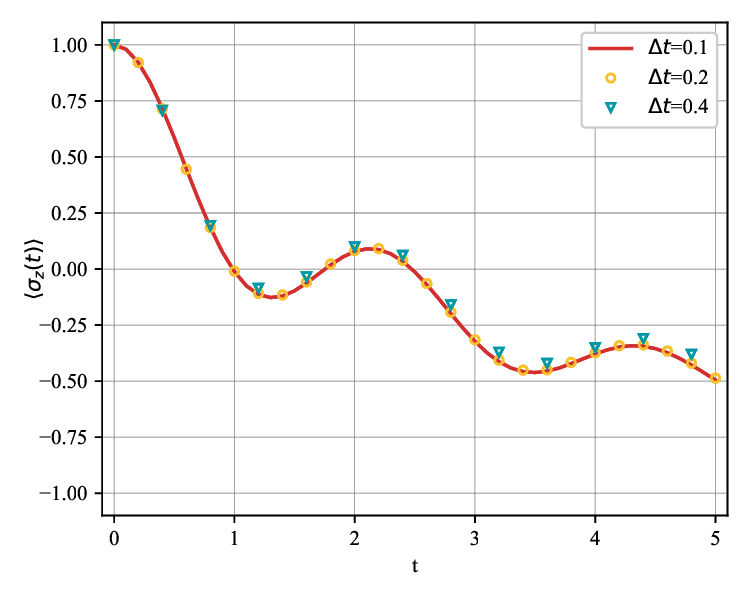}
         \caption{Spins 1 and 5}
         \label{fig_convergence_dt_spin1}
     \end{subfigure}
     \hfill
    \begin{subfigure}[b]{0.3\textwidth}
         \centering
         \includegraphics[width=\textwidth]{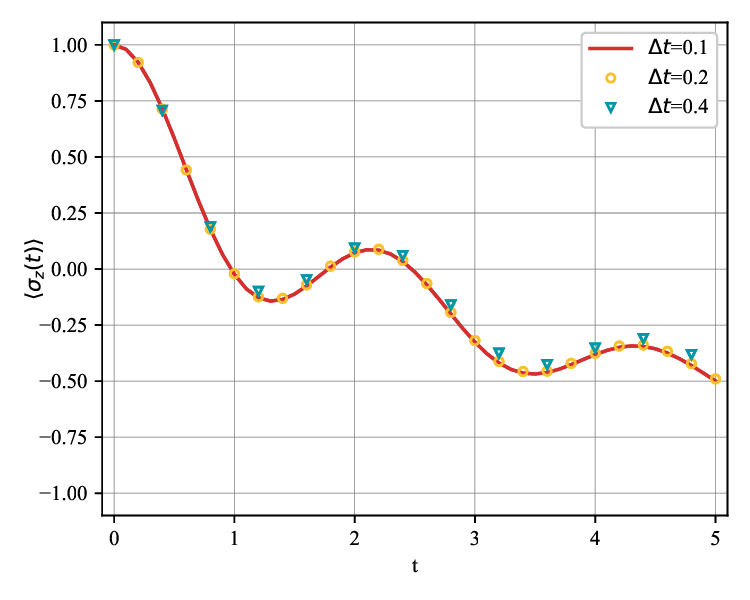}
         \caption{Spins 2 and 4}
         \label{fig_convergence_dt_spin2}
     \end{subfigure}
     \hfill
     \begin{subfigure}[b]{0.3\textwidth}
         \centering
         \includegraphics[width=\textwidth]{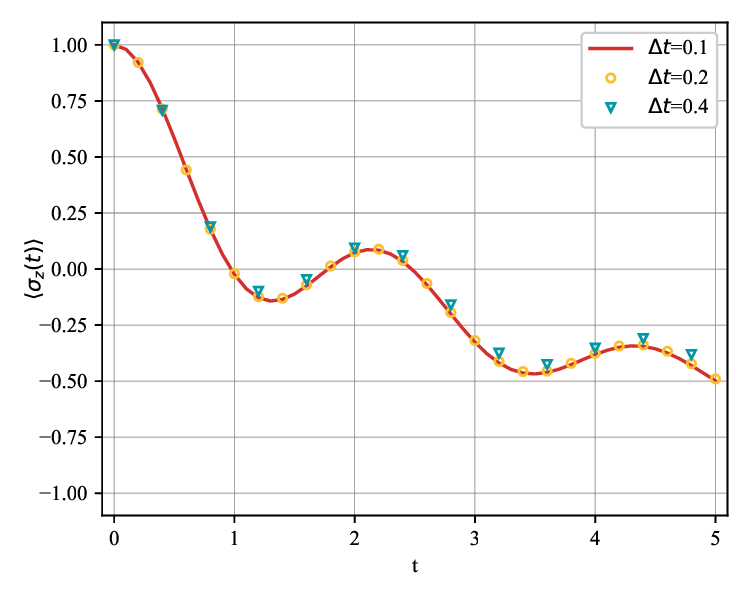}
         \caption{Spin 3}
         \label{fig_convergence_dt_spin3}
     \end{subfigure}
    \caption{Convergence of the method with respect to $\Delta t$.}
    \label{fig_convergence_dt}
\end{figure}

Additionally, we present the numerical results of the convergence test for time step $\Delta t=0.4,0.2,0.1$ with $\bar{M}=3$, $\bar{N}=2$ in \cref{fig_convergence_dt}. The results indicate that the choices of $\Delta t = 0.1$ and $\Delta t = 0.2$ give almost the same accuracy, suggesting that a time step of $\Delta t = 0.2$ provides sufficient accuracy for our simulations.

\begin{figure}
    \centering
    \begin{subfigure}[b]{0.3\textwidth}
         \centering
         \includegraphics[width=\textwidth]{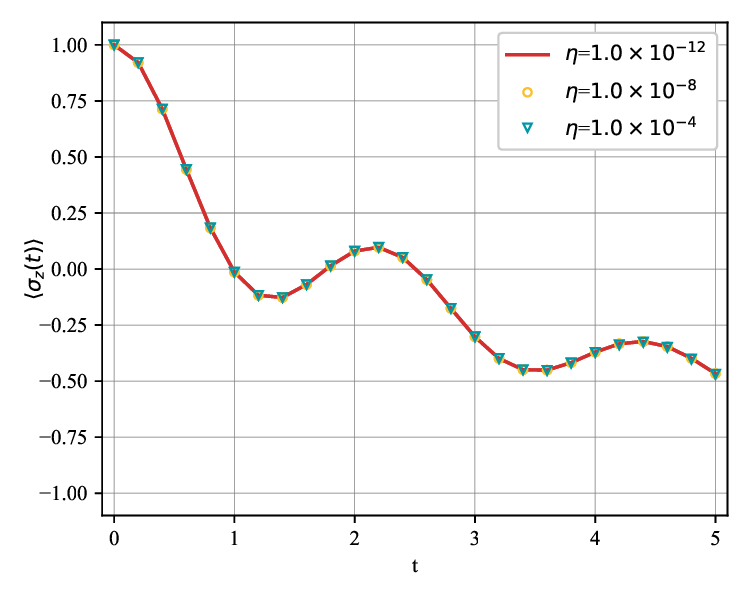}
         \caption{Spins 1 and 10}
         \label{fig_convergence_cutoff_spin1}
     \end{subfigure}
     \hfill
    \begin{subfigure}[b]{0.3\textwidth}
         \centering
         \includegraphics[width=\textwidth]{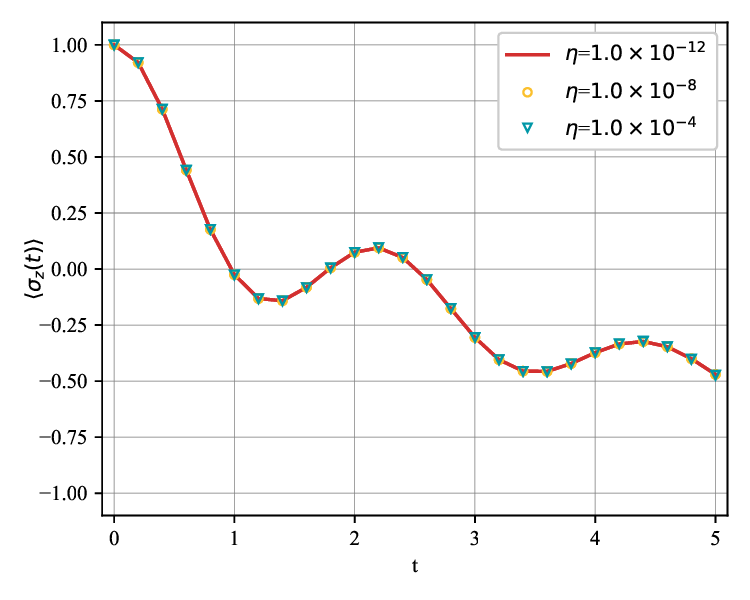}
         \caption{Spins 5 and 6}
         \label{fig_convergence_cutoff_spin5}
     \end{subfigure}
     \hfill
     \begin{subfigure}[b]{0.3\textwidth}
         \centering
         \includegraphics[width=\textwidth]{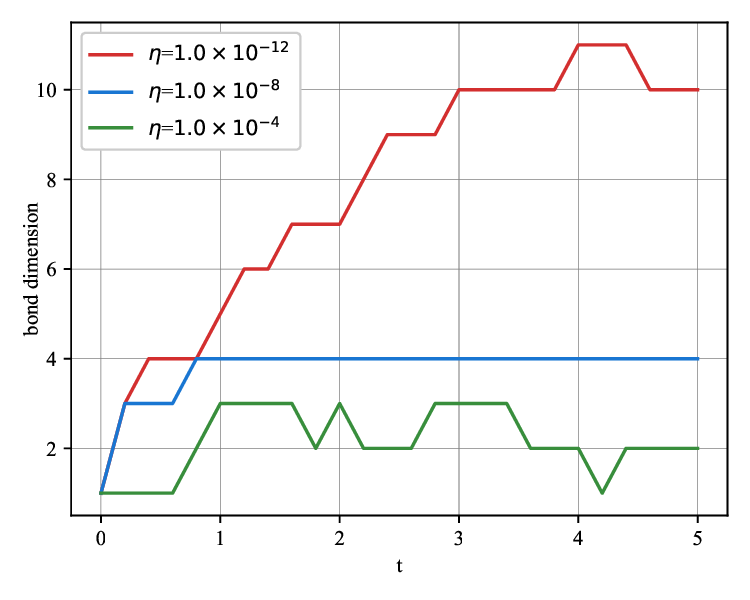}
         \caption{Bond dimension of $\rho_s(t)$}
         \label{fig_sigmaz_10spins_bond_dim}
     \end{subfigure}
    \caption{Comparison of $\eta = 1.0\times 10^{-12}, 1.0\times 10^{-8}$ and $1.0\times 10^{-4}$.}
    \label{fig_convergence_cutoff}
\end{figure}

Beyond the parameters of the inchworm method, we further consider another contribution factor, the accuracy threshold $\eta$ in the compression of MPS. We note that such compression is exclusively applied in the resummation of propagators, where we sum over propagators formatted as MPS and then compress the summed MPS according to the prescribed $\eta$. The parameters for inchworm are set as $\bar{N}=2$, $\bar{M}=3$, $\Delta t=0.2$. For a $10$-spin system, the $\langle O_k(t) \rangle$ for $k=1,5,6,10$ given by $\eta=1.0\times 10^{-4},1.0\times 10^{-8}$ and $1.0\times 10^{-12}$ are provided in \cref{fig_convergence_cutoff_spin1} and \cref{fig_convergence_cutoff_spin5}. The differences between curves are minimal; however, the curve at $\eta = 1.0\times 10^{-4}$ exhibits an extremely slight deviation from the other two. From \cref{fig_sigmaz_10spins_bond_dim}, we observe that as $\eta$ ranges from $1\times 10^{-4}$ to $1\times 10^{-12}$, the maximum bond dimension of $\rho_s(t)$ in time interval $[0,5]$ increases from 3 to 11. Considering that the number of components in a full-tensor reduced density matrix of a $10$-spin system is $4^{10}$, even with the finest $\eta = 1.0 \times 10^{-12}$, the number of components in $\rho_s(t)$ after compression remains at most $4 \times 11^2 \times 10$.



\begin{figure}
    \centering
    \begin{subfigure}[b]{0.3\textwidth}
         \centering
         \includegraphics[width=\textwidth]{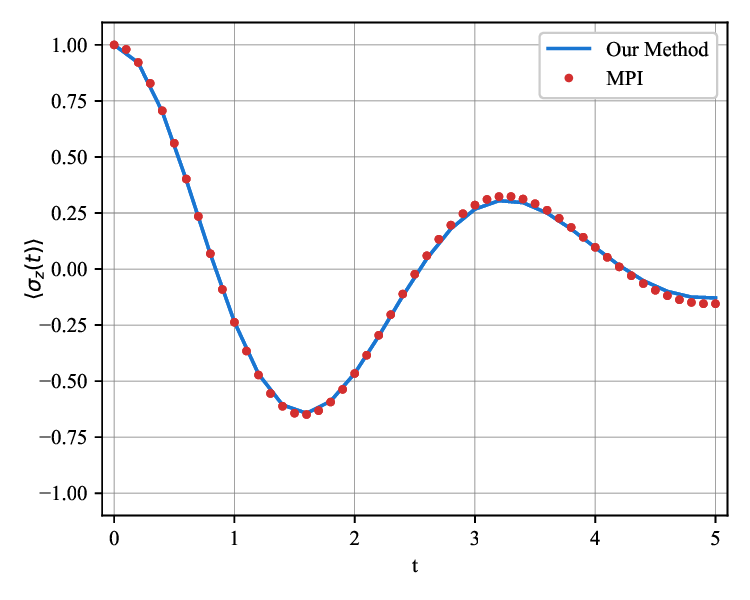}
         \caption{Spins 1 and 5}
         \label{fig_vs_mpi_spin1}
     \end{subfigure}
     \hfill
    \begin{subfigure}[b]{0.3\textwidth}
         \centering
         \includegraphics[width=\textwidth]{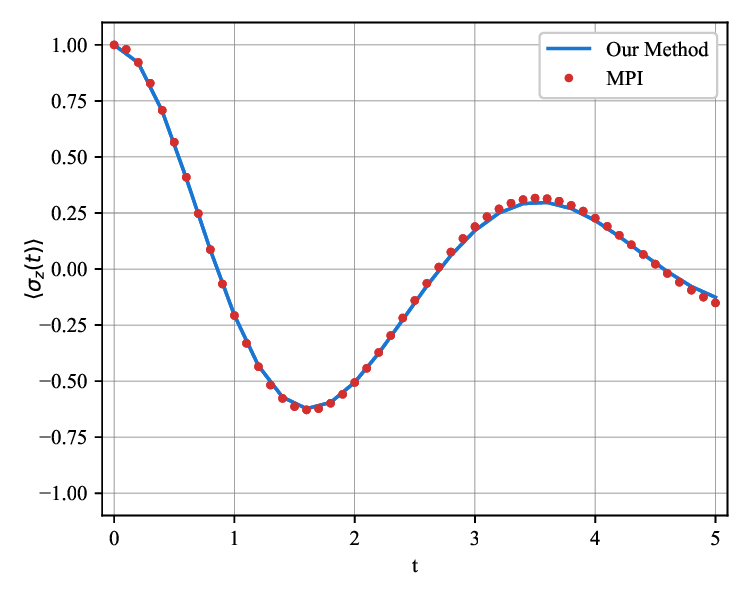}
         \caption{Spins 2 and 4}
         \label{fig_vs_mpi_spin2}
     \end{subfigure}
     \hfill
     \begin{subfigure}[b]{0.3\textwidth}
         \centering
         \includegraphics[width=\textwidth]{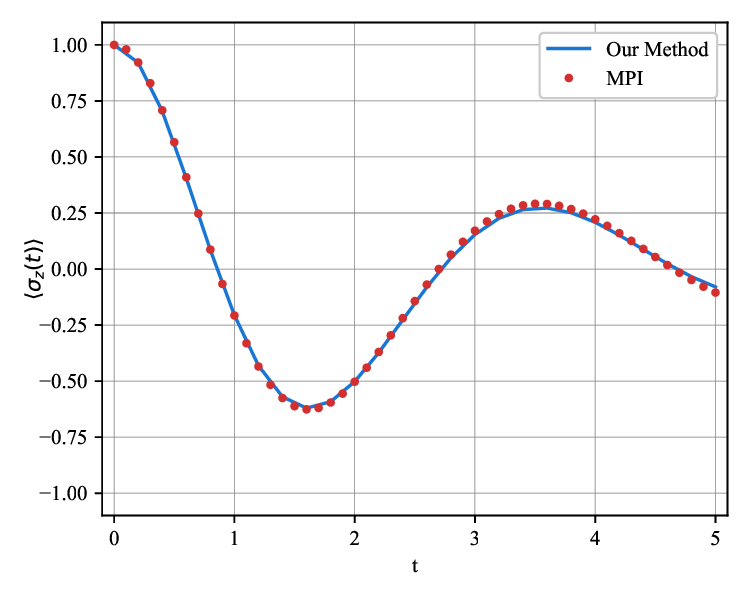}
         \caption{Spin 3}
         \label{fig_vs_mpi_spin3}
     \end{subfigure}
    \caption{Comparison of our method and the MPI method.}
    \label{fig_vs_mpi}
\end{figure}

We compare our results with results given by the MPI method, which is obtained by the \textsc{PathSum} package according to Kundu and Makri's works 
\cite{makri1995numerical,makri2014blip,makri2018modular,makri2018communication,kundu2019modular,makri2020smallMatrixDisentanglement,kundu2020modular,kundu2021efficient,kundu2023pathsum}.
$$\xi = 0.2,\quad \beta = 5, \quad\omega_c = 2.5, \quad \omega_{\text{max}}=4\omega_c,\quad \epsilon^{(k)}=0.0,\quad\Delta = 1.0,\quad J_z^{(k)}=0.2$$
$$ \Delta t = 0.2,\quad \bar{M}=3,\quad \bar{N}=4,\quad \eta=1.0\times 10^{-15},\quad \forall k =1,\ldots,5.$$
The results of all spins are given in \cref{fig_vs_mpi}.
In all these tests, our methods correctly show the physical symmetry. The first and the last spins behave consistently as they are physically identical if we label the spin chain from the last spin to the first.
In addition, our method, designed for the Heisenberg model, performs well in the diagonal coupling Ising model, a special case of the Heisenberg model.
In the following experiments, we also carry out tests on a general Heisenberg model.

\subsection{Convergence tests of TTM}
In this section, we present the numerical results that illustrate the utility of TTM in our Inchworm method. Our goal is to evaluate the accuracy of TTM and show how MPS compression techniques facilitate bond dimension reduction.

The Hamiltonian for this experiment is set as
$$\epsilon^{(k)}=0.0,\quad \Delta ^{(k)}=1.0,\quad V^{(k,k+1)}=J_x^{(k)}\sigma_x^{(k)}\otimes\sigma_x^{(k+1)}+J_y^{(k)}\sigma_y^{(k)}\otimes\sigma_y^{(k+1)}$$
The term $V^{(k,k+1)}$ characterizes the interaction in the Frenkel model \cite{frenkel1931transformation1,frenkel1931transformation2}, a typical model in the Heisenberg chain.
The dynamical maps $\mathcal{E}_n$ is generated using inchworm framework, with the fixed parameters
$$\bar{N}=2,\quad \bar{M}=1, \quad \Delta t = 0.1.$$
Furthermore, a reduced density matrix is generated by the inchworm method using the same parameters and serves as a ``reference solution'' in this experiment. The step size of TTM is set as $\delta t = \Delta t = 0.1$. 

\begin{figure}
    \centering
    \begin{subfigure}[b]{0.45\textwidth}
         \centering
         \includegraphics[width=\textwidth]{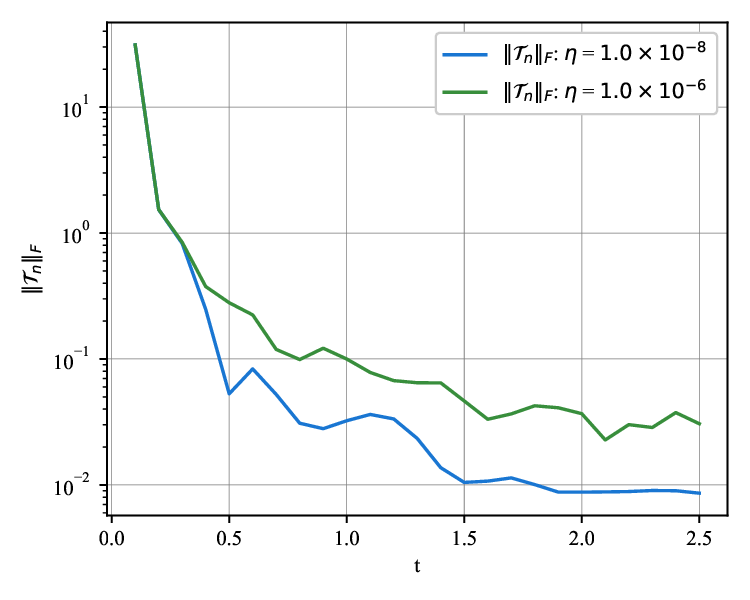}
         \caption{$\|\mathcal{T}_n\|_F$: $\eta=1.0\times10^{-6}$, $10^{-8}$.}
         \label{fig_T_norm_wrt_cutoff_log}
     \end{subfigure}
     \hfill
    \begin{subfigure}[b]{0.45\textwidth}
         \centering
         \includegraphics[width=\textwidth]{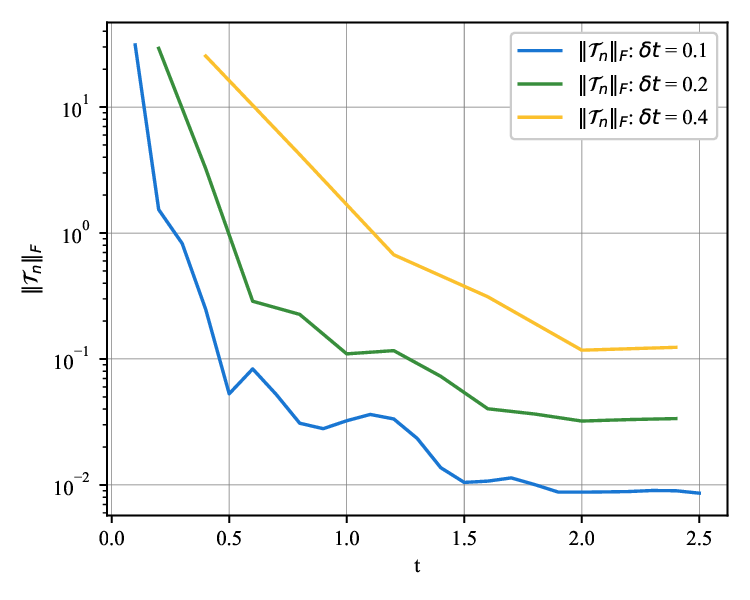}
         \caption{$\|\mathcal{T}_n\|_F$: $\eta=1.0\times 10^{-6}$, $\delta t = 0.1$, $0.2$, $0.4$.}
         \label{fig_T_norm_wrt_tau_log}
     \end{subfigure}
    \caption{Decay of Frobenius norm of transfer tensors $\mathcal{T}_n$.}
    \label{fig_T}
\end{figure}

\begin{figure}
    \centering
    \begin{subfigure}[b]{0.3\textwidth}
         \centering
         \includegraphics[width=\textwidth]{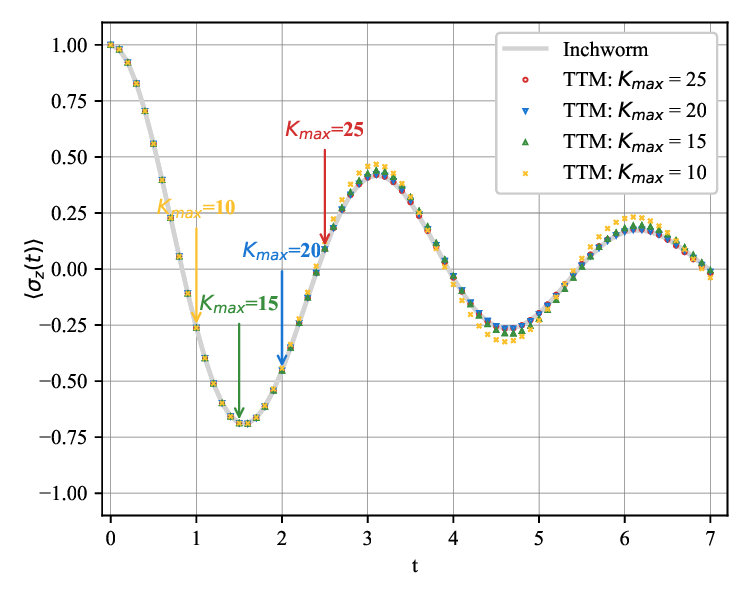}
         \caption{Spins 1 and 5}
         \label{fig_ttm_5spins_wrt_kmax_spin15}
     \end{subfigure}
     \hfill
    \begin{subfigure}[b]{0.3\textwidth}
         \centering
         \includegraphics[width=\textwidth]{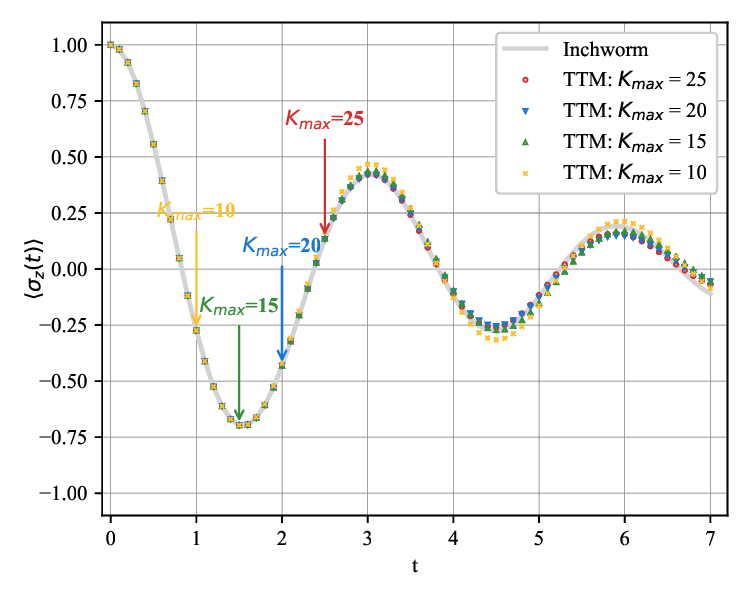}
         \caption{Spins 2 and 4}
         \label{fig_ttm_5spins_wrt_kmax_spin24}
     \end{subfigure}
     \hfill
     \begin{subfigure}[b]{0.3\textwidth}
         \centering
         \includegraphics[width=\textwidth]{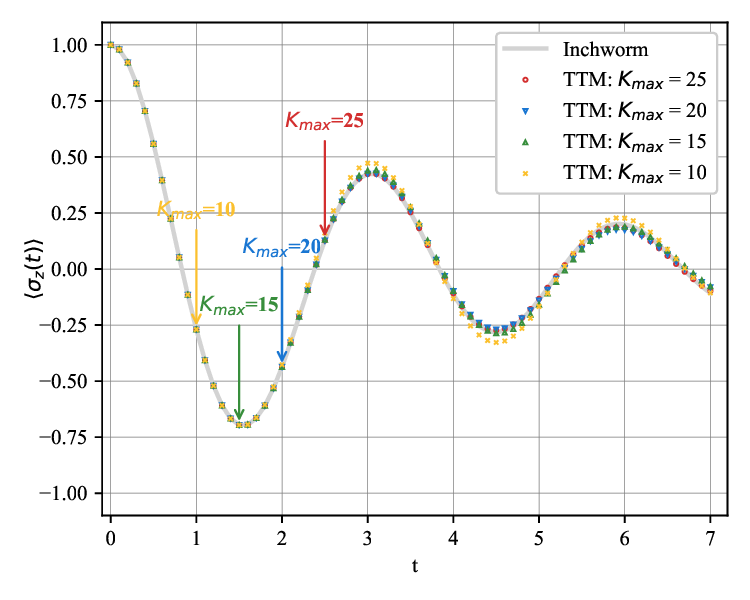}
         \caption{Spin 3}
         \label{fig_ttm_5spins_wrt_kmax_spin3}
     \end{subfigure}
    \caption{Comparison of TTM results with different memory lengths $K_{\mathrm{max}}=10,15,20,25$.}
    \label{fig_ttm_5spins_wrt_kmax}
\end{figure}

For a $5$-spin system, we first estimate the decay of the Frobenius norm of transfer tensors $\mathcal{T}_n$. Setting $\eta = 1.0\times 10^{-8}$ up to $K_{\mathrm{max}}=25$, \cref{fig_T} shows a rapid decay of $\|\mathcal{T}_n\|_F$, suggesting a relatively weak influence from the bath. \Cref{fig_T_norm_wrt_cutoff_log} displays the evolution of the natural logarithm $\ln(\|\mathcal{T}_n\|_F)$ for different values of $\eta$. The blue curve, corresponding to $\eta = 1.0 \times 10^{-8}$, stabilizes around $10^{-2.1}$ between $t=2.0$ and $t=2.5$, suggesting that $\|\mathcal{T}_n\|_F$ effectively reaches a steady state close to zero by $t=2.0$. Two potential sources of error that prevent $\|\mathcal{T}_n\|_F$ from decaying to zero are MPO compression and time discretization. As shown in \cref{fig_T_norm_wrt_cutoff_log}, $\|\mathcal{T}_n\|_F$ with a larger compression accuracy threshold $\eta$ exhibits a more pronounced deviation from zero; \cref{fig_T_norm_wrt_tau_log} illustrates the time discretization error introduced by selecting the TTM step size $\delta t$. Comparatively, $\|\mathcal{T}_n\|_F$ with $\delta t = 0.2$, $0.4$ respectively and $\eta = 1.0 \times 10^{-8}$ for $n=1,\ldots,6$, stabilize around $10^{-1}$ and $10^{-1.5}$, indicating larger errors.  
Moreover, we compare the reduced density matrices generated by these transfer tensors $\mathcal{T}_n$ with different memory lengths $K_{\mathrm{max}}=10,15,20$ and $25$, respectively. The evolution of observable $\langle O_k (t)\rangle$ for $k=1,\ldots,5$ is shown in \cref{fig_ttm_5spins_wrt_kmax}. The gray solid line represents the $\rho_s(t)$ computed by the inchworm method while the colored dotted lines correspond to $\rho_s(t)$ generated by TTM with varying $K_{\mathrm{max}}$ values. Colored vertical arrows mark the start of TTM propagation; namely, the dynamical maps $\mathcal{E}_n$ and transfer tensors $\mathcal{T}_n$ are calculated from $n=1$ to $n=K_{\mathrm{max}}$, and from $t=\delta t K_{\mathrm{max}}$ onwards, TTM propagates using the information captured by $\mathcal{T}_n$ for $n=1,\ldots,K_{\mathrm{max}}$. From \cref{fig_ttm_5spins_wrt_kmax}, we can observe that, similar to the inchworm method, TTM preserves the physical symmetry of the spin chain. As $K_{\mathrm{max}}$ increases, the curves converge towards the inchworm curve, with the blue and red curves, corresponding to $K_{\mathrm{max}}=20$ and  $K_{\mathrm{max}}=25$ respectively, almost overlapping. This overlap is consistent with the finding that $\|\mathcal{T}_n\|_F$ stabilizes around zero from $t=2.0$ to $t=2.5$, indicating that a memory length of $K_{\mathrm{max}}=20$ is sufficient for this system-bath model. Consequently, we have chosen to use a memory length $K_{\mathrm{max}}=20$ for subsequent experiments.

\begin{figure}
    \centering
    \begin{subfigure}[b]{0.3\textwidth}
         \centering
         \includegraphics[width=\textwidth]{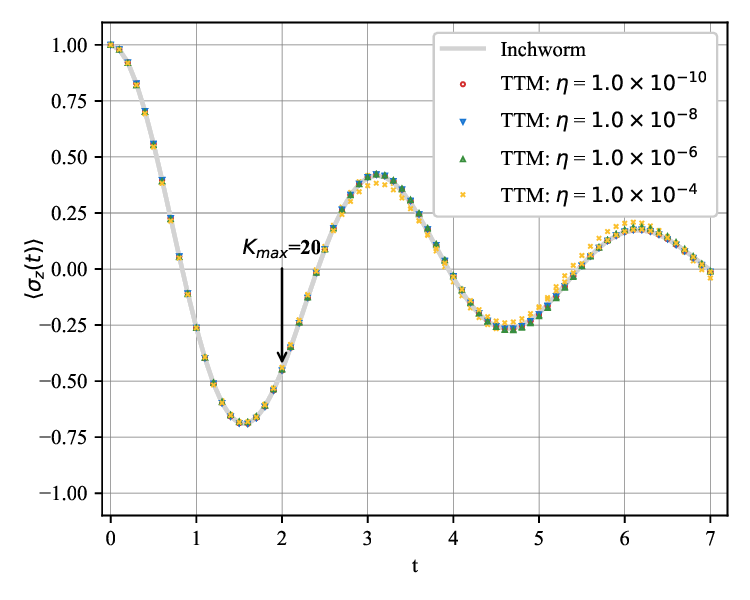}
         \caption{Spins 1 and 5}
         \label{fig_ttm_5spins_wrt_cutoff_spin15}
     \end{subfigure}
     \hfill
    \begin{subfigure}[b]{0.3\textwidth}
         \centering
         \includegraphics[width=\textwidth]{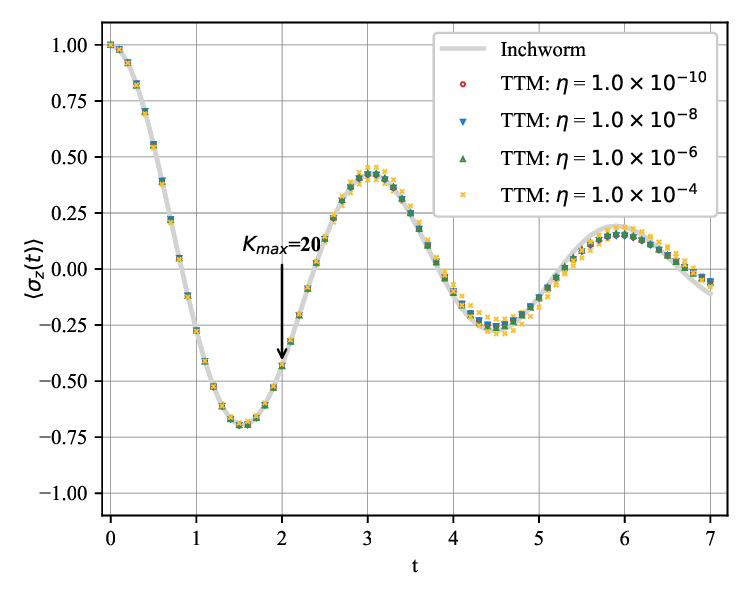}
         \caption{Spins 2 and 4}
         \label{fig_ttm_5spins_wrt_cutoff_spin24}
     \end{subfigure}
     \hfill
     \begin{subfigure}[b]{0.3\textwidth}
         \centering
         \includegraphics[width=\textwidth]{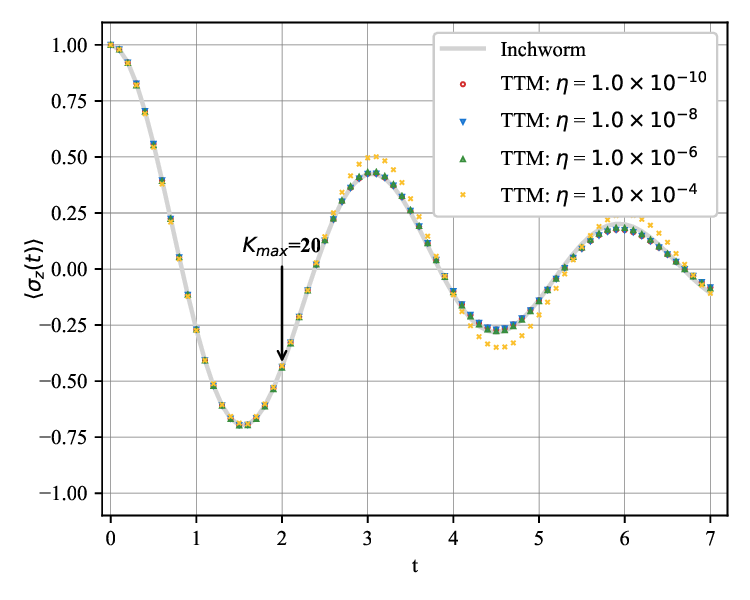}
         \caption{Spin 3}
         \label{fig_ttm_5spins_wrt_cutoff_spin3}
     \end{subfigure}
    \caption{Comparison of TTM results with $\eta = 1.0\times 10^{-10},1.0\times 10^{-8},1.0\times 10^{-6}$ and $1.0\times 10^{-4}$.}
    \label{fig_ttm_5spins_wrt_cutoff}
\end{figure}

We now set the memory length at $K_{\mathrm{max}}=20$ and compare the results computed with different accuracy threshold $\eta$. In calculating $\mathcal{E}_n$, compression occurs after summing MPOs; for $\mathcal{T}_n$, it occurs after both summing and multiplying MPOs; for $\rho_s(t_m)$, it occurs after multiplying $T_k$ to $\rho_s(t_{m-k})$. The evolution of observable, generated with $\eta =1.0\times 10^{-4},1.0\times 10^{-6}, 1.0\times 10^{-8}$ and $1.0\times 10^{-10}$, is displayed in \cref{fig_ttm_5spins_wrt_cutoff}. The yellow curve corresponding to $\eta = 1.0\times 10^{-4}$ shows obvious deviation; physical symmetry is not preserved, indicating that an accuracy threshold of $\eta = 1.0\times 10^{-4}$ is insufficient for the accurate resummation of dynamical-map propagators. $\eta = 1.0\times 10^{-8}$ and $\eta = 1.0\times 10^{-10}$ yield nearly identical results, while the curve $\eta = 1.0\times 10^{-6}$ shows a slight deviation, suggesting an $\eta$ range from $1.0\times 10^{-6}$ to $1.0\times 10^{-8}$ is acceptable for this case.

\begin{figure}
    \centering
    \begin{subfigure}[b]{0.3\textwidth}
         \centering
         \includegraphics[width=\textwidth]{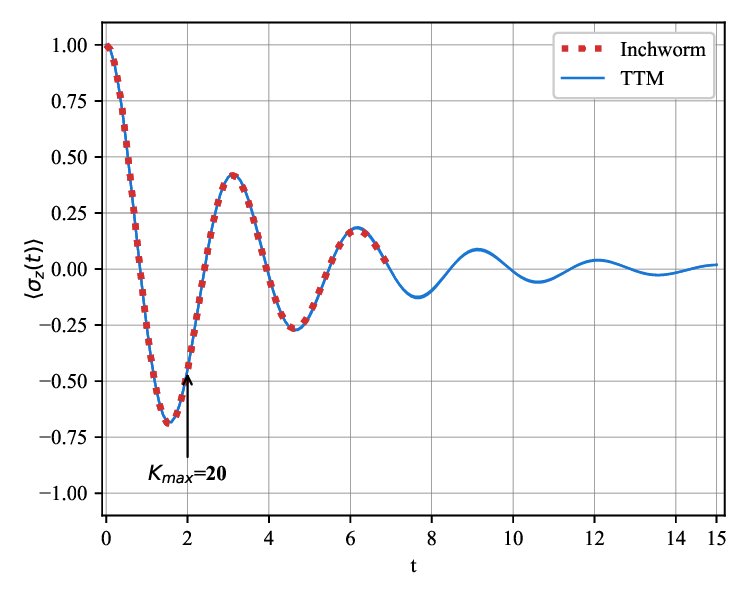}
         \caption{Spins 1 and 5}
         \label{fig_ttm_5spins_150steps_1e6_spin15}
     \end{subfigure}
     \hfill
    \begin{subfigure}[b]{0.3\textwidth}
         \centering
         \includegraphics[width=\textwidth]{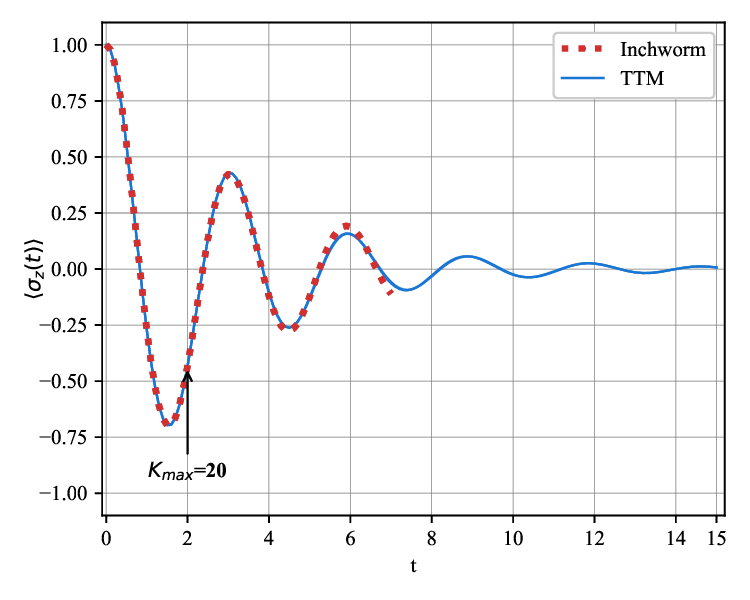}
         \caption{Spins 2 and 4}
         \label{fig_ttm_5spins_150steps_1e6_spin24}
     \end{subfigure}
     \hfill
     \begin{subfigure}[b]{0.3\textwidth}
         \centering
         \includegraphics[width=\textwidth]{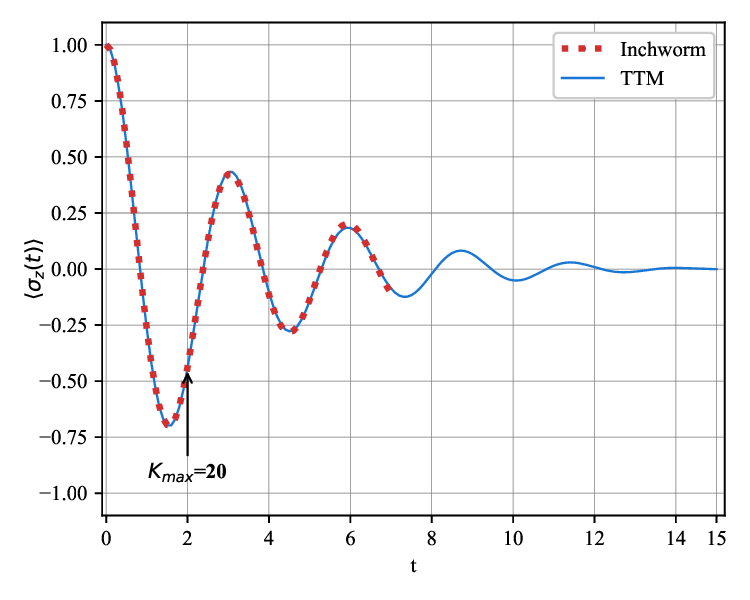}
         \caption{Spin 3}
         \label{fig_ttm_5spins_150steps_1e6_spin3}
     \end{subfigure}
    \caption{Numerical experiment for TTM with $K_{\mathrm{max}}=20$ and $\eta = 1.0\times 10^{-6}$.}
    \label{fig_ttm_5spins_150steps_1e6}
\end{figure}

We set $K_{\mathrm{max}}=20$ and $\eta = 1.0\times 10^{-6}$, use TTM to propagate to a longer time span up to $t=15$. We then compare the result with those using the inchworm method up to $t=7$, as illustrated in \cref{fig_ttm_5spins_150steps_1e6}.

\begin{figure}
    \centering
    \includegraphics[scale = 0.6]{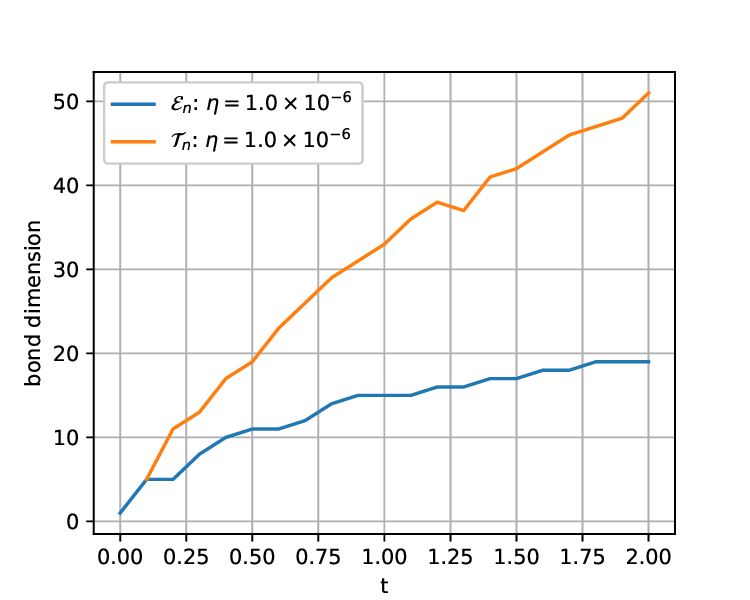}
    \caption{Evolution of the maximum bond dimension of $\mathcal{E}_n$ and $\mathcal{T}_n$ up to $n=20$.}
    \label{fig_ttm_bond_dim_10_spins}
\end{figure}

We also present the results for a $10$-spin system in \cref{fig_ttm_5spins_150steps_1e6}. With $K_{\mathrm{max}} = 20$ and $\eta = 1.0 \times 10^{-6}$, we compare the TTM results to those obtained using the inchworm method up to $t = 5$. The maximum bond dimensions of the dynamical maps $\mathcal{E}_n$ and $\mathcal{T}_n$ for $n = 1, \ldots, 20$ increase from $5$ to $19$ and from $5$ to $51$, respectively as shown in \cref{fig_ttm_bond_dim_10_spins}. 
The maximum of the full storage in \texttt{complex double} required by a single $\mathcal{E}_n$ and $\mathcal{T}_n$ up to $n=20$ is at most $16\times 4^2\times 19^2\times 10\approx0.92$ MB and $16\times 4^2 \times 51^2\times10\approx6.66$ MB, respectively.
This represents a significant reduction from the full-tensor estimate of $17.6$ TB. However, the bond dimension of $\mathcal{T}_n$ increases much more rapidly than that of $\mathcal{E}_n$, which poses a challenge in efficiently evaluating the transfer tensors for long-time simulations.
\begin{figure}
    \centering
    \begin{subfigure}[b]{0.3\textwidth}
         \centering
         \includegraphics[width=\textwidth]{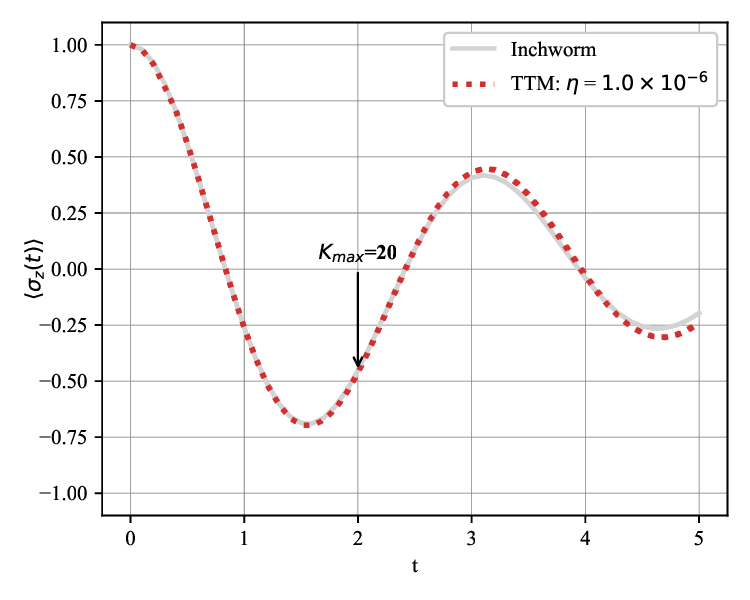}
         \caption{Spins 1 and 10}
         \label{fig_ttm_10spins_50steps_1e6_spin110}
     \end{subfigure}
     \hfill
    \begin{subfigure}[b]{0.3\textwidth}
         \centering
         \includegraphics[width=\textwidth]{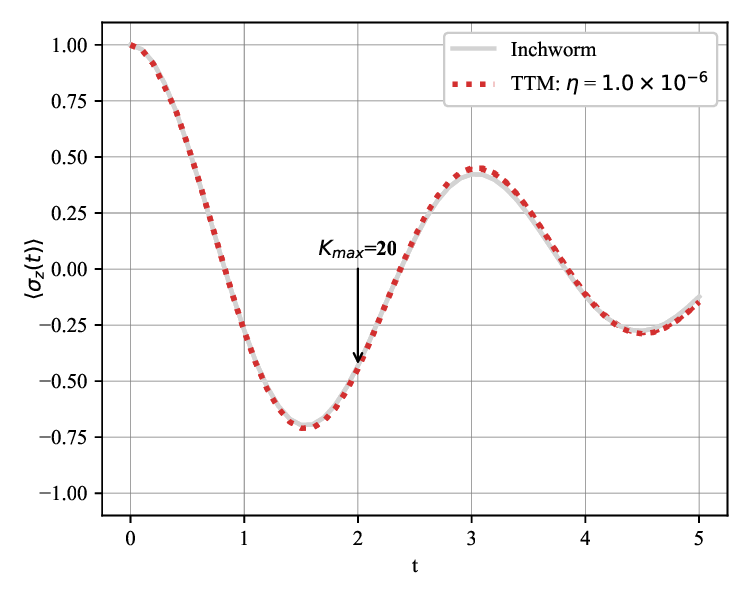}
         \caption{Spins 2 and 9}
         \label{fig_ttm_10spins_50steps_1e6_spin29}
     \end{subfigure}
     \hfill
     \begin{subfigure}[b]{0.3\textwidth}
         \centering
         \includegraphics[width=\textwidth]{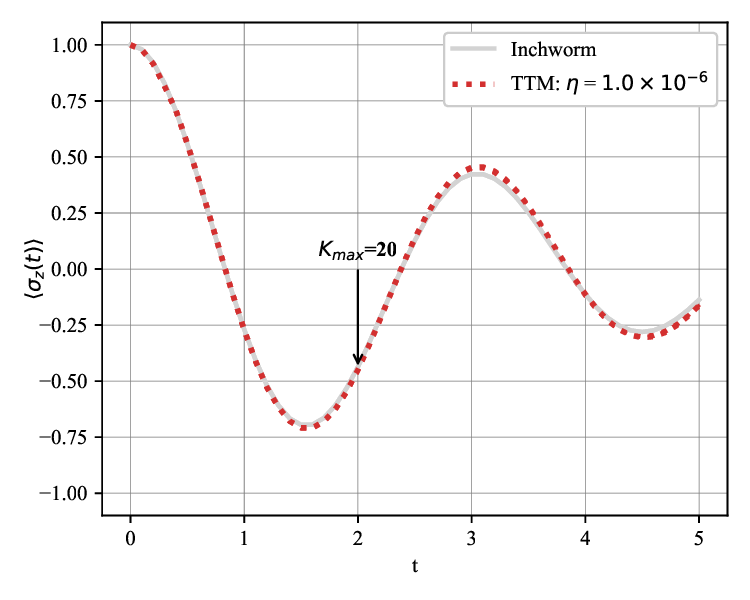}
         \caption{Spins 3 and 8}
         \label{fig_ttm_10spins_50steps_1e6_spin38}
     \end{subfigure}
    \caption{Numerical experiment for TTM with $K_{\mathrm{max}}=20$ and $\eta = 1.0\times 10^{-6}$.}
    \label{fig_ttm_10spins_50steps_1e6}
\end{figure}

\begin{figure}
    \centering
    \begin{subfigure}[b]{0.45\textwidth}
         \centering
         \includegraphics[width=\textwidth]{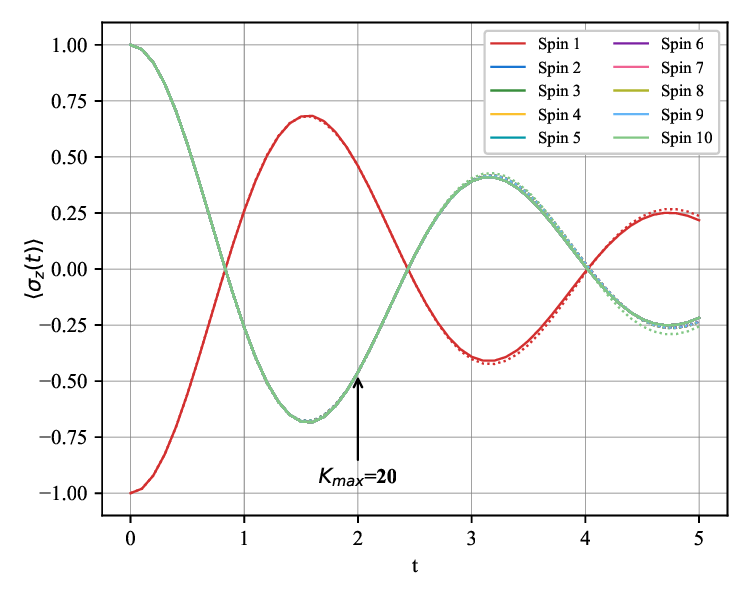}
         \caption{$J_x=J_y=0.0$}
         \label{fig_1stdown_J00}
     \end{subfigure}
     \hfill
    \begin{subfigure}[b]{0.45\textwidth}
         \centering
         \includegraphics[width=\textwidth]{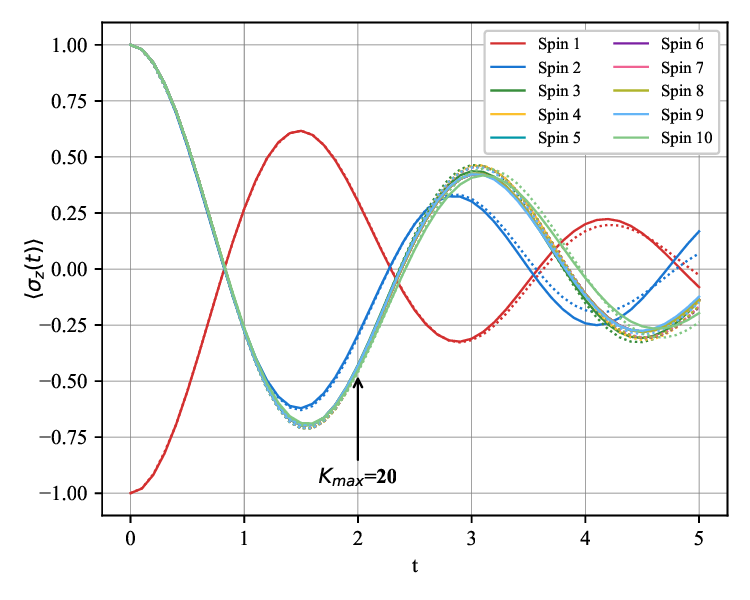}
         \caption{$J_x=J_y=0.1$}
         \label{fig_1stdown_J01}
     \end{subfigure}
    \caption{Evolution of $\langle \sigma_z^{(k)}(t) \rangle$ for a 10-spin chain with the first spin initially in $\ket{-1}$ and others in $\ket{+1}$. Solid lines show results from the inchworm method, and dotted lines are from TTM. Red lines represent $\langle \sigma_z^{(1)}(t) \rangle$, with other spins in different colors.}
    \label{fig_1stdown}
\end{figure}

Additionally, we also carry out an experiment where the first spin is initially in the state $\ket{\zeta^{(1)}}=\ket{-1}$ and all other spins have the initial states $\ket{\zeta^{(k)}}=\ket{+1}$ for $k=2,\ldots,10$. 
The evolution of the observable $\langle\sigma_z^{(k)}(t)\rangle$ is plotted in \cref{fig_1stdown}. Such a spin chain is no longer symmetric. When there is no inter-spin coupling, as shown in \cref{fig_1stdown_J00}, the curves for spins 2 to 10 are identical; however, when the inter-spin coupling is introduced, following the previous example with $J_x=J_y=0.1$,
the behavior of Spin 2 is significantly influenced by Spin 1, as evidenced in \cref{fig_1stdown_J01}. This influence is highlighted by a local minimum around $t=1.5$; the $\langle\sigma_z^{(2)}(t)\rangle$ curve is higher compared to the corresponding curve in \cref{fig_ttm_10spins_50steps_1e6_spin29}. Although we adopted an identical bath parameter setting and the same truncated memory length $K_{\text{max}}=20$, the result by TTM shows larger errors with the inchworm result, which suggests that the current setting of $\eta=1.0\times 10^{-6}$ may not be sufficiently accurate for this dynamic scenario.

\section{Concluding Remarks}
\label{sec_conclusion}

We develop a combined numerical approach to simulate spin chains with non-diagonal interspin interactions coupled to harmonic baths. For the short-time simulation, we extended the previous work on inchworm algorithms for open spin chains with diagonal couplings \cite{wang2023real} to encompass more complex scenarios involving non-diagonal spin couplings. These couplings are decomposed into sums of tensor products of local operators on each spin. Similar to the previous work, we apply the Dyson series to decompose the whole system into spin-boson units, which are then processed using the inchworm algorithm. The entire system is reconstructed by computing truncated summations over differently colored “crosses” that signify spin-spin interactions. The computational cost of this method is estimated as $O(L^{\bar{M}+\bar{N}+2})$, where $L$ represents the number of time steps, and $\bar{M}$ and $\bar{N}$ are the truncation parameters for the series expansions. While the matching process of differently colored “crosses” introduces additional costs proportional to $|A|^{\bar{N}}$, this does not alter the overall scaling with respect to $L$.

For the long-time simulation, we investigated the applicability of the transfer tensor method (TTM), a general method to simulate open quantum systems. We use the inchworm algorithm to compute the dynamical maps. Specifically, we focus on calculating “single dynamical maps” rather than single propagators and concatenate these maps during the summation process. To minimize computational storage demands, we represent dynamical maps, transfer tensors, and reduced density matrices within a tensor network framework. Additionally, we employ compression techniques to control the bond dimensions of MPO/MPS. As a result, for a $K$-spin system, the required storage, initially estimated at $16 \times 2^{4K}$ bytes, can be significantly reduced to $16 \times 4^2 \times L^2 K$ bytes, where $L$ represents the maximum bond dimension of MPOs. 

\begin{acknowledgement}
Z. Cai's work was supported by the Academic Research Fund of the Ministry of Education of Singapore under grant A-8002392-00-00.
\end{acknowledgement}
\section*{Appendix}
\begin{appendix}

\label{apx_tpc}
We present the proof of two-point correlation representation of bath influence functional $\mathcal{L}_b$ here.
According to Wick's theorem, the bath influence functional can be represented in terms of two-point correlation functions:
\begin{equation}
\mathcal{L}_b\left(\tau_1, \ldots, \tau_M\right)= \begin{cases}0, & \text { if } M \text { is odd } \\ \sum_{\mathfrak{q} \in \mathcal{Q}_M} \prod_{\left(i, j\right) \in \mathfrak{q}} \operatorname{tr}_b \left\{\mathcal{T}\left[W_{b,I}(\tau_i)W_{b,I}(\tau_j)\rho_b\right]\right\}, & \text { if } M \text { is even }\end{cases}
\end{equation}
where $W_{b,I}(\tau)=e^{\ii H_b |\tau|}W_b e^{-\ii H_b |\tau|}$.
\begin{equation}
    \begin{split}
        B(x,y)\equiv&  \operatorname{tr}_b \left\{\mathcal{T}\left[e^{\ii H_b x}W_b e^{-\ii H_b x}e^{\ii H_b y}W_b e^{-\ii H_b y}\rho_b\right]\right\}\\
        &=\frac{1}{\pi} \int_0^{\infty} J(\omega)\left[\operatorname{coth}\left(\frac{\beta \omega}{2}\right) \cos (\omega (x-y))-\mathrm{i} \sin (\omega (x-y))\right] \mathrm{d} \omega
    \end{split}
\end{equation}
We define $B^*(x,y)$ as
\begin{equation}
    B^*(x,y)\equiv\frac{1}{\pi} \int_0^{\infty} J(\omega)\left[\operatorname{coth}\left(\frac{\beta \omega}{2}\right) \cos (\omega \Delta t)-\mathrm{i} \sin (\omega \Delta t)\right] \mathrm{d} \omega
\end{equation}
where $\Delta t = |x|-|y|$. 
We consider the following cases:
\begin{itemize}
    \item If $\tau_2\geqslant \tau_1\geqslant 0$,
    \begin{equation}
        \begin{split}
            B(\tau_1,\tau_2) 
            &= \tr_b \left(\mathcal{T}\left[W_{b,I}(\tau_i)W_{b,I}(\tau_j)\rho_b(0)\right]\right) \\
            &= \tr_b \left(
                \e^{\ii H_b \tau_2} W_b \e^{-\ii H_b \tau_2}
                \e^{\ii H_b \tau_1} W_b \e^{-\ii H_b \tau_1}
                \rho_b
            \right) \\
         &=\overline{B^*(\tau_1,\tau_2)}
        \end{split}
    \end{equation}
    \item $\tau_j\geq 0\geq \tau_i$,
    \begin{equation}
        \begin{split}
            \operatorname{tr}_b \left\{\mathcal{T}\left[W_{b,I}(\tau_i)W_{b,I}(\tau_j)\rho_b\right]\right\}&=\operatorname{tr}_b \left\{e^{\ii H_b \tau_j}W_b e^{-\ii H_b \tau_j}\rho_be^{-\ii H_b \tau_i}W_b e^{\ii H_b \tau_i}\right\}\\
            &= \operatorname{tr}_b \left\{e^{-\ii H_b \tau_i}W_b e^{\ii H_b \tau_i} e^{\ii H_b \tau_j}W_b e^{-\ii H_b \tau_j}\rho_b\right\}\\
         &=B(-\tau_i,\tau_j)\\
         &=B^*(\tau_i,\tau_j)
        \end{split}
    \end{equation}
    \item $0\geq \tau_j\geq \tau_i$,
    \begin{equation}
        \begin{split}
            \operatorname{tr}_b \left\{\mathcal{T}\left[W_{b,I}(\tau_i)W_{b,I}(\tau_j)\rho_b\right]\right\}&=\operatorname{tr}_b \left\{\rho_b e^{ -\ii H_b \tau_j}W_b e^{\ii H_b \tau_j}e^{-\ii H_b \tau_i}W_b e^{\ii H_b \tau_i}\right\}\\
            &=B(-\tau_j,-\tau_i)\\
            &=\overline{B^*(\tau_i,\tau_j)}
        \end{split}
    \end{equation}
\end{itemize}
Therefore, 
\begin{equation}
    \operatorname{tr}_b \left\{\mathcal{T}\left[W_{b,I}(\tau_i)W_{b,I}(\tau_j)\rho_b\right]\right\}=\begin{cases}
        \overline{B^*(\tau_i,\tau_j)},&\text{ if } \tau_i\tau_j> 0\\
        B^* (\tau_i,\tau_j),& \text{ if } \tau_i\tau_j \leq 0
    \end{cases}
\end{equation}

\end{appendix}

\bibliography{myBib}

\providecommand{\latin}[1]{#1}
\makeatletter
\providecommand{\doi}
  {\begingroup\let\do\@makeother\dospecials
  \catcode`\{=1 \catcode`\}=2 \doi@aux}
\providecommand{\doi@aux}[1]{\endgroup\texttt{#1}}
\makeatother
\providecommand*\mcitethebibliography{\thebibliography}
\csname @ifundefined\endcsname{endmcitethebibliography}  {\let\endmcitethebibliography\endthebibliography}{}
\begin{mcitethebibliography}{66}
\providecommand*\natexlab[1]{#1}
\providecommand*\mciteSetBstSublistMode[1]{}
\providecommand*\mciteSetBstMaxWidthForm[2]{}
\providecommand*\mciteBstWouldAddEndPuncttrue
  {\def\EndOfBibitem{\unskip.}}
\providecommand*\mciteBstWouldAddEndPunctfalse
  {\let\EndOfBibitem\relax}
\providecommand*\mciteSetBstMidEndSepPunct[3]{}
\providecommand*\mciteSetBstSublistLabelBeginEnd[3]{}
\providecommand*\EndOfBibitem{}
\mciteSetBstSublistMode{f}
\mciteSetBstMaxWidthForm{subitem}{(\alph{mcitesubitemcount})}
\mciteSetBstSublistLabelBeginEnd
  {\mcitemaxwidthsubitemform\space}
  {\relax}
  {\relax}

\bibitem[Esposito \latin{et~al.}(2010)Esposito, Lindenberg, and Van~den Broeck]{esposito2010entropy}
Esposito,~M.; Lindenberg,~K.; Van~den Broeck,~C. Entropy production as correlation between system and reservoir. \emph{New J. Phys.} \textbf{2010}, \emph{12}, 013013\relax
\mciteBstWouldAddEndPuncttrue
\mciteSetBstMidEndSepPunct{\mcitedefaultmidpunct}
{\mcitedefaultendpunct}{\mcitedefaultseppunct}\relax
\EndOfBibitem
\bibitem[Nielsen and Chuang(2002)Nielsen, and Chuang]{nielsen2002quantum}
Nielsen,~M.~A.; Chuang,~I. Quantum computation and quantum information. 2002\relax
\mciteBstWouldAddEndPuncttrue
\mciteSetBstMidEndSepPunct{\mcitedefaultmidpunct}
{\mcitedefaultendpunct}{\mcitedefaultseppunct}\relax
\EndOfBibitem
\bibitem[Correa \latin{et~al.}(2017)Correa, Perarnau-Llobet, Hovhannisyan, Hern{\'a}ndez-Santana, Mehboudi, and Sanpera]{correa2017enhancement}
Correa,~L.~A.; Perarnau-Llobet,~M.; Hovhannisyan,~K.~V.; Hern{\'a}ndez-Santana,~S.; Mehboudi,~M.; Sanpera,~A. Enhancement of low-temperature thermometry by strong coupling. \emph{Phys. Rev. A} \textbf{2017}, \emph{96}, 062103\relax
\mciteBstWouldAddEndPuncttrue
\mciteSetBstMidEndSepPunct{\mcitedefaultmidpunct}
{\mcitedefaultendpunct}{\mcitedefaultseppunct}\relax
\EndOfBibitem
\bibitem[Salado-Mej{\'\i}a \latin{et~al.}(2021)Salado-Mej{\'\i}a, Rom{\'a}n-Ancheyta, Soto-Eguibar, and Moya-Cessa]{salado2021spectroscopy}
Salado-Mej{\'\i}a,~M.; Rom{\'a}n-Ancheyta,~R.; Soto-Eguibar,~F.; Moya-Cessa,~H. Spectroscopy and critical quantum thermometry in the ultrastrong coupling regime. \emph{Quantum Sci. Technol.} \textbf{2021}, \emph{6}, 025010\relax
\mciteBstWouldAddEndPuncttrue
\mciteSetBstMidEndSepPunct{\mcitedefaultmidpunct}
{\mcitedefaultendpunct}{\mcitedefaultseppunct}\relax
\EndOfBibitem
\bibitem[Leggett \latin{et~al.}(1987)Leggett, Chakravarty, Dorsey, Fisher, Garg, and Zwerger]{leggett1987dynamics}
Leggett,~A.~J.; Chakravarty,~S.; Dorsey,~A.~T.; Fisher,~M.~P.; Garg,~A.; Zwerger,~W. Dynamics of the dissipative two-state system. \emph{Rev. Mod. Phys.} \textbf{1987}, \emph{59}, 1\relax
\mciteBstWouldAddEndPuncttrue
\mciteSetBstMidEndSepPunct{\mcitedefaultmidpunct}
{\mcitedefaultendpunct}{\mcitedefaultseppunct}\relax
\EndOfBibitem
\bibitem[Cerrillo and Cao(2014)Cerrillo, and Cao]{cerrillo2014nonmarkovian}
Cerrillo,~J.; Cao,~J. Non-{M}arkovian dynamical maps: numerical processing of open quantum trajectories. \emph{Phys. Rev. Lett.} \textbf{2014}, \emph{112}, 110401\relax
\mciteBstWouldAddEndPuncttrue
\mciteSetBstMidEndSepPunct{\mcitedefaultmidpunct}
{\mcitedefaultendpunct}{\mcitedefaultseppunct}\relax
\EndOfBibitem
\bibitem[Strathearn \latin{et~al.}(2018)Strathearn, Kirton, Kilda, Keeling, and Lovett]{strathearn2018efficient}
Strathearn,~A.; Kirton,~P.; Kilda,~D.; Keeling,~J.; Lovett,~B.~W. Efficient non-{M}arkovian quantum dynamics using time-evolving matrix product operators. \emph{Nat. Commun.} \textbf{2018}, \emph{9}, 3322\relax
\mciteBstWouldAddEndPuncttrue
\mciteSetBstMidEndSepPunct{\mcitedefaultmidpunct}
{\mcitedefaultendpunct}{\mcitedefaultseppunct}\relax
\EndOfBibitem
\bibitem[Lindblad(1976)]{lindblad1976generators}
Lindblad,~G. On the generators of quantum dynamical semigroups. \emph{Commun. Math. Phys.} \textbf{1976}, \emph{48}, 119--130\relax
\mciteBstWouldAddEndPuncttrue
\mciteSetBstMidEndSepPunct{\mcitedefaultmidpunct}
{\mcitedefaultendpunct}{\mcitedefaultseppunct}\relax
\EndOfBibitem
\bibitem[Nakajima(1958)]{nakajima1958quantum}
Nakajima,~S. On quantum theory of transport phenomena: Steady diffusion. \emph{Prog. Theor. Phys.} \textbf{1958}, \emph{20}, 948--959\relax
\mciteBstWouldAddEndPuncttrue
\mciteSetBstMidEndSepPunct{\mcitedefaultmidpunct}
{\mcitedefaultendpunct}{\mcitedefaultseppunct}\relax
\EndOfBibitem
\bibitem[Zwanzig(1960)]{zwanzig1960ensemble}
Zwanzig,~R. Ensemble method in the theory of irreversibility. \emph{J. Chem. Phys.} \textbf{1960}, \emph{33}, 1338--1341\relax
\mciteBstWouldAddEndPuncttrue
\mciteSetBstMidEndSepPunct{\mcitedefaultmidpunct}
{\mcitedefaultendpunct}{\mcitedefaultseppunct}\relax
\EndOfBibitem
\bibitem[Makri(1995)]{makri1995numerical}
Makri,~N. Numerical path integral techniques for long time dynamics of quantum dissipative systems. \emph{J. Math. Phys.} \textbf{1995}, \emph{36}, 2430--2457\relax
\mciteBstWouldAddEndPuncttrue
\mciteSetBstMidEndSepPunct{\mcitedefaultmidpunct}
{\mcitedefaultendpunct}{\mcitedefaultseppunct}\relax
\EndOfBibitem
\bibitem[Makri(1998)]{makri1998quantum}
Makri,~N. Quantum dissipative dynamics: A numerically exact methodology. \emph{J. Phys. Chem. A} \textbf{1998}, \emph{102}, 4414--4427\relax
\mciteBstWouldAddEndPuncttrue
\mciteSetBstMidEndSepPunct{\mcitedefaultmidpunct}
{\mcitedefaultendpunct}{\mcitedefaultseppunct}\relax
\EndOfBibitem
\bibitem[Feynman and Vernon(1963)Feynman, and Vernon]{feynman1963theory}
Feynman,~R.~P.; Vernon,~F. The theory of a general quantum system interacting with a linear dissipative system. \emph{Ann. Phys.} \textbf{1963}, \emph{24}, 118--173\relax
\mciteBstWouldAddEndPuncttrue
\mciteSetBstMidEndSepPunct{\mcitedefaultmidpunct}
{\mcitedefaultendpunct}{\mcitedefaultseppunct}\relax
\EndOfBibitem
\bibitem[Makri(2014)]{makri2014blip}
Makri,~N. Blip decomposition of the path integral: Exponential acceleration of real-time calculations on quantum dissipative systems. \emph{J. Chem. Phys.} \textbf{2014}, \emph{141}, 134117\relax
\mciteBstWouldAddEndPuncttrue
\mciteSetBstMidEndSepPunct{\mcitedefaultmidpunct}
{\mcitedefaultendpunct}{\mcitedefaultseppunct}\relax
\EndOfBibitem
\bibitem[Wang and Cai(2022)Wang, and Cai]{wang2022differential}
Wang,~G.; Cai,~Z. Differential Equation Based Path Integral for Open Quantum Systems. \emph{SIAM J. Sci. Comput.} \textbf{2022}, \emph{44}, B771--B804\relax
\mciteBstWouldAddEndPuncttrue
\mciteSetBstMidEndSepPunct{\mcitedefaultmidpunct}
{\mcitedefaultendpunct}{\mcitedefaultseppunct}\relax
\EndOfBibitem
\bibitem[Makri(2024)]{makri2024kink}
Makri,~N. Kink Sum for Long-Memory Small Matrix Path Integral Dynamics. \emph{J. Phys. Chem. B} \textbf{2024}, \relax
\mciteBstWouldAddEndPunctfalse
\mciteSetBstMidEndSepPunct{\mcitedefaultmidpunct}
{}{\mcitedefaultseppunct}\relax
\EndOfBibitem
\bibitem[Makri(2020)]{makri2020smallMatrixPath}
Makri,~N. Small matrix path integral for system-bath dynamics. \emph{J. Chem. Theory Comput.} \textbf{2020}, \emph{16}, 4038--4049\relax
\mciteBstWouldAddEndPuncttrue
\mciteSetBstMidEndSepPunct{\mcitedefaultmidpunct}
{\mcitedefaultendpunct}{\mcitedefaultseppunct}\relax
\EndOfBibitem
\bibitem[Makri(2021)]{makri2021smallMatrixPathIntegralExtended}
Makri,~N. Small matrix path integral with extended memory. \emph{J. Chem. Theory Comput.} \textbf{2021}, \emph{17}, 1--6\relax
\mciteBstWouldAddEndPuncttrue
\mciteSetBstMidEndSepPunct{\mcitedefaultmidpunct}
{\mcitedefaultendpunct}{\mcitedefaultseppunct}\relax
\EndOfBibitem
\bibitem[Wang and Cai(To appear)Wang, and Cai]{wang2024tree}
Wang,~G.; Cai,~Z. Tree-based Implementation of the Small Matrix Path Integral for System-Bath Dynamics. \emph{Commun. Comput. Phys.} \textbf{To appear}, \relax
\mciteBstWouldAddEndPunctfalse
\mciteSetBstMidEndSepPunct{\mcitedefaultmidpunct}
{}{\mcitedefaultseppunct}\relax
\EndOfBibitem
\bibitem[Erpenbeck \latin{et~al.}(2023)Erpenbeck, Gull, and Cohen]{erpenbeck2023quantum}
Erpenbeck,~A.; Gull,~E.; Cohen,~G. Quantum Monte Carlo method in the steady state. \emph{Phys. Rev. Lett.} \textbf{2023}, \emph{130}, 186301\relax
\mciteBstWouldAddEndPuncttrue
\mciteSetBstMidEndSepPunct{\mcitedefaultmidpunct}
{\mcitedefaultendpunct}{\mcitedefaultseppunct}\relax
\EndOfBibitem
\bibitem[Xu \latin{et~al.}(2017)Xu, Song, Song, and Shi]{xu2017convergence}
Xu,~M.; Song,~L.; Song,~K.; Shi,~Q. Convergence of high order perturbative expansions in open system quantum dynamics. \emph{J. Chem. Phys.} \textbf{2017}, \emph{146}\relax
\mciteBstWouldAddEndPuncttrue
\mciteSetBstMidEndSepPunct{\mcitedefaultmidpunct}
{\mcitedefaultendpunct}{\mcitedefaultseppunct}\relax
\EndOfBibitem
\bibitem[Dyson(1949)]{dyson1949radiation}
Dyson,~F.~J. The radiation theories of {T}omonaga, {S}chwinger, and {F}eynman. \emph{Phys. Rev.} \textbf{1949}, \emph{75}, 486\relax
\mciteBstWouldAddEndPuncttrue
\mciteSetBstMidEndSepPunct{\mcitedefaultmidpunct}
{\mcitedefaultendpunct}{\mcitedefaultseppunct}\relax
\EndOfBibitem
\bibitem[Wick(1950)]{wick1950evaluation}
Wick,~G.-C. The evaluation of the collision matrix. \emph{Phys. Rev.} \textbf{1950}, \emph{80}, 268\relax
\mciteBstWouldAddEndPuncttrue
\mciteSetBstMidEndSepPunct{\mcitedefaultmidpunct}
{\mcitedefaultendpunct}{\mcitedefaultseppunct}\relax
\EndOfBibitem
\bibitem[Jr. \latin{et~al.}(1990)Jr., Gubernatis, Scalettar, White, Scalapino, and Sugar]{loh1990sign}
Jr.,~L.~E.; Gubernatis,~J.; Scalettar,~R.; White,~S.; Scalapino,~D.; Sugar,~R. Sign problem in the numerical simulation of many-electron systems. \emph{Phys. Rev. B} \textbf{1990}, \emph{41}, 9301\relax
\mciteBstWouldAddEndPuncttrue
\mciteSetBstMidEndSepPunct{\mcitedefaultmidpunct}
{\mcitedefaultendpunct}{\mcitedefaultseppunct}\relax
\EndOfBibitem
\bibitem[Cai \latin{et~al.}(2023)Cai, Lu, and Yang]{cai2023numerical}
Cai,~Z.; Lu,~J.; Yang,~S. Numerical analysis for inchworm Monte Carlo method: Sign problem and error growth. \emph{Math. Comput.} \textbf{2023}, \emph{92}, 1141--1209\relax
\mciteBstWouldAddEndPuncttrue
\mciteSetBstMidEndSepPunct{\mcitedefaultmidpunct}
{\mcitedefaultendpunct}{\mcitedefaultseppunct}\relax
\EndOfBibitem
\bibitem[Chen \latin{et~al.}(2017)Chen, Cohen, and Reichman]{chen2017inchwormITheory}
Chen,~H.-T.; Cohen,~G.; Reichman,~D.~R. Inchworm {M}onte {C}arlo for exact non-adiabatic dynamics. I. Theory and algorithms. \emph{J. Chem. Phys.} \textbf{2017}, \emph{146}, 054105\relax
\mciteBstWouldAddEndPuncttrue
\mciteSetBstMidEndSepPunct{\mcitedefaultmidpunct}
{\mcitedefaultendpunct}{\mcitedefaultseppunct}\relax
\EndOfBibitem
\bibitem[Chen \latin{et~al.}(2017)Chen, Cohen, and Reichman]{chen2017inchwormIIBenchmarks}
Chen,~H.-T.; Cohen,~G.; Reichman,~D.~R. Inchworm {M}onte {C}arlo for exact non-adiabatic dynamics. II. Benchmarks and comparison with established methods. \emph{J. Chem. Phys.} \textbf{2017}, \emph{146}, 054106\relax
\mciteBstWouldAddEndPuncttrue
\mciteSetBstMidEndSepPunct{\mcitedefaultmidpunct}
{\mcitedefaultendpunct}{\mcitedefaultseppunct}\relax
\EndOfBibitem
\bibitem[Cai \latin{et~al.}(2020)Cai, Lu, and Yang]{cai2020inchworm}
Cai,~Z.; Lu,~J.; Yang,~S. Inchworm {M}onte {C}arlo method for open quantum systems. \emph{Commun. Pure Appl. Math.} \textbf{2020}, \emph{73}, 2430--2472\relax
\mciteBstWouldAddEndPuncttrue
\mciteSetBstMidEndSepPunct{\mcitedefaultmidpunct}
{\mcitedefaultendpunct}{\mcitedefaultseppunct}\relax
\EndOfBibitem
\bibitem[Prokof’ev and Svistunov(2007)Prokof’ev, and Svistunov]{prokof2007bold}
Prokof’ev,~N.; Svistunov,~B. Bold diagrammatic {M}onte {C}arlo technique: When the sign problem is welcome. \emph{Phys. Rev. Lett.} \textbf{2007}, \emph{99}, 250201\relax
\mciteBstWouldAddEndPuncttrue
\mciteSetBstMidEndSepPunct{\mcitedefaultmidpunct}
{\mcitedefaultendpunct}{\mcitedefaultseppunct}\relax
\EndOfBibitem
\bibitem[Prokof’ev and Svistunov(2008)Prokof’ev, and Svistunov]{prokof2008bold}
Prokof’ev,~N.; Svistunov,~B. Bold diagrammatic {M}onte {C}arlo: A generic sign-problem tolerant technique for polaron models and possibly interacting many-body problems. \emph{Phys. Rev. B} \textbf{2008}, \emph{77}, 125101\relax
\mciteBstWouldAddEndPuncttrue
\mciteSetBstMidEndSepPunct{\mcitedefaultmidpunct}
{\mcitedefaultendpunct}{\mcitedefaultseppunct}\relax
\EndOfBibitem
\bibitem[Cai and Wang(2022)Cai, and Wang]{cai2022numerical}
Cai,~Z.; Wang,~Y. Numerical solver for the {B}oltzmann equation with self-adaptive collision operators. \emph{SIAM J. Sci. Comput.} \textbf{2022}, \emph{44}, B275--B309\relax
\mciteBstWouldAddEndPuncttrue
\mciteSetBstMidEndSepPunct{\mcitedefaultmidpunct}
{\mcitedefaultendpunct}{\mcitedefaultseppunct}\relax
\EndOfBibitem
\bibitem[Boag \latin{et~al.}(2018)Boag, Gull, and Cohen]{boag2018inclusion}
Boag,~A.; Gull,~E.; Cohen,~G. Inclusion-exclusion principle for many-body diagrammatics. \emph{Phys. Rev. B} \textbf{2018}, \emph{98}, 115152\relax
\mciteBstWouldAddEndPuncttrue
\mciteSetBstMidEndSepPunct{\mcitedefaultmidpunct}
{\mcitedefaultendpunct}{\mcitedefaultseppunct}\relax
\EndOfBibitem
\bibitem[Yang \latin{et~al.}(2021)Yang, Cai, and Lu]{yang2021inclusion}
Yang,~S.; Cai,~Z.; Lu,~J. Inclusion--exclusion principle for open quantum systems with bosonic bath. \emph{New J. Phys.} \textbf{2021}, \emph{23}, 063049\relax
\mciteBstWouldAddEndPuncttrue
\mciteSetBstMidEndSepPunct{\mcitedefaultmidpunct}
{\mcitedefaultendpunct}{\mcitedefaultseppunct}\relax
\EndOfBibitem
\bibitem[Cai \latin{et~al.}(2022)Cai, Lu, and Yang]{cai2022fast}
Cai,~Z.; Lu,~J.; Yang,~S. Fast algorithms of bath calculations in simulations of quantum system-bath dynamics. \emph{Comput. Phys. Commun.} \textbf{2022}, 108417\relax
\mciteBstWouldAddEndPuncttrue
\mciteSetBstMidEndSepPunct{\mcitedefaultmidpunct}
{\mcitedefaultendpunct}{\mcitedefaultseppunct}\relax
\EndOfBibitem
\bibitem[Cai \latin{et~al.}(2023)Cai, Wang, and Yang]{cai2023bold}
Cai,~Z.; Wang,~G.; Yang,~S. The bold-thin-bold diagrammatic {M}onte {C}arlo method for open quantum systems. \emph{SIAM J. Sci. Comput.} \textbf{2023}, \emph{45}, A1812--A1843\relax
\mciteBstWouldAddEndPuncttrue
\mciteSetBstMidEndSepPunct{\mcitedefaultmidpunct}
{\mcitedefaultendpunct}{\mcitedefaultseppunct}\relax
\EndOfBibitem
\bibitem[Luck(1993)]{luck1993critical}
Luck,~J.-M. Critical behavior of the aperiodic quantum {I}sing chain in a transverse magnetic field. \emph{J. Stat. Phys.} \textbf{1993}, \emph{72}, 417--458\relax
\mciteBstWouldAddEndPuncttrue
\mciteSetBstMidEndSepPunct{\mcitedefaultmidpunct}
{\mcitedefaultendpunct}{\mcitedefaultseppunct}\relax
\EndOfBibitem
\bibitem[Hopfield(1982)]{hopfield1982neural}
Hopfield,~J.~J. Neural networks and physical systems with emergent collective computational abilities. \emph{Proc. Natl. Acad. Sci. U.S.A.} \textbf{1982}, \emph{79}, 2554--2558\relax
\mciteBstWouldAddEndPuncttrue
\mciteSetBstMidEndSepPunct{\mcitedefaultmidpunct}
{\mcitedefaultendpunct}{\mcitedefaultseppunct}\relax
\EndOfBibitem
\bibitem[Meier \latin{et~al.}(2003)Meier, Levy, and Loss]{meier2003quantum}
Meier,~F.; Levy,~J.; Loss,~D. Quantum computing with spin cluster qubits. \emph{Phys. Rev. Lett.} \textbf{2003}, \emph{90}, 047901\relax
\mciteBstWouldAddEndPuncttrue
\mciteSetBstMidEndSepPunct{\mcitedefaultmidpunct}
{\mcitedefaultendpunct}{\mcitedefaultseppunct}\relax
\EndOfBibitem
\bibitem[Schneidman \latin{et~al.}(2006)Schneidman, Berry, Segev, and Bialek]{schneidman2006weak}
Schneidman,~E.; Berry,~M.~J.; Segev,~R.; Bialek,~W. Weak pairwise correlations imply strongly correlated network states in a neural population. \emph{Nature} \textbf{2006}, \emph{440}, 1007--1012\relax
\mciteBstWouldAddEndPuncttrue
\mciteSetBstMidEndSepPunct{\mcitedefaultmidpunct}
{\mcitedefaultendpunct}{\mcitedefaultseppunct}\relax
\EndOfBibitem
\bibitem[J{\o}rgensen and Pollock(2019)J{\o}rgensen, and Pollock]{jorgensen2019exploiting}
J{\o}rgensen,~M.~R.; Pollock,~F.~A. Exploiting the causal tensor network structure of quantum processes to efficiently simulate non-{M}arkovian path integrals. \emph{Phys. Rev. Lett.} \textbf{2019}, \emph{123}, 240602\relax
\mciteBstWouldAddEndPuncttrue
\mciteSetBstMidEndSepPunct{\mcitedefaultmidpunct}
{\mcitedefaultendpunct}{\mcitedefaultseppunct}\relax
\EndOfBibitem
\bibitem[Ye and Chan(2021)Ye, and Chan]{ye2021constructing}
Ye,~E.; Chan,~G.~K. Constructing tensor network influence functionals for general quantum dynamics. \emph{J. Chem. Phys.} \textbf{2021}, \emph{155}\relax
\mciteBstWouldAddEndPuncttrue
\mciteSetBstMidEndSepPunct{\mcitedefaultmidpunct}
{\mcitedefaultendpunct}{\mcitedefaultseppunct}\relax
\EndOfBibitem
\bibitem[Bose(2022)]{bose2022pairwise}
Bose,~A. Pairwise connected tensor network representation of path integrals. \emph{Phys. Rev. B} \textbf{2022}, \emph{105}, 024309\relax
\mciteBstWouldAddEndPuncttrue
\mciteSetBstMidEndSepPunct{\mcitedefaultmidpunct}
{\mcitedefaultendpunct}{\mcitedefaultseppunct}\relax
\EndOfBibitem
\bibitem[Erpenbeck \latin{et~al.}(2023)Erpenbeck, Lin, Blommel, Zhang, Iskakov, Bernheimer, N{\'u}{\~n}ez-Fern{\'a}ndez, Cohen, Parcollet, Waintal, and Gull]{erpenbeck2023tensor}
Erpenbeck,~A.; Lin,~W.-T.; Blommel,~T.; Zhang,~L.; Iskakov,~S.; Bernheimer,~L.; N{\'u}{\~n}ez-Fern{\'a}ndez,~Y.; Cohen,~G.; Parcollet,~O.; Waintal,~X.; Gull,~E. Tensor train continuous time solver for quantum impurity models. \emph{Phys. Rev. B} \textbf{2023}, \emph{107}, 245135\relax
\mciteBstWouldAddEndPuncttrue
\mciteSetBstMidEndSepPunct{\mcitedefaultmidpunct}
{\mcitedefaultendpunct}{\mcitedefaultseppunct}\relax
\EndOfBibitem
\bibitem[Bose and Walters(2022)Bose, and Walters]{bose2022multisite}
Bose,~A.; Walters,~P.~L. A multisite decomposition of the tensor network path integrals. \emph{J. Chem. Phys.} \textbf{2022}, \emph{156}\relax
\mciteBstWouldAddEndPuncttrue
\mciteSetBstMidEndSepPunct{\mcitedefaultmidpunct}
{\mcitedefaultendpunct}{\mcitedefaultseppunct}\relax
\EndOfBibitem
\bibitem[Bose and Walters(2022)Bose, and Walters]{bose2022tensor}
Bose,~A.; Walters,~P.~L. Tensor network path integral study of dynamics in B850 LH2 ring with atomistically derived vibrations. \emph{J. Chem. Theory Comput.} \textbf{2022}, \emph{18}, 4095--4108\relax
\mciteBstWouldAddEndPuncttrue
\mciteSetBstMidEndSepPunct{\mcitedefaultmidpunct}
{\mcitedefaultendpunct}{\mcitedefaultseppunct}\relax
\EndOfBibitem
\bibitem[White(1992)]{white1992density}
White,~S.~R. Density matrix formulation for quantum renormalization groups. \emph{Phys. Rev. Lett.} \textbf{1992}, \emph{69}, 2863\relax
\mciteBstWouldAddEndPuncttrue
\mciteSetBstMidEndSepPunct{\mcitedefaultmidpunct}
{\mcitedefaultendpunct}{\mcitedefaultseppunct}\relax
\EndOfBibitem
\bibitem[White(1993)]{white1993density}
White,~S.~R. Density-matrix algorithms for quantum renormalization groups. \emph{Phys. Rev. B} \textbf{1993}, \emph{48}, 10345\relax
\mciteBstWouldAddEndPuncttrue
\mciteSetBstMidEndSepPunct{\mcitedefaultmidpunct}
{\mcitedefaultendpunct}{\mcitedefaultseppunct}\relax
\EndOfBibitem
\bibitem[Makri(2018)]{makri2018modular}
Makri,~N. Modular path integral methodology for real-time quantum dynamics. \emph{J. Chem. Phys.} \textbf{2018}, \emph{149}, 214108\relax
\mciteBstWouldAddEndPuncttrue
\mciteSetBstMidEndSepPunct{\mcitedefaultmidpunct}
{\mcitedefaultendpunct}{\mcitedefaultseppunct}\relax
\EndOfBibitem
\bibitem[Makri(2018)]{makri2018communication}
Makri,~N. Communication: Modular path integral: Quantum dynamics via sequential necklace linking. \emph{J. Chem. Phys.} \textbf{2018}, \emph{148}, 101101\relax
\mciteBstWouldAddEndPuncttrue
\mciteSetBstMidEndSepPunct{\mcitedefaultmidpunct}
{\mcitedefaultendpunct}{\mcitedefaultseppunct}\relax
\EndOfBibitem
\bibitem[Kundu and Makri(2019)Kundu, and Makri]{kundu2019modular}
Kundu,~S.; Makri,~N. Modular path integral for discrete systems with non-diagonal couplings. \emph{J. Chem. Phys.} \textbf{2019}, \emph{151}, 074110\relax
\mciteBstWouldAddEndPuncttrue
\mciteSetBstMidEndSepPunct{\mcitedefaultmidpunct}
{\mcitedefaultendpunct}{\mcitedefaultseppunct}\relax
\EndOfBibitem
\bibitem[Kundu and Makri(2020)Kundu, and Makri]{kundu2020modular}
Kundu,~S.; Makri,~N. Modular path integral for finite-temperature dynamics of extended systems with intramolecular vibrations. \emph{J. Chem. Phys.} \textbf{2020}, \emph{153}, 044124\relax
\mciteBstWouldAddEndPuncttrue
\mciteSetBstMidEndSepPunct{\mcitedefaultmidpunct}
{\mcitedefaultendpunct}{\mcitedefaultseppunct}\relax
\EndOfBibitem
\bibitem[Kundu and Makri(2021)Kundu, and Makri]{kundu2021efficient}
Kundu,~S.; Makri,~N. Efficient matrix factorisation of the modular path integral for extended systems. \emph{Mol. Phys.} \textbf{2021}, \emph{119}, e1797200\relax
\mciteBstWouldAddEndPuncttrue
\mciteSetBstMidEndSepPunct{\mcitedefaultmidpunct}
{\mcitedefaultendpunct}{\mcitedefaultseppunct}\relax
\EndOfBibitem
\bibitem[Wang and Cai(2023)Wang, and Cai]{wang2023real}
Wang,~G.; Cai,~Z. Real-Time Simulation of Open Quantum Spin Chains with the Inchworm Method. \emph{J. Chem. Theory Comput.} \textbf{2023}, \emph{19}, 8523--8540\relax
\mciteBstWouldAddEndPuncttrue
\mciteSetBstMidEndSepPunct{\mcitedefaultmidpunct}
{\mcitedefaultendpunct}{\mcitedefaultseppunct}\relax
\EndOfBibitem
\bibitem[Caldeira and Leggett(1983)Caldeira, and Leggett]{caldeira1983path}
Caldeira,~A.~O.; Leggett,~A.~J. Path integral approach to quantum Brownian motion. \emph{Phys. A} \textbf{1983}, \emph{121}, 587--616\relax
\mciteBstWouldAddEndPuncttrue
\mciteSetBstMidEndSepPunct{\mcitedefaultmidpunct}
{\mcitedefaultendpunct}{\mcitedefaultseppunct}\relax
\EndOfBibitem
\bibitem[Ising(1924)]{ising1924beitrag}
Ising,~E. Beitrag zur theorie des ferro-und paramagnetismus. Ph.D.\ thesis, Grefe \& Tiedemann Hamburg, 1924\relax
\mciteBstWouldAddEndPuncttrue
\mciteSetBstMidEndSepPunct{\mcitedefaultmidpunct}
{\mcitedefaultendpunct}{\mcitedefaultseppunct}\relax
\EndOfBibitem
\bibitem[Dani and Makri(2021)Dani, and Makri]{dani2021quantum}
Dani,~R.; Makri,~N. Quantum quench and coherent--incoherent dynamics of {I}sing chains interacting with dissipative baths. \emph{J. Chem. Phys.} \textbf{2021}, \emph{155}, 234705\relax
\mciteBstWouldAddEndPuncttrue
\mciteSetBstMidEndSepPunct{\mcitedefaultmidpunct}
{\mcitedefaultendpunct}{\mcitedefaultseppunct}\relax
\EndOfBibitem
\bibitem[Breuer and Petruccione(2002)Breuer, and Petruccione]{breuer2002theory}
Breuer,~H.-P.; Petruccione,~F. \emph{The theory of open quantum systems}; Oxford University Press on Demand, 2002\relax
\mciteBstWouldAddEndPuncttrue
\mciteSetBstMidEndSepPunct{\mcitedefaultmidpunct}
{\mcitedefaultendpunct}{\mcitedefaultseppunct}\relax
\EndOfBibitem
\bibitem[Negele and Orland(1988)Negele, and Orland]{Negele1988}
Negele,~J.; Orland,~H. \emph{Quantum Many-particle Systems}; Advanced Book Classics; Basic Books, 1988\relax
\mciteBstWouldAddEndPuncttrue
\mciteSetBstMidEndSepPunct{\mcitedefaultmidpunct}
{\mcitedefaultendpunct}{\mcitedefaultseppunct}\relax
\EndOfBibitem
\bibitem[Gelzinis \latin{et~al.}(2017)Gelzinis, Rybakovas, and Valkunas]{gelzinis2017applicability}
Gelzinis,~A.; Rybakovas,~E.; Valkunas,~L. Applicability of transfer tensor method for open quantum system dynamics. \emph{J. Chem. Phys.} \textbf{2017}, \emph{147}, 234108\relax
\mciteBstWouldAddEndPuncttrue
\mciteSetBstMidEndSepPunct{\mcitedefaultmidpunct}
{\mcitedefaultendpunct}{\mcitedefaultseppunct}\relax
\EndOfBibitem
\bibitem[Kananenka \latin{et~al.}(2016)Kananenka, Hsieh, Cao, and Geva]{kananenka2016accurate}
Kananenka,~A.~A.; Hsieh,~C.-Y.; Cao,~J.; Geva,~E. Accurate long-time mixed quantum-classical {L}iouville dynamics via the transfer tensor method. \emph{J. Phys. Chem. Lett.} \textbf{2016}, \emph{7}, 4809--4814\relax
\mciteBstWouldAddEndPuncttrue
\mciteSetBstMidEndSepPunct{\mcitedefaultmidpunct}
{\mcitedefaultendpunct}{\mcitedefaultseppunct}\relax
\EndOfBibitem
\bibitem[Fishman \latin{et~al.}(2022)Fishman, White, and Stoudenmire]{fishman2022itensor}
Fishman,~M.; White,~S.~R.; Stoudenmire,~E.~M. {The ITensor Software Library for Tensor Network Calculations}. \emph{SciPost Phys. Codebases} \textbf{2022}, 4\relax
\mciteBstWouldAddEndPuncttrue
\mciteSetBstMidEndSepPunct{\mcitedefaultmidpunct}
{\mcitedefaultendpunct}{\mcitedefaultseppunct}\relax
\EndOfBibitem
\bibitem[Makri(2020)]{makri2020smallMatrixDisentanglement}
Makri,~N. Small matrix disentanglement of the path integral: overcoming the exponential tensor scaling with memory length. \emph{J. Chem. Phys.} \textbf{2020}, \emph{152}\relax
\mciteBstWouldAddEndPuncttrue
\mciteSetBstMidEndSepPunct{\mcitedefaultmidpunct}
{\mcitedefaultendpunct}{\mcitedefaultseppunct}\relax
\EndOfBibitem
\bibitem[Kundu and Makri(2023)Kundu, and Makri]{kundu2023pathsum}
Kundu,~S.; Makri,~N. PathSum: A {C}++ and Fortran suite of fully quantum mechanical real-time path integral methods for (multi-) system+ bath dynamics. \emph{J. Chem. Phys.} \textbf{2023}, \emph{158}\relax
\mciteBstWouldAddEndPuncttrue
\mciteSetBstMidEndSepPunct{\mcitedefaultmidpunct}
{\mcitedefaultendpunct}{\mcitedefaultseppunct}\relax
\EndOfBibitem
\bibitem[Frenkel(1931)]{frenkel1931transformation1}
Frenkel,~J. On the transformation of light into heat in solids. I. \emph{Phys. Rev.} \textbf{1931}, \emph{37}, 17\relax
\mciteBstWouldAddEndPuncttrue
\mciteSetBstMidEndSepPunct{\mcitedefaultmidpunct}
{\mcitedefaultendpunct}{\mcitedefaultseppunct}\relax
\EndOfBibitem
\bibitem[Frenkel(1931)]{frenkel1931transformation2}
Frenkel,~J. On the transformation of light into heat in solids. II. \emph{Phys. Rev.} \textbf{1931}, \emph{37}, 1276\relax
\mciteBstWouldAddEndPuncttrue
\mciteSetBstMidEndSepPunct{\mcitedefaultmidpunct}
{\mcitedefaultendpunct}{\mcitedefaultseppunct}\relax
\EndOfBibitem
\end{mcitethebibliography}

\end{document}